\newif\ifcomment

\commentfalse  

\ifcomment
    \documentclass[useAMS,usenatbib,onecolumn]{mn2e}
    \usepackage{charter} 
\else
    \documentclass[useAMS,usenatbib]{mn2e}
    \usepackage{times} 
\fi

\usepackage{ifpdf}
\ifpdf
   \usepackage{graphicx}
   \usepackage{epstopdf}
\else
   \usepackage{graphicx}
\fi
\usepackage{color}
\usepackage[usenames,dvipsnames]{xcolor}
\usepackage{calc}
\usepackage{amssymb}
\usepackage{amsmath}
\usepackage{todonotes}
\usepackage[bookmarks=false]{hyperref}
\usepackage{syntonly}
\usepackage{aas_macros}
\bibpunct{(}{)}{;}{a}{}{;}

\title[Linking Stellar to Dark Matter]{Mapping stellar content to dark matter halos
using galaxy clustering and galaxy-galaxy lensing in the SDSS DR7}
\author[Zu \& Mandelbaum]{
\parbox{\textwidth}{
    Ying Zu\thanks{E-mail: yzu@cmu.edu} and Rachel Mandelbaum
}
\vspace*{4pt} \\
McWilliams Center for Cosmology, Department of Physics, Carnegie Mellon University, 5000 Forbes Avenue,
Pittsburgh, PA 15213, USA\\
}

\providecommand{\avg}[1]{\left\langle #1 \right\rangle}

\def\dd{\mathrm{d}}

\def\lcdm{\Lambda\mathrm{CDM}}
\def\hmsol{h^{-1}M_\odot}
\def\ms{M_*}
\def\mh{M_h}
\def\xigt{\xi_{gt}}
\def\xigg{\xi_{gg}}
\def\xigm{\xi_{gm}}
\def\ximm{\xi_{mm}}
\def\mpc{\mathrm{Mpc}}
\def\hmpccubed{h^{3}\mathrm{Mpc}^{-3}}
\def\hmpc{h^{-1}\mathrm{Mpc}}

\def\hkpc{h^{-1}\mathrm{kpc}}

\def\chod{\texttt{cHOD}}
\def\ihod{\texttt{iHOD}}
\def\sig{\sigma_{\ln\ms}}
\def\hhmsol{h^{-2}M_\odot}
\def\ds{\Delta\Sigma}
\def\kms{\mathrm{km}\,s^{-1}}

\ifcomment
\fi

\newlength{\figtextwidth}
\begin{document}

\ifcomment
    \fontsize{12pt}{12pt}\selectfont 
    \setlength{\figtextwidth}{\textwidth + 5.5cm}%
\else
    \setlength{\figtextwidth}{\textwidth + 0.0cm}%
\fi

\date{\today} \maketitle
\begin{abstract}
    The mapping between the distributions of the observed galaxy stellar mass and the underlying dark matter
    halos provides the crucial link from theories of large--scale structure formation to interpreting the
    complex phenomena of galaxy formation and evolution. We develop a novel statistical method, based on the
    Halo Occupation Distribution model~(HOD), to solve for this mapping by jointly fitting the galaxy
    clustering and the galaxy-galaxy lensing measured from the Sloan Digital Sky Survey~(SDSS). The method,
    called the {\ihod} model, extracts maximum information from the survey by including ${\sim}80\%$ more
    galaxies than the traditional HOD methods, and takes into account the incompleteness of the stellar
    mass samples in a statistically consistent manner. The derived stellar-to-halo mass relation not
    only explains the clustering and lensing of SDSS galaxies over almost four decades in stellar mass,
    but also successfully predicts the stellar mass functions observed in SDSS. Due to its capability
    of modelling significantly more galaxies, the {\ihod} is able to break the degeneracy between the
    logarithmic scatter in the stellar mass at fixed halo mass and the slope of the stellar-to-halo
    mass relation at high mass end, without the need to assume a strong prior on the scatter and/or
    use the stellar mass function as an input. We detect a decline of the scatter with halo mass, from
    $0.22_{-0.01}^{+0.02}$ dex at below $10^{12}\hmsol$ to $0.18\pm{0.01}$ dex at $10^{14}\hmsol$. The
    model also enables stringent constraints on the satellite stellar mass functions at fixed halo mass,
    predicting a departure from the Schechter functional form in high mass halos. The {\ihod} model can
    be easily applied to existing and future spectroscopic datasets, greatly improving the statistical
    constraint on the stellar-to-halo mass relation compared to the traditional HOD methods within the
    same survey.
\end{abstract}

\section{Introduction}
\label{sec:intro}

The distribution of galaxy stellar mass in the present--day Universe provides important clues to answering
fundamental questions concerning the cosmic assembly of baryons in the $\Lambda$-cold dark matter~($\lcdm$)
paradigm~\citep{fukugita1998, keres2005, faucher2011, dave2012}. In particular, what fraction of baryons are
condensed into stars as opposed to spreading out in the form of gas and dust in and outside of
galaxies~\citep{cen1999, mcgaugh2010, putman2012, shull2012}?
 Among those locked in stars, how much stellar mass is stored within the central galaxies of the dark matter
 halos~\citep{delucia2007, vonderlinden2007}, and how is the rest distributed among the satellite galaxies
 within their larger host halos~\citep{hansen2009, yang2009, leauthaud2012}?  In this paper we develop a novel
 statistical approach within the Halo Occupation Distribution~\citep[HOD;][]{jing1998, ma2000, peacock2000,
 seljak2000, yang2003, scoccimarro2001, cooray2002, berlind2002, guzik2002, zheng2005, mandelbaum2006,
     vandenbosch2007} framework, to solve for the mapping between the stellar mass content and the dark matter
halos using the spatial clustering and the weak gravitational lensing of the Sloan Digital Sky
Survey~\citep[SDSS;][]{york2000} spectroscopic sample galaxies. The inferred mapping not only explains the
galaxy auto-correlation function~(i.e., clustering) and the galaxy--matter cross-correlation~(i.e., lensing)
successfully, but also correctly predicts the observed stellar mass function~(SMF), placing strong constraints
on the physics that governs the formation and distribution of galaxies within halos.

The most comprehensive way to link the galaxy properties to the dark matter halos is to directly model the
expected physical processes involved in the formation and evolution of stars and gas~(along with metals and
dust), by either running hydrodynamic simulations~\citep[][]{hernquist1989, katz1996, norman1999,
    teyssier2002, oshea2004, springel2005, dimatteo2005, oppenheimer2006, booth2009, vogelsberger2013}
or semi-analytical models~(SAMs) along the halo merger trees of N-body simulations~\citep{baugh2006,
somerville2008, kang2005, bower2006, delucia2006}.  The primary advantage of the latter approach is that it is
computationally much less expensive compared to the hydrodynamic simulations, albeit with a large number of
free parameters.  However, even within the hydrodynamic simulations some of the key processes, like the star
formation from the collapse of molecular clouds, the stellar feedback from supernovae explosions and galactic
winds, and the impact of the active galactic nuclei~(i.e., the AGN feedback), are happening on scales well
below the resolution limit, so the treatment of the physics is often at the ``subgrid'' level, i.e., put in by
hand using empirical scaling relations~\citep[see][for a review]{benson2010}. Both the hydrodynamic
simulations and the SAMs nonetheless have enjoyed great success over the past decade, reproducing a wide range
of the observed galaxy properties~(see \citealt{oppenheimer2008}, \citealt{khandai2014}, and
\citealt{vogelsberger2014} for the latest hydrodynamic simulations and see \citealt{guo2011} for the recent
development in SAM), though still with several possible areas of improvement.  Most importantly, because the
simulated outputs are directly tied to the prescribed subgrid physics and its parameters, by confronting the
predictions of this ``forward modelling'' approach with new observations, we can constantly furnish our
understanding of the galaxy formation physics in a cosmological context, especially the baryonic feedback
mechanisms that help sculpt the galaxy SMF~\citep{fontanot2009}. However, the computational complexities of
hydrodynamic simulations and the SAM degeneracies make it difficult to come to definite conclusions about
galaxy formation theories in certain cases.

Alternatively, one can focus on the {\it statistical} link between just the stellar content and the dark
matter halos, assuming that the enormous diversity in the individual galaxy assembly histories inside halos of
the same mass would reduce to a stochastic scatter about the mean stellar-to-halo mass relation~(SHMR) by
virtue of the central limit theorem. The great success of the HOD framework in explaining the clustering and
lensing of galaxies, and their dependencies on galaxy properties like colour, luminosity, and stellar mass at
different redshifts~\citep[e.g.,][]{zehavi2011, parejko2013, guo2014}, further validates this assumption about
the statistical simplicity of the SHMR.  For a given cosmology, the HOD formalism describes the relationship
between the stellar and dark matter in terms of $\langle N_g(\mh)\rangle$, the average number of galaxies of a
given type~(e.g., central or satellite) within a halo of virial mass $\mh$, along with the spatial and
velocity distributions of galaxies within that halo~(see \citealt{yang2003} for a closely related approach,
the conditional luminosity function).  Furthermore, the HOD is potentially a powerful tool to constrain
cosmology~\citep{yoo2006, zheng2007, vandenbosch2013, cacciato2013, more2014}, by exploiting the average bias
vs. mass relation of the dark matter halos revealed by the galaxies they contain. The standard HOD modelling
is restricted to individual volume--limited samples, each defined by a stringent combination of stellar mass
and redshift cuts, which leaves many observed galaxies unused. Moreover, the corresponding HOD parameters are
inferred separately for each sample, therefore the SHMR is constrained at a disjunct set of loci where the
mean stellar masses of the samples landed.

Recently, \citeauthor{leauthaud2011}~(2011; hereafter L11) proposed a new HOD framework by further
parameterizing $\langle N_g(\mh)\rangle$ as continuous functions of the threshold stellar mass $\ms$, i.e.,
$\langle N_g({>}\ms|\mh)\rangle$. In particular, the mean SHMR of central galaxies and its scatter fully
specify the expected number of central galaxies of any stellar mass at fixed halo mass, whereas the number of
satellite galaxies above a certain stellar mass scales with halo mass in a way that is also halo
mass-dependent.  The advantage of this new framework lies in its unique capability to derive the connection
between galaxies and halos using multiple probes simultaneously within a single global HOD model.
\citeauthor{leauthaud2012-a}~(2012b, hereafter L12) demonstrated the efficacy of the L11 model by inferring the
SHMR across the entire observed stellar mass range, using the combination of the SMF, angular galaxy
clustering, and galaxy-galaxy (g-g) lensing of samples above some critical stellar mass limit~(corresponding
to high completeness) in the COSMOS survey. However, similar to the standard HOD technique, the L11 model
requires volume completeness in the stellar mass samples, thereby losing a lot of data in an intrinsically
flux-limited survey.  Our approach is based on the L11 global HOD framework, and (like in L11) can jointly fit
the galaxy clustering and g-g lensing signals of galaxy samples above some critical stellar mass limit, but
(unlike in L11) can take into account the incompleteness in a statistically self-consistent way.  As a result
of the added flexibility of our model, we are able to include significantly more SDSS galaxies at both lower
stellar mass and higher redshift, equivalent to almost doubling the survey volume.

Meanwhile, \citet[][hereafter M10]{moster2010} and other groups~\citep[e.g.,][]{kravtsov2004, vale2004,
conroy2006, shankar2006, behroozi2010, guo2010} adopted the so--called ``sub-halo abundance matching''~(SHAM)
technique to assign stellar masses or luminosities to individual dark matter halos~(including both main and
subhalos) in the N-body simulations. In its simplest form, the SHAM method assumes a monotonic relation
between the SMF estimated from galaxy surveys and the halo mass function predicted by the $\lcdm$.  Instead of
using the subhalo mass at the current epoch, most SHAM studies employed the ``infall mass'', i.e., the mass of
the subhalo before its accretion onto the main halo, and instead of a monotonic relation they either assumed
some fixed scatter or drew the scatter from external priors. Using the ``infall mass'' as a proxy for the
stellar mass, they found surprisingly good agreement between the predicted and the observed clustering and
lensing statistics.  The underlying assumption behind SHAM is that the satellite galaxies that live in
subhalos were central galaxies of their own halos before accretion. After becoming satellites, the dark matter
masses of their subhalos were reduced due to the combined effect of tidal stripping and dynamical friction,
whereas their stellar masses in the centre of those subhalos were much less affected and are thus more closely
tied to the infall masses~\citep{reddick2013, vandenbosch2005}.  \citet{hearin2013} later extended this
assumption to further relate the halo formation time to the colour of each galaxy to interpret the colour
dependence of the clustering and lensing statistics~\citep[also see][for an interesting proposal of a
two--parameter abundance matching scheme]{kulier2015}.  The key distinction between the SHAM and the HOD
models is that, by using the infall mass the SHAM method evades the need to parameterize the HOD of the
satellite galaxies~\citep[but see][on potential numerical issues in tracking subhalos]{guo2011}, which is
usually regarded as a nuance in traditional HOD modelling, at the expense of assuming the same SHMRs for the
central and the satellite galaxies.  However, as we will demonstrate in this study, the capability of our
modified HOD model to maximally extract the information from data enables us to place tight constraint on the
satellite HOD as well, thus shedding new insight into the formation and evolution of the satellite galaxies
without imposing the central SHMR on the satellites.  Moreover, the model breaks the degeneracy between the
scatter and the slope of the SHMR, self--consistently deriving the scatter without the need to assume fixed
values or external priors.

The paper is organised as follows. Section~\ref{sec:data} describes the SDSS data, including the large--scale
structure galaxy catalogue and the matched stellar mass catalogue, and the two sets of galaxy samples selected
for our analyses. In Section~\ref{sec:measurement}, we describe the methods we use to measure galaxy
clustering and g-g lensing for these samples. Section~\ref{sec:model} introduces our new variant of the HOD
model that we use to interpret the clustering and lensing signals, and Section~\ref{sec:pred} describes the
method to theoretically predict these signals using our HOD model.  In Section~\ref{sec:constraints} we
present the results of our constraint from a Bayesian framework via the new HOD modelling technique.  We
examine the SMF, the SHMR of the central galaxies, and the conditional SMFs of the satellites predicted by our
best-fit model in Section~\ref{sec:smc}. We summarise our main findings and discuss future applications of the
new HOD model in Section~\ref{sec:end}.

Throughout this paper, we assume a $\lcdm$ cosmology with $(\Omega_m, \Omega_{\Lambda}, \sigma_8, h)\,{=}\,(0.26,
0.74, 0.77, 0.72)$.  All the length and mass units in this paper are scaled as if the Hubble constant were
$100\,\kms\mpc^{-1}$. In particular, all the separations are co-moving distances in units of either $\hkpc$ or
$\hmpc$, and the stellar mass and halo mass are in units of $\hhmsol$ and $\hmsol$, respectively.  The halo
mass is defined by $\mh\,{\equiv}\,M_{200m}\,{=}\,200\bar{\rho}_m(4\pi/3)r_{200m}^3$, where $r_{200m}$ is the
corresponding halo radius within which the average density of the enclosed mass is $200$ times the mean matter
density of the Universe, $\bar{\rho}_m$.  For the sake of simplicity, $\ln x{=}\log_e x$ is used for the
natural logarithm, and $\lg x{=}\log_{10} x$ is used for the base-$10$ logarithm.

\section{Description of the Data}
\label{sec:data}

Here we describe the spectroscopic data used to define the clustering and the lens samples, and the imaging
data used to define the source samples for g-g lensing.

\subsection{SDSS Main Galaxy Sample and NYU-VAGC}
\label{subsec:mgs}

This study is based on the final data release of the SDSS~\citep[DR7;][]{abazajian2009}, which contains the
completed data set of the SDSS-I and the SDSS-II. The survey imaged a quarter of the sky using a drift-scan
camera~\citep{gunn1998} in five photometric bandpasses~\citep[$u$, $g$, $r$, $i$, $z$;][]{fukugita1996} to a
limiting magnitude of ${\simeq}22.5$ in the $r$ band. The imaging data were photometrically and
astrometrically calibrated~\citep{padmanabhan2008}, and from this imaging data targets were selected for
spectroscopic follow-ups with a fibre--fed double spectrograph~\citep{gunn2006}. One of the spectroscopic
products is ``the main galaxy sample''~\citep[MGS;][]{strauss2002} that we use in this study as both the
tracers of stellar mass clustering and the lenses of background sources. In particular, we obtain the MGS data
from the \texttt{dr72} large--scale structure sample \texttt{bright0} of the ``New York University Value Added
Catalogue''~(NYU--VAGC), constructed as described in~\citet{blanton2005}. The \texttt{bright0} sample includes
galaxies with $10{<}m_r{<}17.6$, where $m_r$ is the $r$-band Petrosian apparent magnitude, corrected for
Galactic extinction. We choose a more relaxed bright limit than the commonly used \texttt{safe0}
sample~($14.5{<}m_r{<}17.6$) to allow higher completeness on the high stellar mass end, where constraint on
the mapping between galaxies and halos is particularly lacking.

Due to the finite size of the fibre plugs, no two targets on the same plate can be closer than $55''$,
resulting in ${\sim}7\%$ of the MGS galaxies with unknown redshifts.
 A simple remedy for these fibre collisions, as used in the
\texttt{bright0} sample, is to assign these galaxies the redshifts of their nearest neighbours on the sky, thus
exactly preserving the angular clustering signal. For our purpose of measuring the projected correlation $w_p$
at a given redshift $z$, it would overestimate the clustering signal below the projected physical scale
corresponding to the fibre size at that redshift, $r_p^{\mathrm{fb}}(z)$, by including physically distant
pairs of galaxies that happen to align within $55''$ on the sky. Above $r_p^{\mathrm{fb}}(z)$, however, this
``nearest--neighbour'' redshift assignment scheme recovers the underlying $w_p$ remarkably
well~\citep{zehavi2005}.  We thus limit our $w_p$ measurement to those scales above. We do not consider other
more sophisticated corrections that would potentially recover the signals below the fibre
radius~\citep{guo2012}, as $r_p^{\mathrm{fb}}$ is always below $0.17\hmpc$ across the redshift range of our
samples ~($z_{\mathrm{max}}{=}0.30$).  Furthermore, even at the highest redshifts where $r_p^{\mathrm{fb}}$ is
relatively large, the samples are progressively dominated by more massive galaxies that have a larger
correlation length, so the one-halo term is still well resolved in the clustering measurement above
$r_p^{\mathrm{fb}}$. For the g-g lensing measurement, this correction using the nearest neighbour slightly
blurs the signals around those ``collided'' lenses in the transverse direction because of the inaccurate
conversion from angles to physical separations.  This blurring effect can be safely ignored in our analysis as
it does not reduce the overall amplitude of the average lensing signal.

In addition to the angular positions and redshifts of the galaxies, the NYU--VAGC also calculated
k--corrections for all the MGS galaxies using templates based on stellar population synthesis~(SPS) models,
while providing approximate stellar mass estimates and star formation histories for all the \texttt{bright0}
galaxies~\citep{blanton2007}. Comparison with the stellar mass estimates from \citet{kauffmann2003} shows a
scatter of only $0.1$ dex and mean biases of less than $0.2$ dex depending on stellar mass. This tight
relationship, as we show later in Section~\ref{subsec:mstar}, provides crucial information for estimating
stellar masses for galaxies that do not have valid entries in the MPA/JHU stellar mass catalogue (an updated
version of the \citealt{kauffmann2003} catalogue) that we adopted.

To minimise any potential artefacts in the measurements due to irregular survey geometry and areas with low
spectroscopic coverage, we use data exclusively within the contiguous area in the North Galactic Cap and from
sectors~(each defined by a region covered by a unique set of plates) where the angular completeness is greater
than $0.8$.  The final sample used for the galaxy clustering analysis includes $513{,}150$ galaxies over a sky
area of $6395.49$ deg$^2$. A further 5 per cent of the area is eliminated for the lensing analysis, due to the
absence of source galaxies in that area (for example, because of poor imaging quality or more conservative
masking around bright stars).

\subsection{Stellar Mass Estimate}
\label{subsec:mstar}

We employ the stellar mass estimates from the latest MPA/JHU value-added galaxy
catalogue\footnote{\url{http://home.strw.leidenuniv.nl/~jarle/SDSS/}}.  The stellar masses were estimated
based on fits to the SDSS photometry following the philosophy of~\citet{kauffmann2003} and~\citet{salim2007},
and assuming the Chabrier~\citep{chabrier2003} initial mass function~(IMF) and the~\citet{bruzual2003} SPS
model. The estimation is very similar to the spectroscopic fits as described in~\citet{kauffmann2003}, but
instead of spectral features like the D4000 index and the H$\delta$ absorption lines, the broad-band
photometry~(after correction for emission lines) is used for the fits. The key difference between the
photometry--based estimates and the spectroscopic ones is that the stellar mass-to-light ratio~($\ms/L$)
estimated from spectroscopic features is only representative for the region sampled by the fibre, and a
constant $\ms/L$ must then be assumed to extrapolate to a total stellar mass. To avoid extrapolation, we
employ the total mass estimates obtained through direct fits to the SDSS total photometry, i.e., the
\texttt{cmodel} magnitudes.

We then match the MPA/JHU stellar mass catalogue to the NYU-VAGC \texttt{bright0} sample and identify valid,
unambiguous MPA/JHU stellar mass estimates for all but $32{,}327$~($6.3\%$) of the MGS galaxies.  Although it is
unclear what causes the failure in estimating stellar mass for these systems, they are statistically
indistinguishable from the matched population in redshift, luminosity, flux, and colour. Therefore, given the
tight scaling relationship between the NYU-VAGC stellar mass $M_*^{N}$ and the MPA/JHU mass $M_*^{M}$, we can
recover the individual photometric stellar mass estimates for these unmatched galaxies in a way that is
statistically consistent with the matched ones. We first divide the \texttt{bright0} galaxies into red and
blue populations based on their $g{-}r$ colours~($>$ or $<$ $0.8$; k-corrected to $z{=}0.1$), because the scaling
between the two types of estimates is colour--dependent and is much tighter within each single--colour
population than the entire sample~\citep{bell2001}.  Within each colour, we then compute the probability
distribution of $M_*^M$ at each fixed $M_*^N$ using the successfully matched galaxies, and assign each
unmatched galaxy with $M_*^N$ a random stellar mass $\ms^M$ drawn from that distribution.

Stellar mass estimation is subject to various theoretical uncertainties in predicting the $\ms/L$, due to the
choice of the SPS model, dust extinction law, stellar evolution model, and most importantly, the form of the
IMF~(see \citealt{conroy2013} for a general review on these topics and L12 for a detailed discussion of the
various systematic uncertainties in stellar mass estimation).  Assuming no trend in IMF with galaxy type,
environment, or redshift\footnote{For evidence suggesting non-universal IMFs, however,
see~\citet{vandokkum2010}, \citet{conroy2012}, \citet{cappellari2012}, and \citet{ferreras2013}.}, the range
of different IMFs causes uncertainty in the absolute normalisation of the $\ms/L$ of factors of two to several
depending on the passband~\citep{bell2001}, thus shifting the inferred SMF horizontally while maintaining the
same shape. The second major source of uncertainty lies in the difficulty in correctly measuring the total
fluxes of individual galaxies from aperture photometry. The SDSS pipeline returns a fraction of the total
light even when the \texttt{cmodel} magnitudes are used~\citep{abazajian2009}, due to a combination of
over-subtraction of the sky background and poor model fits to the light profiles~\citep{bernardi2013}. This
issue is particularly acute for massive galaxies, where the measured number density of galaxies above
$3\times10^{11}\hhmsol$ may be underestimated by as much as a factor of five.  Unfortunately, there is no
well-calibrated correction that can be applied to the MPA/JHU catalogue.  Therefore, we use the MPA/JHU
catalogue as-is and focus on the particular mapping between the MPA/JHU stellar masses to dark matter halos.
For the purpose of this mapping, both types of systematic errors on the stellar mass are rather benign, as
long as the ranking order of galaxies in their stellar mass estimates is largely unperturbed.  Unless
otherwise noted, we will refer to the MPA/JHU stellar mass estimate simply as the ``stellar mass'' and denote
it with $M_*$ for the rest of the paper.

\subsection{Stellar Mass Sample Selection}
\label{subsec:sample}

\begin{figure*}
    \centering\resizebox{0.95\figtextwidth}{!}{\includegraphics{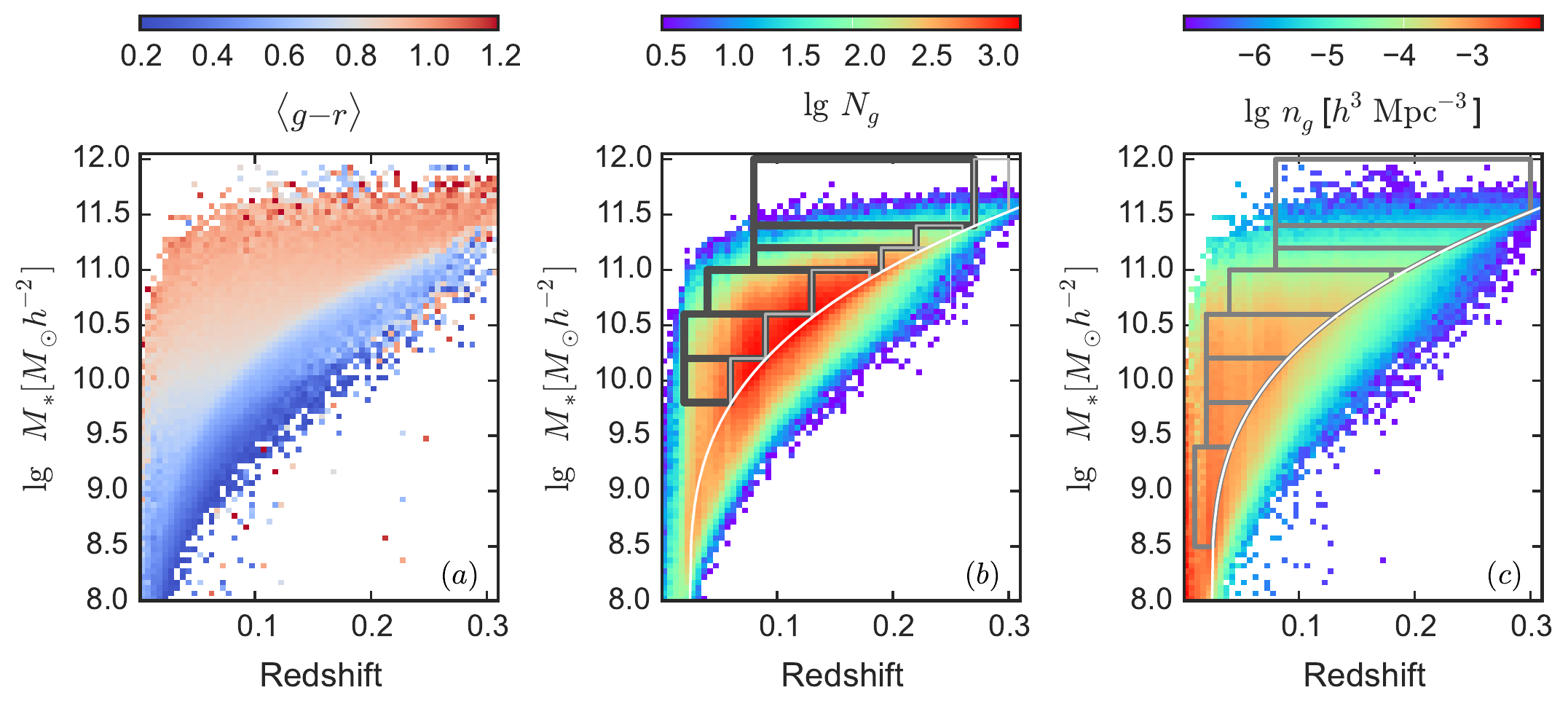}} \caption{Distributions
    of the mean $g-r$ rest--frame colour~(panel $a$), the observed number counts~(panel $b$), and the co-moving
    number density~(panel $c$) of SDSS DR7 galaxies on the stellar mass vs.\ redshift plane, with colourbars
    showing on the top. In the middle panel, the thick ``boxes'' indicate the galaxy samples selected for the
    traditional HOD analysis~({\chod}), while the thin ``wedges'' represent the omitted region where galaxies
    are the most abundant. The gray regions in the right panel are the galaxy samples that were used for our
    fiducial {\ihod} analysis. The curved lower boundary of the selection is determined by the mixture limit
    below which the red galaxy population is severely underrepresented in the spectroscopic survey, and is
    derived from the distinctive white stripe~($\langle g-r\rangle=0.8$) in the left panel~(see
    Equation~\ref{eqn:ql}).}
\label{fig:mz}
\end{figure*}

\begin{table*}
\centering \caption{The two sets of stellar mass bins used for the {\ihod} and {\chod} analyses, corresponding
    to the dark and gray thick selections in Figure~\ref{fig:mz}b and~\ref{fig:mz}c, respectively.  The
    {\ihod} analysis includes two extra stellar mass samples below $\lg\ms{=}9.8$, while the higher stellar
    mass samples have the same binning in $\lg\ms$ and the same minimum redshifts $z_{\mathrm{min}}$ for the
    two analyses. However, all the {\ihod} samples have more extended redshift ranges, thresholded by the
    mixture limit defined in Equation~\eqref{eqn:ql} on the far side, so the maximum redshift
    $z_{\mathrm{max}}$ listed for each sample is the maximum redshift of any galaxies in that sample.  For the
    $z_{\mathrm{max}}$ and $N_g$ we list the corresponding numbers for the {\chod} analysis in parenthesis.
    The average large-scale bias for all, central, and satellite galaxies, the average logarithmic halo mass
    corresponding to the central
    galaxies, and the satellite fraction of each {\ihod} sample~(with $68\%$ uncertainties) are derived from
    the fiducial {\ihod} constraint for the corresponding {\ihod} samples.
}
\begin{tabular}{ccccccccc}
\hline
\hline
$\lg\ms/\hhmsol$  & $z_{\mathrm{min}}$   & $z_{\mathrm{max}}$   & $N_g$  & $f_{\mathrm{sat}}$  &
$\avg{b_{\mathrm{all}}}$  &
$\avg{b_{\mathrm{sat}}}$ & $\avg{b_{\mathrm{cen}}}$  & $\avg{\lg\mh^{\mathrm{cen}}/\hmsol}$ \\
\hline
8.5-9.4   & 0.01       &0.04        & $13{,}616$              & $0.42_{-0.04}^{+0.04}$ & $1.12_{-0.02}^{+0.02}$ & $1.63_{-0.07}^{+0.07}$ & $0.74_{-0.02}^{+0.02}$ & $11.16_{-0.20}^{+0.16}$  \\
9.4-9.8   & 0.02       &0.06        & $16{,}247$              & $0.42_{-0.04}^{+0.03}$ & $1.15_{-0.02}^{+0.01}$ & $1.66_{-0.06}^{+0.06}$ & $0.77_{-0.01}^{+0.01}$ & $11.44_{-0.10}^{+0.09}$  \\
9.8-10.2  & 0.02       &0.09~(0.06) & $46{,}910$~($22{,}409$) & $0.42_{-0.03}^{+0.03}$ & $1.19_{-0.01}^{+0.01}$ & $1.71_{-0.05}^{+0.05}$ & $0.82_{-0.01}^{+0.01}$ & $11.74_{-0.06}^{+0.05}$  \\
10.2-10.6 & 0.02       &0.13~(0.09) & $96{,}946$~($58{,}209$) & $0.37_{-0.02}^{+0.02}$ & $1.26_{-0.01}^{+0.01}$ & $1.83_{-0.04}^{+0.04}$ & $0.93_{-0.01}^{+0.01}$ & $12.15_{-0.04}^{+0.03}$  \\
10.6-11.0 & 0.04       &0.18~(0.13) & $102{,}307$~($60{,}283$)& $0.26_{-0.01}^{+0.01}$ & $1.40_{-0.01}^{+0.01}$ & $2.11_{-0.04}^{+0.05}$ & $1.15_{-0.01}^{+0.01}$ & $12.68_{-0.03}^{+0.02}$  \\
11.0-11.2 & 0.08       &0.22~(0.19) & $24{,}908$~($19{,}506$) & $0.17_{-0.01}^{+0.01}$ & $1.73_{-0.01}^{+0.01}$ & $2.65_{-0.06}^{+0.07}$ & $1.54_{-0.02}^{+0.02}$ & $13.21_{-0.02}^{+0.02}$  \\
11.2-11.4 & 0.08       &0.26~(0.22) & $10{,}231$~($7{,}427$)  & $0.11_{-0.01}^{+0.02}$ & $2.13_{-0.02}^{+0.02}$ & $3.27_{-0.08}^{+0.10}$ & $1.99_{-0.03}^{+0.02}$ & $13.58_{-0.02}^{+0.02}$  \\
11.4-12.0 & 0.08       &0.30~(0.27) & $3{,}137$~($2{,}649$)   & $0.05_{-0.01}^{+0.01}$ & $2.84_{-0.06}^{+0.06}$ & $4.28_{-0.13}^{+0.17}$ & $2.76_{-0.06}^{+0.06}$ & $13.96_{-0.03}^{+0.03}$  \\
\hline
\end{tabular}
\label{tab:smbins}
\end{table*}
The flux--limited nature of surveys like the SDSS makes it very difficult to select volume--limited galaxy
samples thresholded or binned in stellar mass, due to the large spread in $\ms/L$ at fixed $\ms$.  However,
the $\ms/L$ distribution of the red, quiescent galaxies is much narrower than that of the blue, star--forming
galaxies~\citep{gallazzi2009}, therefore at any given redshift the observed red population has a sharp cutoff
at low stellar mass, where the observed galaxies are dominated by the blue population near the flux limit.
This phenomenon is best illustrated in Figure~\ref{fig:mz}a, where we show the average $g{-}r$ colour at each
($\ms$, $z$) location, colour--coded by the colourbar on top. The map is well separated into two regimes based
on the mixing of galaxy colours by the narrow streak of the $\langle g{-}r \rangle=0.8$ population, which we
hereafter refer to as the ``mixture'' limit, analogous and related to the ``flux'' limit on a luminosity vs.\
$z$ diagram.

In Figure~\ref{fig:mz}a, above the mixture limit the variation of the average $g{-}r$ colour with $\ms$ is
largely uniform across all redshifts below $0.1$, with a gradual and smooth transition at
$\ms\sim10^{10}\hhmsol$ as a result of the galaxy evolution physics~\citep[a.k.a, downsizing;][]{cowie1996}.
The sharp colour transition across the mixture limit, however, is purely a manifestation of the flux-limited
nature of the sample, combined with the fact that quiescent galaxies produce much less light than their
star-forming counterparts at the same $\ms$ due to their higher $\ms/L$ ratio. The sharpness of the transition
corresponds to the narrow width of the $\ms/L$ distribution of these quiescent galaxies.  Therefore, the
mixture limit provides a simple empirical guideline for selecting stellar mass samples of relatively high
volume-completeness, circumventing the problem of theoretically determining the maximum $\ms/L$ ratio for
galaxies at the flux limit of each redshift.

Since our goal is to infer the stellar-to-halo mass mapping for the average galaxy population with a fair
mix of both the quiescent and the star--forming galaxies, which occupy dark matter halos in different ways,
we will restrict our analysis to the galaxies above the ``mixture'' limit~(see Section~\ref{subsec:cvsi}
for an additional factor for making this choice of restriction).  The functional form we adopt to describe
the mixture limit $M_*^{\mathrm{mix}}(z)$ is
\begin{equation}
    \lg\left(\frac{M_*^{\mathrm{mix}}}{\hhmsol}\right) = 5.4 \times (z-0.025)^{0.33} + 8.0, \label{eqn:ql}
\end{equation}
i.e., the white curves shown in Figure~\ref{fig:mz}b and Figure~\ref{fig:mz}c, slightly more conservative than
the $\langle g{-}r \rangle{=}0.8$ streak in Figure~\ref{fig:mz}a. We then define different stellar mass
samples above this mixture limit for the clustering and lensing measurements in Figure~\ref{fig:mz}b
and~\ref{fig:mz}c, which show the distributions of galaxy number counts and number densities, respectively, on
the $\ms$--$z$ diagram. The thick black boxes in Figure~\ref{fig:mz}b represent a typical sample selection
scheme adopted by the traditional HOD analysis. In order to predict the clustering and lensing signals, the
HOD model has to assume either volume--completeness of the sample within those boxes, or an ad hoc
prescription describing the completeness as function of $\ms$ and/or $z$~\citep[e.g.,][]{miyatake2013}.  Both
assumptions are less than ideal due to our ignorance about the $\ms/L$ ratio distribution of galaxies.
Furthermore, the rectangular shape of the selections, imposed merely for the sake of modelling convenience,
inevitably misses the regions where the galaxies are the most abundant~(thin gray ``wedges'') on the 2D
histogram because they occupy a much larger co-moving volume per unit redshift compared to the lower redshift
``boxes''.  In principle the traditional HOD analysis can include more galaxies inside those ``wedges'' by
adopting finer stellar mass bins with fewer galaxy per sample, but the measurement signal from each individual
sample would be very noisy, rendering this scheme highly impractical.  In this study we develop two novel
improvements over the traditional HOD approach: 1) to statistically account for the sample incompleteness in a
self--consistent way, and 2) to be able to predict the clustering and lensing signal for all the galaxies
above the mixture limit~(thick gray selections in Figure~\ref{fig:mz}c).  We hereafter refer to the
traditional HOD approach as the {\chod} and our improved version the {\ihod}, where \texttt{c} and \texttt{i}
are loosely tied to ``completeness'' and ``incompleteness'', respectively.  Table~\ref{tab:smbins} summarises
the basic information of the two sets of
sample selections used by the two modelling methods. In total, we select $314{,}302$~($61\%$ of the
\texttt{bright0} sample) galaxies for the {\ihod} analysis, and from them $170{,}483$~($54\%$ of the {\ihod}
galaxies) are used for the {\chod} analysis.

\subsection{Source catalogue}

As sources for the g-g lensing measurement, we use a catalogue of background galaxies
\citep{2012MNRAS.425.2610R} with a number density of 1.2 arcmin$^{-2}$ with weak lensing shears estimated
using the re-Gaussianization method \citep{2003MNRAS.343..459H} and photometric redshifts from Zurich
Extragalactic Bayesian Redshift Analyzer \citep[ZEBRA,][]{2006MNRAS.372..565F}.   The catalogue was
characterised in several papers that describe the data, and use both the data and simulations to estimate
systematic errors \citep[see ][]{2012MNRAS.425.2610R,2012MNRAS.420.1518M,2012MNRAS.420.3240N,mandelbaum2013}.

\section{Measuring Galaxy Clustering and Galaxy-Galaxy Lensing}
\label{sec:measurement}

\begin{figure}
    \centering\resizebox{0.48\figtextwidth}{!}{\includegraphics{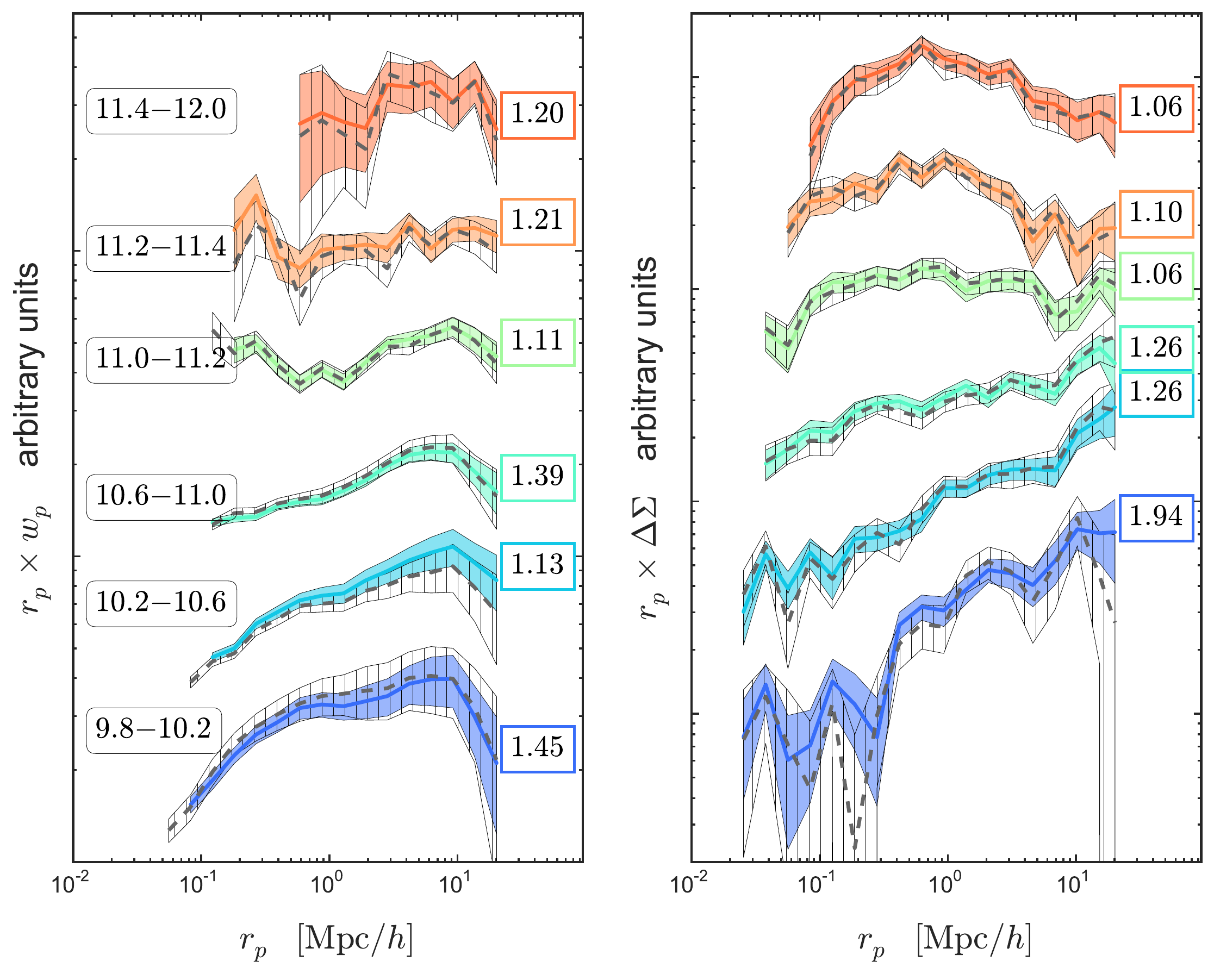}} \caption{Comparison
    between the clustering~(left) and lensing~(right) measurements for the {\ihod}~(coloured solid) and
    {\chod}~(gray dashed) samples of the same stellar mass ranges, marked at the beginning of each pair of
    curves in the left panel.  In each panel, the coloured and hatched bands represent the sizes of the
    uncertainties in the corresponding measurements, and the number marked at the end of each pair of bands is
    the ratio between the average signal-to-noise ratios of the {\ihod} and the {\chod} measurements, all
    ${>}1$ because of the extra galaxies included in the {\ihod} samples. To avoid clutter, the units of the
    projected correlation function~(left) and the g-g lensing signals~(right) for each stellar mass bin are
    scaled arbitrarily.
}\label{fig:compsn}
\end{figure}

Figure~\ref{fig:compsn} compares the clustering and g-g lensing signals measured for the six pairs of
{\ihod}~(solid) and {\chod}~(dashed) samples that share the same stellar mass ranges~(marked in the left
panel). In each panel, the coloured and the hatched bands illustrate the sizes of the measurement
uncertainties~(i.e., the diagonal of the error matrices) in the {\ihod} and the {\chod} cases, respectively.
At the end of each pair of uncertainty bands, we mark the ratio between the average signal-to-noise
ratios~($S/N$) of the {\ihod} and the {\chod} measurements. The improvement in the $S/N$ of the g-g lensing
measurements~(right panel) is consistent with the reduction in the Poisson errors by factors of
$\sqrt{N_g^i/N_g^c}$, where $N_g^{i}$ and $N_g^{c}$ are the numbers of galaxies in the {\ihod} and the {\chod}
samples, respectively~(the 4th column of Table~\ref{tab:smbins}). For the clustering measurements~(left
panel), the improvement for the three higher mass bins is consistent with the expected decrease in the Poisson
error by $N_g^i/N_g^c$, while for the three lower mass bins the improvement is smaller than expected from
assuming the errors are pure Poisson, due to the strong cosmic variance in these samples. These improvements
in the $S/N$ of the measurements are crucial to the success of the {\ihod} model in robustly solving the
mapping between galaxies and halos without resorting to any external information~(e.g., the SMF) or strong
priors on the model parameters~(e.g., the scatter in the SHMR).

We now describe how we measure the projected correlation function from galaxy pair counting and the surface
density contrast from g-g lensing, and the reader who are familiar with the technical details can skip the
remainder of this section.

\subsection{Projected Galaxy Correlation Function}
\label{subsec:wpdata}

We measure the projected correlation function $w_p$ for each galaxy sample by integrating the 2D
redshift--space correlation function $\xi^s$,
\begin{equation}
    w_p(r_p) = \int_{-r_{\pi}^{\mathrm{max}}}^{+r_{\pi}^{\mathrm{max}}} \xi^{s}(r_p, r_\pi) d r_\pi,
    \label{eqn:wp}
\end{equation}
where $r_p$ and $r_\pi$ are the projected and the line-of-sight~(LOS) co--moving distances between two
galaxies. We measure the $w_p$ signal out to a maximum projected distance of $r_p^{\mathrm{max}}=20\hmpc$,
where the galaxy bias is approximately linear. For the integration limit, we adopt a maximum LOS distance of
$r_{\pi}^{\mathrm{max}}{=}60\hmpc$, where the average pairwise peculiar velocity is very
small~\citep{tinker2008}. Based on a mildly modified linear Kaiser formalism, \citet{vandenbosch2013} showed
that the $w_p$
measured with $r_{\pi}^{\mathrm{max}}{=}60\hmpc$ should be boosted by a few to $8\%$ depending on $r_p$,
in order to correct for the residual redshift-space distortion~(RRSD) effect due to the small but non-zero
peculiar velocity beyond $r_{\pi}{>}60\hmpc$~(see their figure 6). The boost was calibrated for a mock galaxy
sample binned in luminosity with different cosmology, therefore it is not suitable to directly apply it to our
measurement. However, the impact of the RRSD effect on our result, especially on the SHMR, can be roughly
estimated as follows: in the most extreme case that all the constraining power comes from the clustering at
large scales, based on the average bias vs. halo mass relation for the central galaxies listed in
Table~\ref{tab:smbins}, we can
infer that a $5\%$ boost in $w_p$ translates to $2.5\%$ in galaxy bias, which translates to
to ${\sim}0.015$ dex in halo mass, shifting the SHMR toward the lower mass end by $0.015$ dex. This shift is
much smaller than the scatter in the halo mass at fixed stellar mass and our constraint is not solely coming
from $w_p$, therefore the impact of ignoring RRSD should be negligible.

We compute the 2D correlation using
the Landy--Szalay estimator~\citep{landy1993},
\begin{equation}
    \xi^s(r_p, r_\pi) = \frac{DD - 2DR + RR}{RR},
    \label{eqn:landy}
\end{equation}
where $DD$, $DR$, and $RR$ represent the number counts of pairs of two data galaxies, one data and one random
galaxies, and two random galaxies, respectively. The Landy--Szalay estimator has minimal variance and is
insensitive to the number of random points~\citep{kerscher2000}.
For each given galaxy, we find all the neighbouring galaxies within a cylinder of radius $r_p^{\mathrm{max}}$ and
height $2 r_\pi^{max}$ centered on that galaxy, using the \texttt{libkd} package within the
\texttt{Astrometry.net} software~\citep{lang2010}. We then count all the pairs in each cell centered on $(r_p,
r_\pi)$ and the three sets of pair counts thus directly give the value of $\xi^s(r_p, r_\pi)$ via
Equation~\eqref{eqn:landy}.

The error covariance matrix for each $w_p$ measurement is estimated via the jackknife resampling
technique~\citep{norberg2009}. We divide the entire footprint into $200$ spatially contiguous, roughly
equal--size patches on the sky and compute the $w_p$ for each of the $200$ jackknife subsamples by leaving out
one patch at a time. For each stellar mass sample, we adopt the sample mean of the $200$ subsample
measurements as our final estimate of $w_p$, and the sample covariance matrix as an approximate to the
underlying error covariance.

The random galaxy catalogue was constructed following two steps. We first generate a sample of random
positions on the sky with $10$ times the size of the data catalogue, using the \texttt{mangle}
software~\citep{hamilton2004, swanson2008} and
the \texttt{bright0} angular selection function provided on the NYU--VAGC website. Secondly, we calculate the
2D joint distribution of stellar mass and redshift from the data catalogue, and draw ${5131,500}$ random pairs
of $(\ms, z)$ values to assign to the stored random positions. In this way, we ensure the random galaxy
catalogue has exactly the same angular, radial, and stellar mass joint selection functions as the data
galaxies. We apply the same sample selection criterion to the random catalogue and the data
sample. As mentioned in Section~\ref{subsec:mgs}, we only use the $w_p$ values down to the
physical distance that corresponds to the fibre radius at the maximum redshift of that sample.

The $w_p$ measurements for the eight {\ihod} stellar mass samples are shown as the solid circles with
errorbars in the top sub-panels in Figure~\ref{fig:datavspred}. The errorbars reflect the diagonal components
of the jackknife covariance matrices. Due to the strong cosmic variance effects in the low stellar mass
samples~($\lg\ms{<}10.6$), overall the off-diagonal components~(not shown here) are strong and persist on both
small and larges scales, while for the high stellar mass ones~($\lg\ms{>}11.0$) they are only prominent on
scales larger than $5\hmpc$ and between two adjacent distance bins, i.e., along the diagonal blocks.  We will
refer back to Figure~\ref{fig:datavspred} and discuss in more detail the comparison between the measurements
and the predictions from our best-fit model in Section~\ref{sec:constraints}.

\subsection{Surface Density Contrast from Galaxy--Galaxy Lensing}

Here we describe how we measure the surface density contrast from g-g lensing.  The lensing
measurement begins with identification of background source galaxies around each lens
 (with photometric redshift
larger than the lens spectroscopic redshift).  Inverse variance weights are assigned to each lens-source pair,
including both shape noise and measurement error terms in the variance:
\begin{equation}
w_{ls} = \frac{1}{\Sigma_{\rm crit}^{2}(\sigma_e^2 + \sigma_{\rm SN}^2)},
\label{eqn:wls}
\end{equation}
where $\sigma_e^2$ is the shape measurement error due to pixel
  noise, and $\sigma_{\rm SN}^2$ is the RMS intrinsic ellipticity (both quantities are per component, rather
  than total; the latter is fixed to $0.365$ following \citealt{2012MNRAS.425.2610R}).
$\Sigma_{\rm crit}$ is the critical surface mass density defined by
\begin{equation}
\Sigma_{\rm
crit}^{-1}(z_l,z_s)\equiv \frac{4\pi G}{ c^{2}}\frac{D_{ls}D_l(1+z_l)^2}{D_s},
\end{equation}
where $D_l$ and $D_s$ are the angular diameter distances to lens and source, and $D_{ls}$ is the distance
between them.  We use the estimated photometric redshift each source to compute $D_s$ and $D_{ls}$.  The
factor of $(1+z_l)^2$ comes from our use of co-moving coordinates.

The projected mass density  in each radial  bin can be computed via a summation over lens-source pairs
``$ls$'' and random lens-source pairs ``$rs$'':
\begin{equation}\label{eq:ds-estimator}
\Delta\Sigma(r_p) = \frac{\sum_{ls} w_{ls} e_t^{(ls)}
\Sigma_{{\rm crit}}(z_l,z_s)}{2 {\cal
    R}\sum_{rs} w_{rs}},
\end{equation}
where $e_t$ is the tangential ellipticity component of the source galaxy with respect to the lens position,
the factor of $2{\cal R}$ converts our definition of ellipticity to the tangential shear $\gamma_t$, and $r_p$
is the co-moving projected radius from the lens.  The division by $\sum w_{rs}$ accounts for the fact that some
of our `sources' are physically associated with the lens, and therefore not lensed by it \citep[see,
e.g.,][]{2004AJ....127.2544S}.  Finally, we subtract off a similar signal measured around random lenses, to
subtract off any coherent systematic shear contributions \citep{2005MNRAS.361.1287M}; this signal is
statistically consistent with zero for all scales used in this work.

To calculate the error bars, we also used the jackknife resampling method.  As shown in
\cite{2005MNRAS.361.1287M}, internal estimators of error bars (in that case, bootstrap rather than jackknife)
perform consistently with external estimators of errorbars for $\Delta\Sigma$ due to its being dominated by
shape noise.

Use of photometric redshifts, which have nonzero bias and significant scatter, gives rise to a bias in the
signals that can be easily corrected using the method from \cite{2012MNRAS.420.3240N}.  This bias is a
function of lens redshift, and is properly calculated including all weight factors for each lens sample taking
into account its redshift distribution.  For typical lens samples in this work, the bias for which we apply a
correction is of order 1 per cent, far below the statistical errors; the maximum is slightly below 10 per
cent.

\section{Model for Mapping the Stellar Content to Halos}
\label{sec:model}

\subsection{A Tale of Two HODs: {\chod} vs. {\ihod}}
\label{subsec:cvsi}

The key to statistically solving the mapping between stellar content and dark matter halos is $p(\ms, \mh)$,
the 2D joint probability density distribution of a galaxy with stellar mass $\ms$ sitting in a halo of mass
$\mh$, normalised so that $\iint p(\ms, \mh)\,\dd\ms\,\dd\mh{=}1$.  For analysing a galaxy sample with stellar
mass range $[\ms^0, \ms^1]$ and redshift range $[z_0, z_1]$, it is common practice for HOD models to directly
parameterize the occupation number as function of halo mass for the entire sample, $\avg{N_g(\mh)}$, which is
related to $p(\ms, \mh)$ via
\begin{multline}
    \avg{N_g(\mh)} = \frac{n_g}{z_1 - z_0} \left(\frac{\mathrm{d}n}{\mathrm{d}\mh}\right)^{-1}\\\int_{z_0}^{z_1}\int_{\ms^{0}}^{\ms^{1}} p(\ms, \mh)
    f_{\mathrm{obs}}(\ms | \mh, z) \,\mathrm{d}z \,\mathrm{d}\ms,
    \label{eqn:nmh}
\end{multline}
where $\mathrm{d}n/\mathrm{d}\mh$ is the halo mass function, $n_g$ is the total galaxy number density in the
Universe~(both observed and unobserved), and $f_{{\mathrm{obs}}}$ is the detection rate that varies between
$0$ and $1$. At any given redshift, the detection rate can be explicitly written as
\begin{equation}
    f_{{\mathrm{obs}}}(\ms | \mh, z)
    = \int_{0}^{\Gamma_{\mathrm{max}}{\equiv}\ms/L_{{\mathrm{min}}}(z)}g(\Gamma | \ms, \mh)
    \,\mathrm{d}\Gamma, \label{eqn:fobs}
\end{equation}
where $g(\Gamma|\ms, \mh)$ is the distribution of the stellar mass-to-light ratio $\Gamma{\equiv}\ms/L$ of
galaxies at fixed $\ms$ within halos of mass $\mh$, and $L_{{\mathrm{min}}}(z)$ is the luminosity threshold
corresponding to the flux limit at $z$.  At low redshift, $L_{\mathrm{min}}(z)$ is low~(large
$\Gamma_{\mathrm{max}}$), so the integral extends over a wide range of $\Gamma$ values and gives a result that
approaches $1$, while at higher redshift, $L_{\mathrm{min}}(z)$ is high~(small $\Gamma_{\mathrm{max}}$),
limiting the range of accessible $\Gamma$ values severely and thus lowering $f_{\mathrm{obs}}$.

In order to describe the sample using a single HOD, $f_{{\mathrm{obs}}}$ must be uniform across the
redshift range, i.e., $f_{{\mathrm{obs}}}(\ms |\mh, z) \equiv f_{\mathrm{obs}}(\ms | \mh)$, which happens
{\it if and only if} $f_{\mathrm{obs}}(\ms | \mh)\equiv 1$, so that
\begin{equation}
    \avg{N_g(\mh)} = n_g \left(\frac{\mathrm{d}n}{\mathrm{d}\mh}\right)^{-1}\int_{\ms^{0}}^{\ms^{1}}
    p(\ms, \mh) \,\mathrm{d}\ms.  \label{eqn:nmhchod}
\end{equation}
Following the standard procedure in HOD modelling, we adopt Equation~\eqref{eqn:nmhchod} for the {\chod}
model, writing separate contributions to $p(\ms, \mh)$ for central and satellite galaxies, and derive
constraint by applying it to the samples defined in Figure~\ref{fig:mz}b. For the sake of comparison to
the {\ihod} constraint, we do not use the number density of sample galaxies as an input to the {\chod}
constraints.

However, due to the increase of the luminosity threshold with redshift for a flux-limited sample, the
$\ms$--selected galaxy populations are strictly stratified in $z$ and the HODs at different redshifts must be
treated separately. At fixed redshift $z$, the HOD is
\begin{multline}
    \avg{N_g(\mh|z)} = \\ n_g\left(\frac{\mathrm{d}n}{\mathrm{d}\mh}\right)^{-1} \int_{\ms^{0}}^{\ms^{1}}
    p(\ms, \mh) f_{\mathrm{obs}}(\ms | \mh , z) \,\mathrm{d}\ms .  \label{eqn:nmhz_old}
\end{multline}
Therefore, in order to predict $\avg{N_g(\mh|z)}$ from $p(\ms, \mh)$ we need to know $f_{\mathrm{obs}}(\ms
| \mh , z)$, which is inaccessible to us because of our ignorance of $g(\Gamma | \ms, \mh)$. However,
we can rewrite the above Equation as
\begin{multline}
    \avg{N_g(\mh|z)} = \\
     \left(\frac{\mathrm{d}n}{\mathrm{d}\mh}\right)^{-1} \int_{\ms^{0}}^{\ms^{1}} p(\mh | \ms)
    \left[\Phi(\ms)f_{\mathrm{obs}}(\ms | \mh , z)\right] \,\mathrm{d}\ms,
    \label{eqn:nmhz}
\end{multline}
by using Bayes' Theorem $p(\mh|\ms){=}p(\ms,\mh)/p(\ms)$ and the definition of the {\it parent} SMF
$\Phi(\ms){\equiv}n_g\,p(\ms)$.  To make further progress in our predictions, we adopt the {\it ansatz}
that above the mixture limit $\ms{>}\ms^{\mathrm{mix}}(z)$, the dependence of $f_{\mathrm{obs}}(\ms | \mh,
z)$ on the halo mass is very weak, i.e., $f_{\mathrm{obs}}(\ms|\mh , z){\sim}f_{\mathrm{obs}}(\ms |
z)$. Applying this ansatz to Equation~\eqref{eqn:nmhz} we arrive at
\begin{equation}
    \avg{N_g(\mh | z)} = \left(\frac{\mathrm{d}n}{\mathrm{d}\mh}\right)^{-1}\int_{\ms^{0}}^{\ms^{1}} p(\mh |
    \ms) \Phi_{\mathrm{obs}}(\ms|z) \mathrm{d}\ms,
    \label{eqn:nmhihod}
\end{equation}
where $\Phi_{\mathrm{obs}}(\ms|z){=}\Phi(\ms)f_{\mathrm{obs}}(\ms|z)$ is the {\it observed} SMF at redshift
$z$, directly accessible from the survey. For modelling the  samples defined in Figure~\ref{fig:mz}c for the
{\ihod} analysis, we measure the observed galaxy SMF at each redshift, and then obtain the HOD for that
redshift slice using Equation~\eqref{eqn:nmhihod}. In this way, we avoid the need to explicitly model the
incompleteness as a function of $\ms$ and/or $\mh$. Although it appears from Equation~\eqref{eqn:nmhihod} that
both the amplitude and the shape of $\Phi_{\mathrm{obs}}$ are used to derive $\avg{N_g(\mh | z)}$, in essence
we only use the shape as an input to the {\ihod} constraint, because the normalisation of $\avg{N_g(\mh | z)}$
is irrelevant to the prediction of the clustering and lensing signals.

Before going any further, here we will lay out the theoretical arguments leading to the ansatz and defer
the detailed discussion on its validation using consistency checks later in Section~\ref{subsec:post}
and~\ref{subsec:smf}. Splitting the galaxies into red and blue populations explicitly in
Equation~\eqref{eqn:fobs}, we have
\begin{multline}
    f_{{\mathrm{obs}}}(\ms | \mh,z) = \int_{0}^{\Gamma_{\mathrm{max}}(z)}
    \left\{ f_{\mathrm{red}}(\ms | \mh) g_{\mathrm{red}}(\Gamma | \ms, \mh) \right. \\
    + \left.\left[1 - f_{\mathrm{red}}(\ms | \mh)\right] g_{\mathrm{blue}}(\Gamma | \ms, \mh) \right\} \,\mathrm{d}\Gamma,
    \label{eqn:fobscolor}
\end{multline}
where $f_{\mathrm{red}}(\ms | \mh)$ is the intrinsic fraction of red galaxies at given $\ms$ within halos of
mass $\mh$. To understand the potential dependence of $f_{\mathrm{obs}}(\ms|\mh, z)$ on the halo mass $\mh$,
we first examine the variations of $f_{\mathrm{red}}(\ms | \mh)$ and $g_{\mathrm{red/blue}}(\Gamma | \ms,
\mh)$ with $\mh$ separately and then combine them using the above Equation.  Since the $\ms/L$ ratio $\Gamma$
is very tightly correlated with galaxy colour $c$~\citep{bell2001}, we can instead look at the colour
distributions of the two populations, $g_{\mathrm{red/blue}}(c | \ms, \mh)$, each with a centroid position
$\bar{c}$ and a spread $\Delta c$. By analysing the variation in the galaxy colour bimodality with stellar
mass and projected neighbour density using SDSS, \citet{baldry2006} found that $\bar{c}$ and $\Delta c$ of the
red and blue populations are stable across different environments of void, field, and groups and
clusters~(i.e., halo masses), while the red fraction $f_{\mathrm{red}}$ increases continuously with $\ms$ and
the local density of the environment, i.e., the so-called ``mass'' and ``environmental'' quenching,
respectively~\citep{peng2012}. This stability in the colour distribution within each coloured population
against the environment implies $g_{\mathrm{blue}}(\Gamma | \ms, \mh){\equiv}g_{\mathrm{blue}}(\Gamma | \ms)$
and $g_{\mathrm{red}}(\Gamma | \ms, \mh){\equiv}g_{\mathrm{red}}(\Gamma | \ms)$, so that
Equation~\eqref{eqn:fobscolor} can be simplified as
\begin{multline}
    f_{{\mathrm{obs}}}(\ms| \mh,z) = f_{\mathrm{red}}(\ms| \mh) G_{\mathrm{red}}(\ms| z) \\
    + \left[1 - f_{\mathrm{red}}(\ms| \mh)\right] G_{\mathrm{blue}}(\ms | z),
    \label{eqn:fobscolor2}
\end{multline}
where
\begin{align}
G_{\mathrm{red}}(\ms | z) &= \int_{0}^{\Gamma_{\mathrm{max}}(z)}g_{\mathrm{red}}(\Gamma | \ms)
\,\mathrm{d}\Gamma \nonumber\\
G_{\mathrm{blue}}(\ms | z) &= \int_{0}^{\Gamma_{\mathrm{max}}(z)}g_{\mathrm{blue}}(\Gamma | \ms)
\,\mathrm{d}\Gamma.
\end{align}
are the fractions (from 0 to 1) of the red/blue galaxies with $\ms$ at $z$ that will be observed given the
flux limit and the separate red/blue distributions of $\ms/L$ ratios.  Above the mixture limit, the
completeness of both the red and blue galaxies are relatively high, so that $G_{\mathrm{red}}(\ms |
z){\approx}G_{\mathrm{blue}}(\ms | z){=}f_{\mathrm{obs}}(\ms|z)$ and Equation~\eqref{eqn:fobscolor2} gives
$f_{{\mathrm{obs}}}(\ms| \mh,z){=}f_{\mathrm{obs}}(\ms | z)$. Below the mixture limit, however, the observed
red galaxies are so scarce that $G_{\mathrm{red}}(\ms| z){\sim}0$, yielding a halo mass--dependent
$f_{\mathrm{obs}}(\ms| \mh,z){\simeq}\left[1 - f_{\mathrm{red}}(\ms| \mh)\right] G_{\mathrm{blue}}(\ms | z)$
because $f_{\mathrm{red}}(\ms|\mh)$ is sensitive to $\mh$~\citep{george2011}.

We expect the ansatz to be largely valid except for the low mass samples~($\ms{<}10^{10}\hhmsol$) where the
colour bimodality might shift with halo mass because of galaxy evolution~\citep{taylor2015}.  We nonetheless
proceed to apply this ansatz to the entire stellar mass range, relying on the fact that the statistical power
of the low mass samples is quite low, and discuss the possible impact on our results later in
Section~\ref{subsec:smf}. Finally, we emphasise that this ansatz assuming $f_{\mathrm{obs}}(\ms| \mh ,
z){\sim}f_{\mathrm{obs}}(\ms | z)$, explicitly made in the {\ihod} model to account for the fact that for any
given $\ms$ we observed fewer intrinsically high--$\ms/L$ systems at higher redshifts, is a much weaker
assumption than required by the traditional HOD models, which assume all the high--$\ms/L$ systems were
detected in the sample, i.e., $f_{\mathrm{obs}}(\ms | \mh, z){=}1$. In other words, since
$f_{\mathrm{obs}}(\ms | \mh, z){=}1$ is a sufficient condition for $f_{\mathrm{obs}}(\ms | \mh,
z){=}f_{\mathrm{obs}}(\ms | z)$, the {\ihod} model includes the {\chod} model as a subset.

It is important to point out that during the constraint, the {\ihod} model remains entirely agnostic of the
overall amplitude of the SMF, whether it be the parent SMF~(normalised by $n_g$) or the observed
ones~(normalised by the product of $n_g$ and $f_{\mathrm{obs}}(\ms|z)$), because neither $n_g$ nor
$f_{\mathrm{obs}}(\ms|z)$ is known {\it a priori}. However, the {\ihod} model is built on top of the halo mass
function, which has a fixed normalisation in any given $\lcdm$ cosmology.  Thus, once constrained by the
clustering and lensing data, the best-fit {\ihod} model would give an explicit prediction for both the shape
and the amplitude of the parent SMF.  This prediction automatically gives an estimate of
$f_{\mathrm{obs}}(\ms|z)$ when compared to the observed SMF at redshift $z$. The estimated
$f_{\mathrm{obs}}(\ms|z)$ thus provides a useful consistency check of the {\ihod} model. Any significant
departure of $f_{\mathrm{obs}}$ from unity at stellar mass above the mixture limit would indicate failure of
our model assumptions (e.g., the ansatz about the weak dependence of $f_{\mathrm{obs}}(\ms|z)$ on $\mh$)
and/or degeneracies in the model parameters~(e.g., residual degeneracy between scatter and amplitude of the
SHMR), while a successful model, as we will discover in Section~\ref{subsec:smf}, would reproduce
$f_{\mathrm{obs}}$ curves that are slightly below unity.  Another consistency check is the comparison between
the SHMR constraints from the {\chod} and the {\ihod} models on the high mass end --- in the regime of
$\ms{\gg}\ms^{\mathrm{mix}}(z)$ where $f_{\mathrm{obs}}(\ms | \mh, z){\rightarrow}1$, the {\ihod} model
becomes identical to the {\chod} model and the two set of constraints, as will be shown later in
Section~\ref{subsec:post}, should be consistent with each other.

In summary, both the {\chod} and {\ihod} models rely on the prediction of $p(\ms, \mh)$. The primary
difference is that the {\chod} model employs a single HOD by assuming that the galaxy sample is
volume-complete in stellar mass, while the {\ihod} model is able to self--consistently take into account the
redshift--dependent selection function of the samples by working in narrow redshift slices and using
information from the shape of the observed galaxy SMF at each slice. This improvement in {\ihod} enables us to
include $84\%$ more galaxies than used in {\chod} by adding the low mass galaxies~($\ms{<}10^{9.8}\hhmsol$) in
the local universe as well as more distant galaxies, hence the significant improvement in the $S/N$ of the
measurements shown in Figure~\ref{fig:compsn}.  This process cannot be extended indefinitely, since at lower
stellar mass (below the mixture limit for a given redshift) the basic assumption behind {\ihod} fails, but
the range of stellar masses where it is applicable is still large enough for significant improvements.

\subsection{Deriving the Two HODs}
\label{subsec:n2p}

Here we describe the theoretical framework for predicting the HODs for both the {\chod} and {\ihod} models. In
particular, we first model the total number of galaxies~(both observed and unobserved) per log-stellar mass
within halos of some fixed mass, $\mathrm{d} N(\ms |\mh)/\mathrm{d}\lg\ms$, and then compute the joint
probability as
\begin{equation}
    p(\ms, \mh) = \frac{\lg e}{\ms n_g}\frac{\mathrm{d}N(\ms|\mh)}{\mathrm{d}\lg\ms}\frac{\mathrm{d}n}{\mathrm{d}\mh}.
\label{eqn:p2d}
\end{equation}
To conform to the traditional HOD notation, we hereafter refer to $\mathrm{d} N(\ms |\mh)/\mathrm{d}\lg\ms$
simply as $\avg{N(\ms |\mh)}$ by implicitly assuming $\Delta\lg\ms\equiv 0.02$ throughout the paper.  Given
$p(\ms, \mh)$, we can use Equation~\eqref{eqn:nmhchod} to specify a single HOD for each of the {\chod}
samples. For the {\ihod} analysis, however, we also need to compute the {\it parent} stellar mass function
\begin{equation}
\Phi(\ms) = n_g \int_0^{+\infty} p(\ms, \mh) \,\mathrm{d}\mh,
\label{eqn:smf}
\end{equation}
and extract the {\it observed} stellar mass function $\Phi_{\mathrm{obs}}(\ms)$ directly from the data. The
distribution of host halo mass for galaxies at fixed stellar mass is simply
\begin{equation}
p(\mh|\ms) = \frac{p(\ms, \mh)}{p(\ms)} = \frac{n_g\, p(\ms, \mh)}{ \Phi(\ms)},
\label{eqn:pmh}
\end{equation}
and we can obtain the HODs for individual redshift slices within each {\ihod} sample using
Equation~\eqref{eqn:nmhihod}. Since the redshift range~($0.02$--$0.30$) spans ${\sim}3$ billion years
during which only $8\%$ of the total stellar mass observed today formed~\citep[cf. equation 15 in][]{madau2014},
 we assume $p(\ms | \mh)$ to be constant with redshift. Therefore, all the redshift evolution in
the theoretical model comes from the cosmic growth in the halo mass function. To speed up the calculation
without loss of accuracy, we adopt the same halo mass function for all the redshift slices within each sample,
calculated at the volume-averaged redshift of that sample. We have tested this approximation by comparing the
predicted signals with those computed from integrating the halo mass functions over all the redshift slices
and the difference is negligible.

\subsection{Parameterizing $\avg{N(\ms |\mh)}$ }
\label{subsec:l10}

\begin{figure*}
    \centering\resizebox{0.95\figtextwidth}{!}{\includegraphics{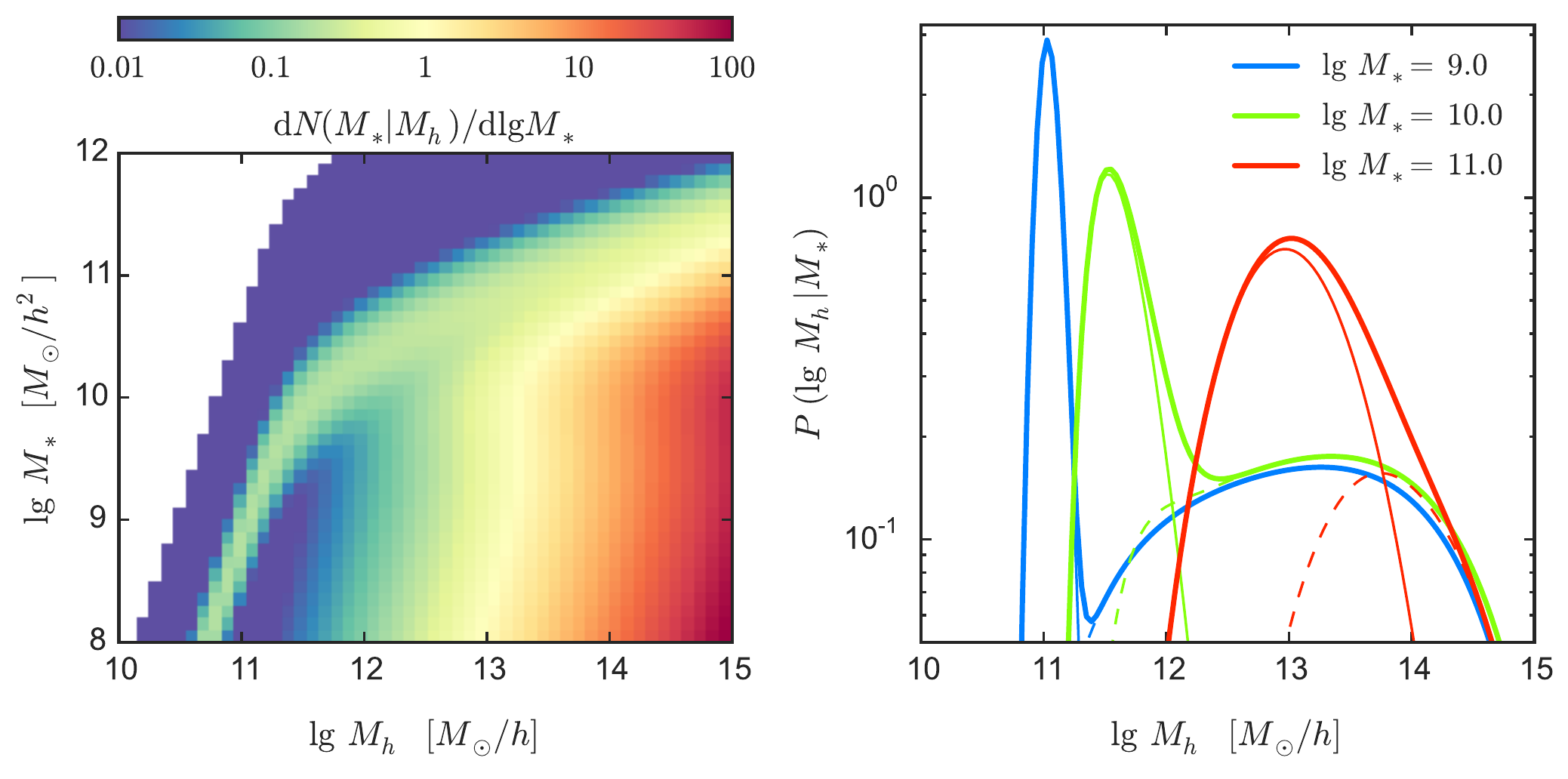}} \caption{
	{\it Left Panel:} The 2D HOD, i.e., the average number of galaxies per dex in stellar mass within halos
    at fixed mass, predicted by the best-fit {\ihod} model. {\it Right Panel:} The probability distributions of
	the host halo mass for galaxies at three fixed log--stellar mass listed on the top right.  Thick
	solid, thin solid, and thin dashed lines represent contributions from all, central, and satellite
	galaxies, respectively.  Galaxies of higher stellar mass are more likely to be central galaxies
	sitting in high mass halos, whereas low mass galaxies are as likely to be satellites of high mass
	halos as centrals of their own halos.}
    \label{fig:h2p}
\end{figure*}

Our analytic model for $\avg{N(\ms|\mh)}$ has two components: 1) the mean and the scatter of the SHMR for
the central galaxies, the combination of which automatically specifies $\avg{N_{\mathrm{cen}}(\ms|\mh)}$, and
2) the mean number of satellite galaxies with stellar mass $\ms$ inside halos of mass $\mh$,
$\avg{N_{\mathrm{sat}}(\ms|\mh)}$. We adopt a similar parameterization for the two components as in L11, but
allow more of them to vary during the constraint, and predict the signals differently. Here we will briefly
describe the functional forms and the model parameterizations.

At fixed halo mass, we assume a log--normal probability distribution for the stellar mass of the central
galaxies, with a log--normal scatter. The mean SHMR is then the sliding mean of the log--normal distribution
as a function of the halo mass, $f_{\mathrm{SHMR}}{\equiv}\exp\avg{\ln\ms(\mh)}$. The L11 functional form for $f_{\mathrm{SHMR}}$ that
we adopt for our analysis is defined by \citet{behroozi2010} via its inverse function,
\begin{equation}
    \mh = M_1  m^\beta  \exp\left(\frac{m^\delta}{1 + m^{-\gamma}}- \frac{1}{2}\right) ,
    \label{eqn:shmr}
\end{equation}
where $m\equiv\ms/M_{*,0}$. Among the five parameters that describe $f_{\mathrm{SHMR}}$, $M_1$ and $M_{*,0}$ are the
characteristic halo mass and stellar mass that separate the behaviours in the low and high mass
ends~($f_{\mathrm{SHMR}}(M_1) = \ln M_{*,0}$). The inverse function starts with a low--mass end slope $\beta$, crosses a
transitional regime around ($M_{*,0}$, $M_1$) dictated by $\gamma$, and reaches a high--mass end slope
$\beta+\delta$. The Figure 1 in L11 illustrates the different responses of $f_{\mathrm{SHMR}}$ to the changes in each of
the five parameters.

The log--normal scatter at fixed halo mass is the quadratic sum of the intrinsic scatter and the measurement
error. It is important to keep in mind that the intrinsic part of this scatter must be the same for all
studies, while studies with different datasets or stellar mass determination methods may have differing
measurement error contributions and thus a different total log--normal scatter.  L12 considered two models for
the scatter, one that is constant and another that includes an empirical stellar mass--dependence in the
measurement error across the entire mass range.  They found that since the constraint on the overall scatter
is primarily driven by the high mass end where the slope of $f_{\mathrm{SHMR}}$ is much shallower, the mass--dependence
of the scatter at the low mass end has little impact on their results.  In light of this finding and to focus
on the scatter at the high mass end, we keep the scatter independent of halo mass below $M_1$, but allow more
freedom in the scatter above the characteristic mass scale, with an extra component that is linear in
$\lg\mh$:
\begin{equation}
    \sig(\mh) = \left\{ \begin{array}{ll}
        \sig,&\mbox{ $\mh<M_1$} \\
            \sig + \eta \lg \frac{\mh}{M_1},&\mbox{ $\mh\geq M_1$}
\end{array} \right.
\label{eqn:sigmh}
\end{equation}

The motivation behind this additional degree of freedom is two--fold: 1) the average measurement uncertainty
of the stellar mass estimates decrease with $\ms$~(hence $\mh$) in the MPA/JHU catalogue, and 2) recently,
there is evidence suggesting a smaller scatter at the high mass end of the SHMR~\citep[e.g.,][]{shankar2014},
although in Equation~\eqref{eqn:sigmh} the scatter is allowed to either increase or decrease with halo mass.
The combination of the mean and the scatter fully specifies the HOD of central galaxies,
\begin{multline}
    \avg{N_{\mathrm{cen}}(\ms|\mh)} = \\
    \frac{1}{\sig(\mh)\sqrt{2\pi}} \exp\left[-\frac{\left[\ln \ms -
    \ln f_{\mathrm{SHMR}}(\mh)\right]^2}{2\sig^2(\mh)}\right].
    \label{eqn:ncen}
\end{multline}

Follow L11, we model $\avg{N_{\mathrm{sat}}(\ms|\mh)}$ as the derivative of the satellite occupation number in
stellar mass--thresholded samples, $\avg{N_{\mathrm{sat}}(>\ms|\mh)}$, which is parameterized as a power of
halo mass with an exponential cutoff and scaled to $\avg{N_{cen}(>\ms|\mh)}$ as follows,
\begin{multline}
    \avg{N_{\mathrm{sat}}(>\ms|\mh)} = \\ \avg{N_{\mathrm{cen}}(>\ms|\mh)} \left(\frac{\mh}{M_{\mathrm{sat}}}\right)^{\alpha_{\mathrm{sat}}}
    \exp\left(\frac{-M_{\mathrm{cut}}}{\mh}\right).
    \label{eqn:nsat}
\end{multline}
Instead of fixing $\alpha_{\mathrm{sat}}$ to be $1$ as in L11, we allow it to vary during the fit. We
parameterize both the characteristic mass of a single satellite-hosting halo, $M_{\mathrm{sat}}$, and the
cutoff mass scale, $M_{\mathrm{cut}}$, as simple power law functions of the threshold stellar mass, so that
\begin{equation}
    \frac{M_{\mathrm{sat}}}{10^{12}\hhmsol} = B_{\mathrm{sat}}
    \left(\frac{f_{\mathrm{SHMR}}^{-1}(\ms)}{10^{12}\hhmsol}\right)^{\beta_{\mathrm{sat}}},
\end{equation}
and
\begin{equation}
    \frac{M_{\mathrm{cut}}}{10^{12}\hhmsol} = B_{\mathrm{cut}}
    \left(\frac{f_{\mathrm{SHMR}}^{-1}(\ms)}{10^{12}\hhmsol}\right)^{\beta_{\mathrm{cut}}},
\end{equation}
respectively. In practice, we choose a $0.02$ dex bin size in stellar mass for the  numerical differentiation
of $\avg{N_{\mathrm{sat}}(>\ms|\mh)}$.

The left panel of Figure~\ref{fig:h2p} displays the $\avg{N(\ms|\mh)}$ map predicted by the best--fit {\ihod}
model in our analysis. The SHMR of the central galaxies is clearly seen as the ``main sequence'' enveloping
the ``cloud'' occupied by the satellite galaxies.  The right panel shows the probability distribution of host
halo masses for galaxies at $M_*=10^{10}\hhmsol$~(blue), $10^{11}\hhmsol$~(green), and $10^{12}\hhmsol$~(red),
each computed from the map in the left panel using the combination of Equations~\eqref{eqn:p2d},
\eqref{eqn:smf}, and \eqref{eqn:pmh}. The thin solid and the dashed curves of each colour show the
contributions to the total probability distribution from the central and the satellite galaxies, respectively.
The central galaxy contribution shows an increasing logarithmic scatter in halo mass with stellar mass, mainly
due to the change of slope in the SHMR when going to the high mass end~(where $\sig$ shrinks slightly as well,
i.e., $\eta<0$). The lower mass satellites, however, populate halos of much greater diversity than their high
mass counterparts. Meanwhile, the satellite fraction increases toward lower stellar masses, but the
total number of galaxies is dominated by central galaxies at all stellar masses.

\subsection{Spatial Distribution of Galaxies within Halos}
\label{subsec:onehalo}

In addition to the parameterization of $\avg{N(\ms|\mh)}$, we also need to model the spatial distribution of
galaxies within dark matter halos for the small-scale clustering and lensing predictions. We assume the
isotropic Navarro-Frenk-White~\citep[NFW:][]{navarro1997} density profile for halos with a concentration--mass
relation $c_{\mathrm{dm}}(\mh)$ calibrated from simulations~(described further below).  Although the NFW
density profile is regarded as ``universal'' only in pure dark matter simulations where the effects of baryons
are absent\footnote{However, see~\citet{gao2008} and~\citet{dutton2014} regarding potentially better
``universality'' when using the~\citet{einasto1965} profile.}, recent studies of nearby rich clusters found
that, despite the fact that the baryons tend to flatten the dark matter distribution in the cluster centre,
the total matter density~(i.e., baryon and dark matter combined) still maintains a NFW shape from the scales
of the central brightest cluster galaxy~(BCG) out to the virial radius~\citep{newman2013}, probably due to the
significant mixing between stars and dark matter as a result of frequent mergers~\citep{laporte2014}. For
smaller systems like the group and galaxy-scale halos with $\mh{<}10^{14}\hmsol$, however, the sum of an NFW
and a stellar mass component is required to explain the inner slope of the observed total matter density
profiles~\citep{mandelbaum2006b, newman2015}. In light of these observational findings, we add a
model--independent stellar mass component as a point source in the halo centre to the g-g lensing predictions
for all the galaxy samples, so that
\begin{equation}
    \Delta\Sigma_{\mathrm{stellar}} = \frac{\langle\ms\rangle}{\pi r_p^2} \label{eqn:dsstar}
\end{equation}
where $\langle\ms\rangle$ is the average stellar mass of each stellar mass sample. Although this extra stellar
component is not necessary for the central galaxies of very rich clusters, i.e., some galaxies within the
highest stellar mass sample, the minimum fitting scale for that sample is above $0.1\hmpc$~(see
Appendix~\ref{app:smallscale}), where the stellar contribution to $\ds$ calculated via
Equation~\eqref{eqn:dsstar} is negligible.

Central galaxies are placed at the centres of the NFW halos. For the modelling of $w_p$, we do not consider the
miscentering effect, which is likely to be important only for around $30\%$ of the BCGs~\citep{george2012}. In
our analysis the galaxy clustering for BCGs is measured down to ${\sim}0.2\hmpc$, therefore largely immune to
the miscentering effect which has a kernel ${\simeq}75\hkpc$~\citep{george2012}.  It is worth noting that the
miscentering issue in our case is more benign than in other recent papers that have explicitly modelled it
\citep[e.g.,][]{miyatake2013, more2014}.  In those papers, the samples that were being modelled had strict
colour and luminosity selection, such that the central galaxy in the halo might not be present in the sample.
In our case, with a flux-limited sample that goes to a relatively low flux limit, the central galaxies in the
vast majority of group- and cluster-size halos should be present in the sample, so we only have to contend
with small offsets of the central galaxy from the halo centre (rather than a complete misidentification of the
central galaxy).  For the g-g lensing, the miscentering effect is compounded and somewhat cancelled by the
contribution from the subhalo of satellite galaxies~\citep{yoo2006}, which is more important for low stellar
mass samples. We will discuss the modelling of the combined effects further in Section~\ref{subsec:subhalo}.

We assume an NFW profile for the satellite distribution as well, but with a different amplitude of the
concentration--mass relation than the dark matter. In particular, we set $c_{\mathrm{sat}}(\mh)\equiv f_{c}
\times c_{\mathrm{dm}}(\mh)$, where $f_c$ characterises the spatial distribution of satellite galaxies
relative to the dark matter within halos.

To summarise, we have in total 13 model parameters. Among them $\{\lg\mh^1, \lg\ms^0, \beta, \delta, \gamma\}$
describe the mean SHMR, $\{B_{\mathrm{sat}}, \beta_{\mathrm{sat}}, B_{\mathrm{cut}}, \beta_{\mathrm{cut}},
\alpha_{\mathrm{sat}}\}$ describe the parent HOD of satellite galaxies, $\{\sig, \eta\}$ describe the
logarithmic scatter about the mean SHMR, and $f_c$ is the ratio between the concentrations of the satellite
distribution and the dark matter profile.

\section{Predicting Signals from the Halo Model}
\label{sec:pred}

\begin{figure*}
    \centering\resizebox{0.95\figtextwidth}{!}{\includegraphics{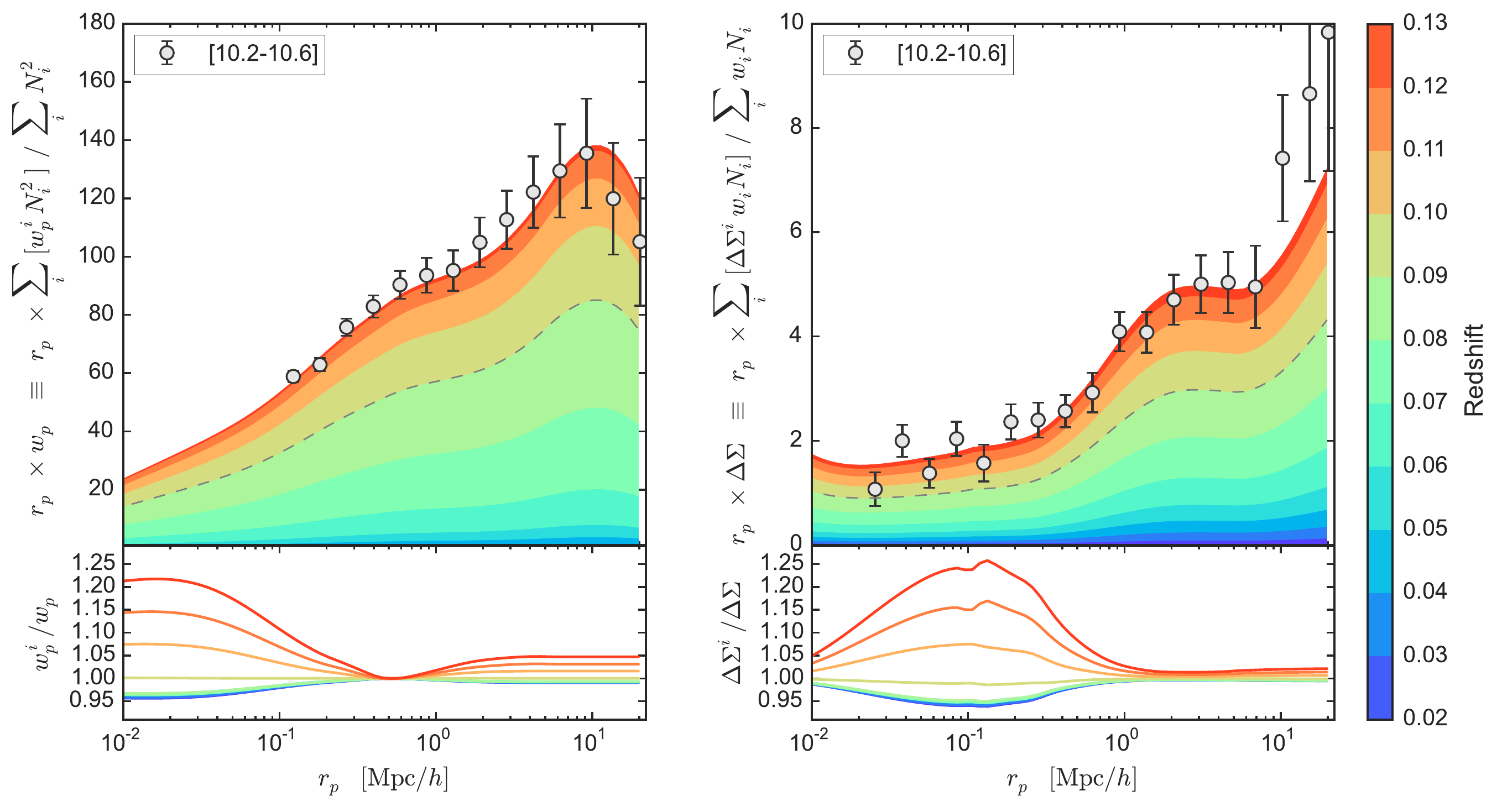}}
    \caption{Contributions from individual redshift slices to the predicted clustering~(left) and
    lensing~(right) signals of the {\ihod} sample with $\lg\ms{=}[10.2, 10.6]$. In each panel, the top
    sub-panel compares the measured signal~(circles with errorbars) to the predicted total signal, decomposed
    into the contributions from each of the 11 redshift slices~(colour--coded by the colourbar on the right).
    The dashed curve in each stack marks the maximum redshift~($0.09$) of the {\chod} sample with the same
    stellar mass range, indicating a ${\sim}40\%$ gain in the total signal by switching from {\chod} to {\ihod}.
    Due to the difference in the weighting schemes, the fractional contributions of the
    same redshift slice to the total signals are different in the two cases. The bottom sub-panels show the
    ratio of the signal predicted for each slice over the signal for the whole sample. In both the clustering and
    lensing cases, the fractional variation is coherent with redshift and below $25\%$ on all scales.}\label{fig:signalslice}
\end{figure*}

\subsection{Prerequisites and Approximations}
\label{subsec:pre}

The goal of our analysis is to infer the SHMR of central galaxies and the HOD of satellite galaxies using the
galaxy clustering and the g-g lensing. From the inferred $\avg{N(\ms|\mh)}$, we can predict the parent stellar
mass function to compare to those empirically reconstructed from the $V_{\mathrm{max}}$ method.  This route is
exactly the reverse of the methodology employed in the SHAM studies~(e.g., M10), which infer the SHMR by
abundance matching to the $V_{\mathrm{max}}$--estimated SMF and then predict the clustering and/or lensing as
a cross--check.

To facilitate a direct comparison with the SHAM results, we adopt the same flat $\lcdm$ cosmology model as in
M10~(listed in Section~\ref{sec:intro}).  For the linear matter power spectrum, we use the low--baryon
transfer function of Eisenstein \& Hu (1998), which is a good approximation to the full transfer function on
scales well below the BAO scale.  To compute the non-linear matter correlation function $\ximm$, we use the
prescription from~\citet{takahashi2012}~(an updated version of the \texttt{halofit} prescription
from~\citealt{smith2003}) to generate the non--linear power spectrum for Fourier transforming to $\ximm$. The
halo mass function and the halo bias function are from~\citet{tinker2008} and~\citet{tinker2010},
respectively.  The halo mass--concentration relationship $c_{\mathrm{dm}}(\mh)$ is from the fitting formula
of~\citet{zhao2009}, which accurately recovers the flattening of halo concentration at high masses.  (In DR7
we do not expect an upturn, which only shows up at redshifts beyond $1$; see~\citealt{prada2012}).

For the {\chod} samples the signal prediction is relatively straightforward, as only one HOD is required to
describe each sample. We describe the prediction of $w_p$ and $\ds$ from a single HOD in
Section~\ref{subsec:xids}. Complexity arises when predicting the signals for the {\ihod} samples, each of
which contains multiple HODs that describe the stratification of galaxy populations due to the
redshift--dependent selection function. We divide each {\ihod} sample into multiple redshift slices of $\Delta
z = 0.01$, which corresponds to a co-moving width of $\sim 30\hmpc$ across the redshift range of the MGS.
This redshift bin size is chosen to make sure the slices are: 1) thin enough for the redshift selection
function to be uniform, and 2) thick enough to include both the one and two--halo terms in the LOS direction
internally.  Within each redshift slice $i$ we predict the clustering and lensing signals, $w_p^i$ and
$\ds^i$, in the same way as in the {\chod} analysis, and combine them to obtain the final predictions for the
full sample,
\begin{equation}
    w_p = \left(\sum\limits_{i} N_i^2 w_p^i \right) \Big/  \left(\sum\limits_{i} N_i^2 \right),
    \label{eqn:wpi}
\end{equation}
and
\begin{equation}
    \ds = \left(\sum\limits_{i} N_i w_i \ds^i \right) \Big/ \left( \sum\limits_{i} N_i w_i\right),
    \label{eqn:dsi}
\end{equation}
where $N_i$ is the number of {\it observed} galaxies in slice $i$ and $w_i$ is the per-galaxy lensing weight
at each redshift. $w_i$ is essentially related to the inverse variance weights $w_{ls}$ defined in
Equation~\eqref{eqn:wls}, but integrated over the source redshift distribution.  It also includes the
geometric factor ($1/\Sigma_c^2$) as well as the fact that an annular bin with a fixed centre value of $r_p$
and width $\Delta r_p$ contains more source galaxies at lower redshift, due to its larger area on the sky.
Since the source catalogue is the same for the g-g lensing in each redshift slice, the pair counting weight
for $\ds^i$ is ${\propto}N_i$ instead of ${\propto}N_i^2$.

Figure~\ref{fig:signalslice} illustrates the fractional contributions of each individual redshift slice to the
predicted total signal~(calculated from the two equations above; top sub-panels), and compares the fractional
variations among the predicted signals of all redshift slices~(bottom sub-panels), using the {\ihod} sample
with $\lg\ms{=}[10.2, 10.6]$ as an example. The dashed curve in each top sub-panel indicates the maximum
redshift~($0.09$) of the corresponding {\chod} sample with the same stellar mass range. Clearly, the three
extra redshift slices above $z=0.09$ included by the {\ihod} analysis contribute ${\sim}40\%$ of the total
signal from the $11$ slices, because of the much higher weights associated with those higher redshift slices.

Since we have used a line-of-sight integration limit of $60\hmpc$ (roughly twice the size of the slice width)
to get $w_p$ from $\xi$, the measured $w_p$ signal will include galaxy pairs that straddle different slices.
This effect is not directly reflected in Equation~\eqref{eqn:wpi}, so the cross--correlation between different
slices might require a separate treatment. However, because the galaxy correlation functions vary very slowly
and smoothly between adjacent slices~(a few per cent on large scales; see the bottom sub-panels of
Figure~\ref{fig:signalslice}), the cross terms are most sensitive to the product of the numbers of galaxies in
each pair of slices, which is correctly accounted for in Equation~\eqref{eqn:wpi}.

In particular, for two slices $i$ and $j$, Equation~\eqref{eqn:wpi} makes the assumption that the 3D
cross-correlation $\xi_{ij}$ is approximately $(N_i^2\xi_i + N_j^2\xi_j)/(2 N_i N_j)$, while the more correct
form should be $\sqrt{\xi_i \xi_j}$, as the cross--correlation is close to the geometric mean of the two
auto--correlation functions $\xi_i$ and $\xi_j$~\citep{zehavi2011}.  When $\Delta\xi/\xi$ is small, the
fractional difference between the two forms is of the order ${\sim} (N_j - N_i)^2 / (N_i N_j) {\equiv} (\Delta
N / N)^2$. Moreover, after integrating $\xi$ via Equation~\eqref{eqn:wp}, the $w_p$ signal is always dominated
by the $\xi^s$ values at $r_{\pi}<30\hmpc$ at any fixed $r_p$.
  We also construct a mock galaxy sample with a similar redshift distribution
and bias evolution as the stellar mass samples in the data, and compare the projected correlation functions
measured from our approximation in Equation~\eqref{eqn:wpi} to that directly from the pair counting described
in Section~\ref{subsec:wpdata}. We find that the error induced by adopting Equation~\eqref{eqn:wpi} is no more
than $2\%$ on all scales and is thus negligible in our analysis.

It is important to note that the theoretical predictions for $w_p$ and $\ds$ via Equation~\eqref{eqn:wpi} and
~\eqref{eqn:dsi} are independent of the total number of observed galaxies in each sample, i.e., the
normalisation of the observed stellar mass function --- if we analyse only half of the galaxies randomly drawn
from the entire \texttt{bright0} sample, the prediction from the same set of model parameters would not
change, and the constraints on the model parameters would stay the same, albeit with larger uncertainties.

\subsection{Predicting $w_p$ and $\ds$ from a single HOD}
\label{subsec:xids}

The analytic model for deriving the galaxy clustering and the g-g lensing signals from a single HOD is based
on the prescriptions given in a series of papers, including~\citet{tinker2005},~\citet{zheng2007},
and~\citet{yoo2006}, with improved treatment of the scale--dependent bias, halo exclusion effect,
over/under-concentrated satellite distributions from earlier works~\citep[e.g.][]{berlind2002, guzik2002,
mandelbaum2006}. We describe the prescription briefly below and refer the interested readers to the
aforementioned three papers for details.

The signals of $w_p$ and $\ds$ are obtained by projecting the 3D real--space galaxy auto--correlation function
$\xigg$ and the galaxy--matter cross--correlation function $\xigm$, respectively. The projection of $\xigg$ to
$w_p$ is given by Equation~\eqref{eqn:wp}, while for the g-g lensing it is via
\begin{equation}
    \ds(r_p) = \langle\Sigma(<r_p)\rangle - \Sigma(r_p),
\label{eqn:ds}
\end{equation}
where
\begin{equation}
   \Sigma(r_p) =  \bar{\rho}_m \int_{-\infty}^{+\infty} \left[1 + \xigm(r_p, r_\pi)\right]\,\dd r_\pi ,
\label{eqn:ds2}
\end{equation}
and
\begin{equation}
   \langle\Sigma(<r_p)\rangle = \frac{2}{r_p^2} \int_0^{r_p} r_p^\prime \Sigma(r_p^\prime)\,\dd r_p^{\prime}.
\label{eqn:ds1}
\end{equation}
We are ignoring the effects from the radial window, which is broad enough that it is not relevant at galaxy
scales~\citep{baldauf2010}.  The formalisms for predicting $\xigm$ and $\xigg$ are very similar, as the
satellite distribution is merely a discrete realisation of NFW tracers with a different concentration than the
dark matter. There is, however, one extra contribution to $\xigm$ and $\ds$ at small scales from the matter
retained by the subhalos that host the satellite galaxies~(beside the model--independent stellar mass
component; see Equation~\ref{eqn:dsstar}). We will discuss this ``subhalo'' lensing term in
more detail in Section~\ref{subsec:subhalo} and focus on the similar terms shared by $\xigg$ and $\xigm$ here.

Let us consider a general scenario in which the correlation is between a primary galaxy catalogue and a
secondary catalogue consisting of tracer particles $t$, whether it be the dark matter~($\xigm$) or the same
galaxies as the primaries~($\xigg$).  The correlation signal between the primary and the secondary  can be
decomposed into two components,
\begin{equation}
    \xigt(r) + 1 = \left[ \xigt^{1h}(r) + 1 \right] +  \left[ \xigt^{2h}(r) + 1 \right],
\end{equation}
where ``1h'' and ``2h'' refer to the so--called  ``one-halo'' and ``two-halo'' terms, respectively, and the $+1$
following each $\xigt$ term is to relate the correlation piece to the corresponding number counts.
  The ``1h'' term can be theoretically derived via the simple \citet{peebles1974} estimator,
\begin{equation}
    \xigt^{1h}(r) + 1 = \frac{DD^{1h}(r)}{RR(r)}.
\end{equation}
The $RR(r)$ term is the expected number of pairs consisting of randomly-distributed galaxies and $t$ particles
separated by distance between $r$ and $r+\mathrm{d}r$,
\begin{equation}
    RR(r) = g (4\pi r^2 \mathrm{d}r) \overline{n}_g \bar{\rho}_t V,
\end{equation}
where $\overline{n}_g$ and $\bar{\rho}_t$ are the mean densities of sample galaxies and $t$ particles,
respectively, and $V$ is the survey volume. The prefactor $g$ is $1$ if $t$ is a different species than the
primary galaxies~($\xigm$, with $\bar{\rho}_t\equiv\bar{\rho}_m$), and $1/2$ if the two are
identical~($\xigg$, with $\bar{\rho}_t\equiv\overline{n}_g$). The $DD(r)$ term requires separate treatment of
central and satellite galaxies within each halo first, and then integration over the halo mass function,
\begin{multline}
    DD^{1h}(r) = V \mathrm{d}r \int \frac{\mathrm{d}n}{\mathrm{d}\mh} \left[ DD_{{\mathrm{cen}}, t}(\mh) F_{{\mathrm{cen}}, t}^\prime\left( r | c_t\right)\right.  \\ +
    \left. DD_{{\mathrm{sat}}, t}(\mh) F_{{\mathrm{sat}}, t}^\prime\left( r | c_g, c_t\right) \right]
    \mathrm{d}\mh,
    \label{eqn:dd1h}
\end{multline}
where $DD_{{\mathrm{cen}}, t}$ and $DD_{{\mathrm{sat}}, t}$ are the total numbers of the ${\mathrm{cen}}$--$t$
and ${\mathrm{sat}}$--$t$ type of pairs expected within a halo of mass $\mh$, respectively, and
$F_{{\mathrm{cen}}, t}(r)$ or $F_{{\mathrm{sat}}, t}(r)$ is the cumulative probability distribution of the
numbers of each of these two pair types within that halo~(although Equation~\eqref{eqn:dd1h} requires their
respective differential forms, described further below).  For $\xigg$,
\begin{equation}
  DD_{{\mathrm{cen}}, g}(\mh) = \avg{N_{\mathrm{cen}}(\mh)N_{\mathrm{sat}}(\mh)},
\end{equation}
and
\begin{equation}
  DD_{{\mathrm{sat}}, g}(\mh) = \avg{N_{\mathrm{sat}}(\mh)(N_{\mathrm{sat}}(\mh)-1)}/2,
\end{equation}
while $F^\prime_{{\mathrm{cen}}, g}(r)$ and $F^\prime_{{\mathrm{cen}}, g}(r)$ are the galaxy NFW profile of
concentration $c_g(\mh)$ and the convolution of that galaxy NFW profile with itself, respectively.  Similarly
for $\xigm$,
\begin{equation}
DD_{{\mathrm{cen}}, m}(\mh) = \avg{N_{\mathrm{cen}}(\mh)} \mh,
\end{equation}
and
\begin{equation}
DD_{{\mathrm{sat}}, m}(\mh) = \avg{N_{\mathrm{sat}}(\mh)}\mh,
\end{equation}
while $F^\prime_{{\mathrm{cen}}, m}(r)$ and $F^\prime_{\mathrm{ sat }, m}(r)$ are the dark matter NFW profile
of concentration $c_{dm}(\mh)$ and the convolution of that dark matter profile with the galaxy NFW profile of
concentration $c_g(\mh)$, respectively.  We use the analytic formula in ~\citet{zheng2007} for the
convolution of two NFW profiles, either with the same or different concentration parameters. All the $F(r)$
functions are normalised so that $F(r=2 r_{200m})$ is unity, i.e., we assume both the satellite galaxies and
dark matter are contained within the virial radii of their host halos. Recently, the study by
\citet{vandaalen2015} suggested that it is important to account for the matter outside halos when estimating
the small scale matter power spectrum using the halo model. However, this missing power problem does not exist
for predicting galaxy clustering and g-g lensing, as the (sub)halos occupied by galaxies are generally massive
enough to be fully accounted in the calculations.

The ``two-halo'' terms are relatively straightforward to calculate above the halo exclusion scales~(%
$r_{\mathrm{ex}}{\sim}3\hmpc$, described further below), via
\begin{equation}
    \xigg^{2h}(r) = b_g^2 \zeta^2(r)\ximm(r),
    \label{eqn:xigg2h}
\end{equation}
and
\begin{equation}
    \xigm^{2h}(r) = b_g \zeta(r) r_{\mathrm{cc}}(r) \ximm(r),
    \label{eqn:xigm2h}
\end{equation}
where $b_g$ is the galaxy linear bias, $\zeta(r)$ characterises the fractional scale--dependence in the large
scale halo bias, and $r_{\mathrm{cc}}(r)$ is the cross--correlation coefficient between the galaxies and the dark
matter on relevant scales, defined by
\begin{equation}
    r_{cc}(r) = \frac{\xigm(r)}{\sqrt{\xigg(r) \ximm(r)}}.
    \label{eqn:rcc}
\end{equation}
To a good approximation $r_{\mathrm{cc}}(r)$ is close to unity on scales above $r_{\mathrm{ex}}$~\citep{guzik2001,
weinberg2004, baldauf2010}, and so we set $r_{\mathrm{cc}}(r)\equiv 1$ and use the empirical fitting function of
$\zeta(r)$ from~\citet{tinker2005}. We computed $b_g$ as the galaxy occupation number-weighted halo bias,
\begin{equation}
    b_g = \overline{n}_g^{-1} \int b(\mh) \frac{\mathrm{d}n}{\mathrm{d}\mh} \mathrm{d}\mh,
\end{equation}
where $b(\mh)$ is the halo bias function~(see the sixth column of Table~\ref{tab:smbins} for the average
linear bias of each {\ihod} sample).

Within the halo exclusion regime~\citep{cacciato2009}, however, Equations~\eqref{eqn:xigg2h}
and~\eqref{eqn:xigm2h} are no longer valid because the centre of one halo cannot sit within the virial radius
of another halo, i.e., $r_{\mathrm{ex}}{<}2{\times}r_{\mathrm{200m}}^{\mathrm{max}}$, where
$r_{\mathrm{200m}}^{\mathrm{max}}$ is the maximum halo radius in the calculation.  Therefore, the ``two-halo''
terms at a given distance require explicit integration over pairs of halos that are too small to run into the
centre of each other when separated by that distance. We follow the prescription described in
\citet{tinker2005} for the treatment of halo exclusion and adopt the method in \citet{yoo2006} to circumvent
the issue of unsatisfied integral constraints in the halo mass function and halo bias function~(see Equation
16 of \citealt{yoo2006} for details).

\subsection{Subhalo Contribution to $\ds$ in MassiveBlack-II}
\label{subsec:subhalo}

\begin{figure*}
    \centering\resizebox{0.95\figtextwidth}{!}{\includegraphics{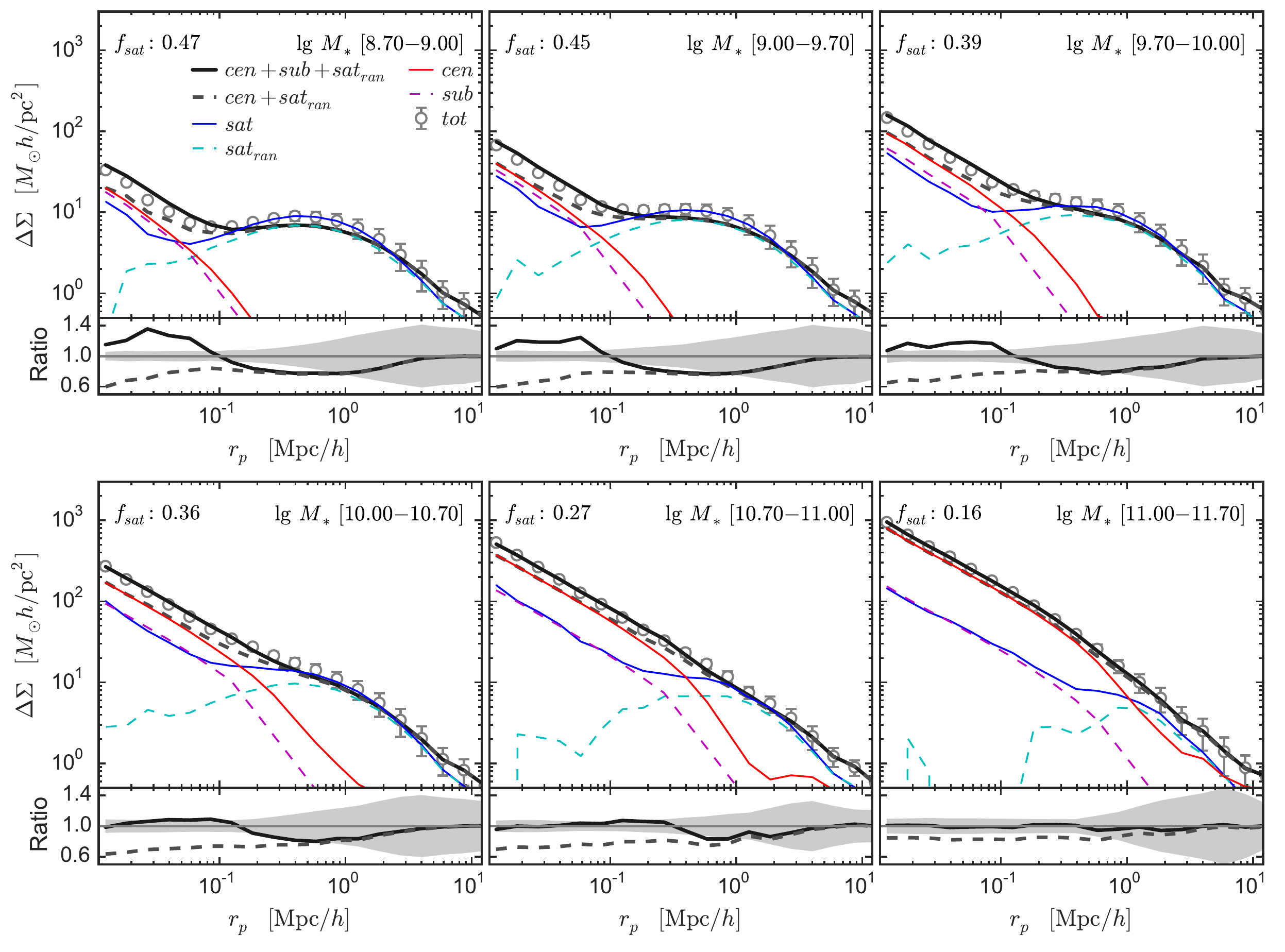}}
    \caption{
    The g-g lensing signal of six different stellar mass samples in the MB-II simulation at $z{=}0.06$, highlighting
    the importance of including the subhalo contribution in the modelling of the surface density contrast $\ds$.
    In each upper sub-panel, open circles are the total $\ds$ signal measured directly from the simulation,
    which is the sum of the red and the blue solid curves, indicating the contributions from the centrals and
    the satellites, respectively. The cyan dashed curve represents the satellite contribution measured after
    randomising the position angles of the simulated satellites~(distances to the halo centre kept fixed). The
    magenta dashed curve accounts for the subhalo contribution by re-scaling the central contribution according
    to the satellite fraction listed on the top left and truncating the 3D density profile at $0.4
    r_{\mathrm{200m}}$. The black solid curve is the sum of the
    subhalo~(magenta), the central~(red), and the randomised satellite~(cyan) curves, and the black dashed
    curve the sum of only the latter two. The ratio of the black solid and the black dashed curves to the open
    circles is shown on each lower sub-panel. Ignoring the subhalo contribution results in an underestimate of
    the total signal by $15\%\text{--}35\%$ on small scales, roughly proportional to the satellite fraction in
    each sample. For the three low mass samples, using the re-scaled and truncated central curve as a rough estimate for the
    subhalo contribution overestimates the total signal by $10\%\text{--}35\%$ on small scales, and by
    $\sim15\%$ in the transitional regime, albeit within the measurement uncertainties indicated by the gray
    band. For the high mass samples the black curves agree with the direct measurements within the
    uncertainties across all scales.
}\label{fig:sublens}
\end{figure*}

The extra ``subhalo'' lensing term, as mentioned in Section~\ref{subsec:xids}, arises because the satellite
galaxies are sitting at the local density peaks~(i.e., their subhalos), rather than random positions within
their main halos. From investigating a suite of N-body cosmological simulations, \citet{mandelbaum2005} found
that the fractional contribution of the subhalo term on scales below $0.1\hmpc$ is roughly equal to the
satellite fraction within the sample, but is cut off on slightly larger scales because of the tidal truncation
of those subhalos inside larger halos. They discovered that a truncation of the subhalos at ${\sim}0.4$ times
the virial radii gives a good match to the lensing signal measured from the simulations.  A subsequent
exploration by Yoo et al. (2006), however, estimated that the error induced by ignoring the subhalo
contribution is below $10\%$ on the relevant scales using a suite of relatively small Smoothed Particle
Hydrodynamic~(SPH) simulations. It is still controversial as to whether the modelling of the subhalo lensing
is necessary.  While some studies~\citep{velander2014, hudson2015} adopted the ``tidally stripped'' subhalo
lensing model proposed in~\citet{mandelbaum2005}, other studies~\citep[e.g.,][]{leauthaud2012-a, coupon2015}
ignored the lensing from subhalos, either based on the result of Yoo et al. (2006), and/or the fact that the
measured total lensing signal is much less constraining than other probes used in their papers~(SMF and galaxy
clustering).

We revisit the subhalo lensing issue by employing the
MassiveBlack-II\footnote{\url{http://mbii.phys.cmu.edu/}} SPH
  simulation~\citep[MB-II;][]{khandai2014}, a P-GADGET cosmological hydrodynamic simulation evolved with a
  total of $2\times 1792^3$ dark matter
and gas particles~(mass resolution $\sim$ several times $10^6\hmsol$) inside a cubic volume of $100^3
\hmpccubed$. MB-II is one of the highest resolution simulations of this size which includes a self-consistent
model for star formation, black hole accretion and associated feedback.  Thanks to its exquisite capability of
resolving subhalos down to $10^9\hmsol$ with realistic baryonic effects~(for the details on the HOD of
simulated galaxies in MB-II, see Tucker et al, {\it in prep}), we quantify the impact of the often--ignored
subhalo lensing term, by comparing the g-g lensing signals measured from the MB-II galaxies with subhalos and
from a sample of mock galaxies without. Following~\citet{yoo2006}, in order to construct a subhalo-less mock
galaxy sample, we identify the satellite galaxies in each main halo, and randomise their position angles
relative to the halo centre while keeping their halo-centric distances fixed.  In this way, we ensure that the
only difference between the lensing signals measured from the MB-II galaxies and the mock galaxies on small
scales is induced by the presence vs.\ absence of dark matter within the subhalos.

Figure~\ref{fig:sublens} shows the results
of this experiment in six different stellar mass bins at $z{=}0.06$, with $\ms$ increasing from left to right and top to
bottom. In each top sub-panel, the satellite fraction and the stellar mass range are marked on the top. The
open circles with errorbars indicate the measurement for the total g-g lensing signal for that stellar mass
bin, decomposed into the central~(red solid) and satellite~(blue solid) terms. The errorbars are derived from
jackknife resampling of the simulation volume.  The cyan dashed line shows the satellite term measured from
the mock galaxies, which decreases rapidly on small scales due to the lack of subhalos. The magenta dashed
line is a crude estimate of the subhalo lensing term via re-scaling the amplitude and truncating the central
term~(red solid) at $0.4\,r_{\mathrm{200m}}$. The amplitude is re-scaled by $f_{\mathrm{sat}}/(1 -
f_{\mathrm{sat}})$, assuming the subhalos that host satellites share the same inner matter density profile as
those host central galaxies of the same stellar mass, analogous to the ansatz employed in the SHAM
technique. The black solid line is the sum of this estimated subhalo term, the directly measured central term,
and the satellite term from the mock sample, serving as a rough estimate of the predicted total signal with
the subhalo lensing contribution, whereas the black dashed line is the sum of central and mock satellite terms
without any subhalo contribution. The difference between the open circles and the black dashed lines
represents the magnitude of the error on g-g lensing due to ignoring the subhalo lensing piece, and the
difference between the open circles and the black solid lines represents the deficiency of the
overly--simplified ``re-scaled central'' subhalo lensing model in over-predicting or under-predicting the
signal at various scales.

The effects of ignoring the subhalo lensing term are better illustrated by the bottom sub-panels of
Figure~\ref{fig:sublens} where the black solid and dashed lines shown above are divided by the measured total
signals. The gray shaded region indicates the typical uncertainties on the ratios propagated from the
measurement uncertainties on $\ds$.  The ratio plots clearly demonstrate
that, ignoring the subhalo lensing term causes $15\%$--$35\%$ systematic under-prediction of the total signal
on scales below $\sim 0.2\hmpc$, and the deviation is proportional to the fraction of satellite galaxies in
the sample. This contradicts the result from \citet{yoo2006} who found the effect generally below $10\%$ at
all radius. The opposite conclusions drawn from the two sets of simulations could be due to the drastic
difference in the resolution and size --- MBII has $2000$ times more particles in $8$ times the volume of
the simulation used by \citet{yoo2006}, and/or the average mass of the subhalos used in the two
experiments~(i.e., \citealt{yoo2006} could only go to subhalos above $10^{12}\hmsol$, hence much lower satellite
fraction).  Once the simple ``re-scaled central'' subhalo lensing  term is added, the predicted signal agrees
with the direct measurement within the uncertainties for the three high stellar mass bins in
the bottom row, but over-predict the total signal by $15\%$--$35\%$ on the relevant scales for the three low
stellar mass bins in the top row.  This discrepancy between the simple remedy and the simulation in the low
mass samples is caused by the enhanced tidal attenuation effect on the low mass subhalos from their host
halos, besides the usual tidal truncation effect seen for subhalos of all masses. This attenuation effect is
consistent with the findings in~\citet{li2014}, who measured the subhalo lensing signal using the satellite
lens galaxies selected from the SDSS group catalogue constructed from a redshift--space Friends-of-Friends
algorithm~\citep{yang2007}. They found significant $\ds$ signal below projected radius~$\sim 0.1\hmpc$ for
satellite lenses located $0.1$--$0.5\hmpc$ away from their respective group centres, and the amplitude of the
subhalo lensing signal can be explained by a truncated and attenuated version of those NFW halos that host
central galaxies of the same stellar masses.

Drawing from these findings in the MB--II simulation and the SDSS groups, we model the subhalo contribution to
the galaxy--matter cross--correlation function $\xigm^{\mathrm{sub}}$ as an attenuated and truncated version
of the central term $\xigm^{\mathrm{cen}}$,
\begin{equation}
    \xigm^{\mathrm{sub}}(r|\ms) = \left\{ \begin{array}{ll}
            f_{t} \times \left[\frac{f_{\mathrm{sat}}}{1-f_{\mathrm{sat}}}\xigm^{\mathrm{cen}}(r|\ms)\right],&\mbox{ $r<r_{t}$} \\ 0,&\mbox{ $r\geq r_{t}$,}
\end{array} \right.
\label{eqn:sublens}
\end{equation}
where $f_t$ and $r_t$ is the attenuation factor and the truncation radius, respectively.  We adopt a
truncation radius of $r_t{=}0.4\,r_{200m}(\ms)$, according to the findings in \citet{mandelbaum2006} and
Figure~\ref{fig:sublens}.  The value of $f_t$ should be stellar mass dependent --- we adopt $f_t$ values of
$0.5$ and $1.0$ for stellar mass samples below and above $10^{10}\hhmsol$, respectively, informed by the
results in Figure~\ref{fig:sublens}.

\begin{figure*}
    \centering\resizebox{0.95\figtextwidth}{!}{\includegraphics{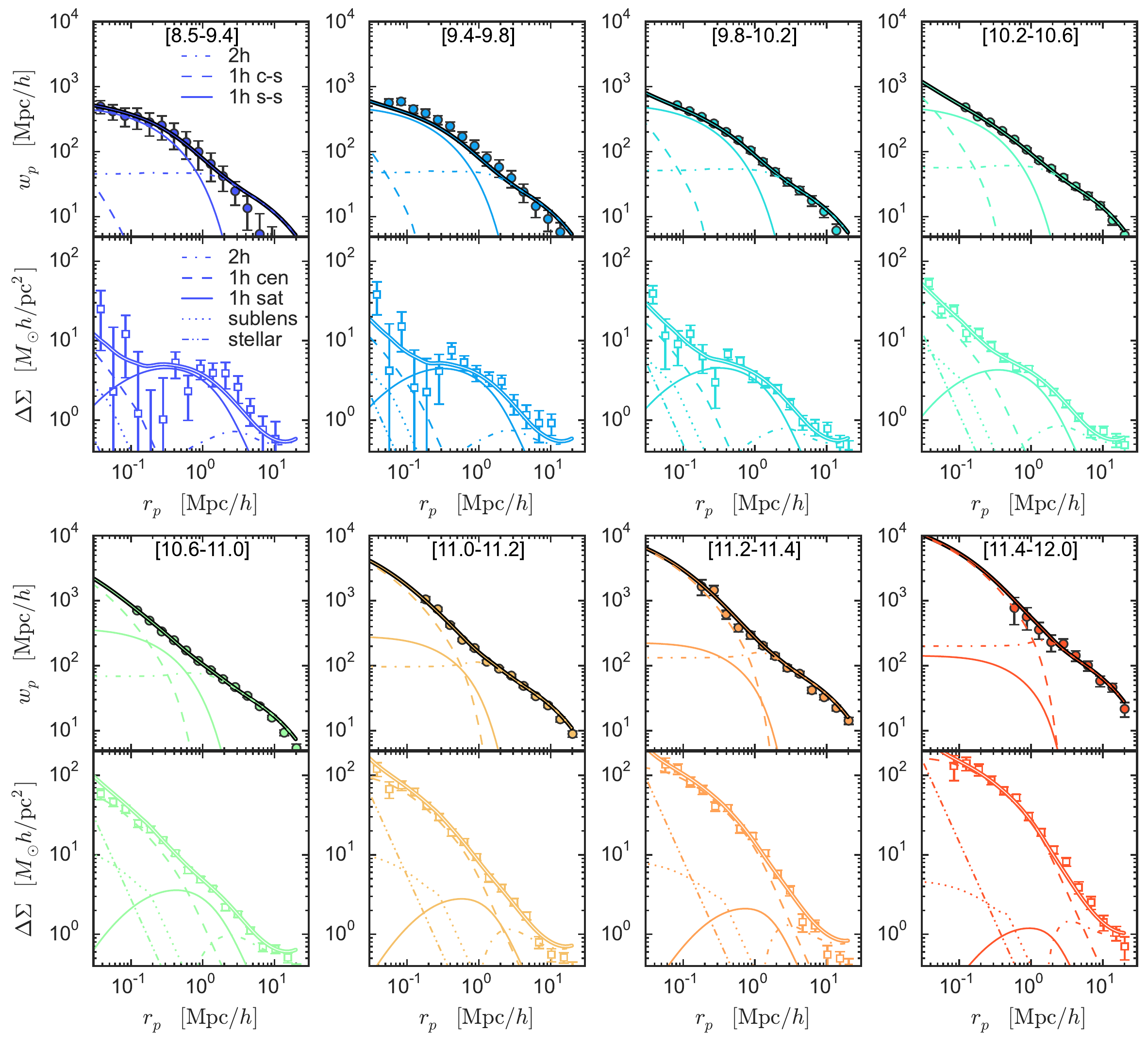}} \caption{ Comparison
    between the galaxy clustering and g-g lensing measurements from SDSS with the signals predicted by the
    best-fit {\ihod} model, for the eight stellar mass-selected samples. For each sample, the top and bottom
    panels show the projected correlation function $w_p$ and g-g lensing signal $\ds$, respectively. In each
    panel, the data points with errorbars are the measurements, and the thick line is the best-fit signal,
    which is decomposed into the 2-halo~(thin solid), the 1-halo central~(thin dashed), and the 1-halo
    satellite~(thin dot--dashed) contributions shown underneath. The dotted and the dot--dot--dashed lines in
    the lower sub-panels represent the lensing contributions from the subhalo dark matter and the galaxy
    stellar mass, respectively. The x-/y-axis ranges are uniform across all panels. The model provides
    excellent fit to both the clustering and lensing signals of galaxies over four decades in stellar mass.
}\label{fig:datavspred}
\end{figure*}

However, as mentioned in Section~\ref{subsec:onehalo}, the small--scale lensing modelling is further
complicated by the miscentering effect, which is absent from Figure~\ref{fig:sublens}. As pointed
out in
\citet{yoo2006}, the miscentering effect smooths the overall $\ds$ signal on small scales with a kernel of
length $\sim 0.1\hmpc$, and is likely to have a large impact for the higher stellar mass bins, i.e., within
groups and clusters. The smoothing effect will effectively reduce the
predicted signal for the central galaxies at the $10\%$--$15\%$ level on scales below $0.1\hmpc$, where the
subhalo lensing effect operates in the opposite direction. Since the satellite fraction at the high mass
bins is $\sim 20\%$, this reduction can be effectively absorbed by a low $f_t$ value of $\sim 0.5$ in
Equation~\eqref{eqn:sublens} for the high mass bins. Therefore, to simultaneously capture both the effects of
subhalo lensing at low $\ms$ and miscentering at high $\ms$ on scales below $0.1\hmpc$, we adopt a constant
attenuation factor of $f_t=0.5$ in our analysis, regardless of the stellar mass of the sample. We have tried
different values of $f_t=0.5\pm 0.2$ and found that our results are robust to those changes, due to the
relatively large statistical uncertainties of the g-g lensing measurements on small scales~(see Figure~\ref{fig:datavspred}).

For each redshift slice within the {\ihod} samples and for each individual {\chod} sample, the predictions for
the $w_p$ and $\ds$ signals are obtained by projecting $\xigg$ and $\xigm$ to 2D, according to
Equation~\eqref{eqn:wp} and~\eqref{eqn:ds}, respectively. We adopt the same $r_\pi^{\mathrm{max}}=60\hmpc$ for
the theoretical predictions of $w_p$ as for the measurements.
Finally, for each {\ihod} sample we combine the predictions from all
its redshift slices to obtain the predictions for that entire sample via Equations~\eqref{eqn:wpi}
and~\eqref{eqn:dsi}.

Figure~\ref{fig:datavspred} compares the predictions from our best--fit model~(thick lines) to the
measurements from data for the eight {\ihod} stellar mass samples. The galaxy clustering and g-g lensing
results are shown in each pair of the top and bottom sub-panels, respectively. Each best-fit curve is
decomposed into the two--halo~(thin dot--dashed), the one--halo central~(thin dashed), and the one--halo
satellite~(thin solid) contributions. For the best-fit g-g lensing signal of each sample, we also show the
contributions from the subhalo lensing term and the ``point source'' stellar mass as the thin dotted and
dot--dot--dashed lines, respectively. The subhalo lensing term begins to dominate the one--halo satellite
contribution at around half the virial radius of the average main halo corresponding to that sample, and then
the stellar mass term takes over at the galactic scales below tens of $\hkpc$.  We will come back to
Figure~\ref{fig:datavspred} for the detailed comparison between the model fits and the measurements in
Section~\ref{subsec:post}.

\section{Parameter Constraints}
\label{sec:constraints}

\subsection{Likelihood Model and Bayesian Inference}
\label{subsec:mcmc}

\begin{figure*}
    \centering\resizebox{0.95\figtextwidth}{!}{\includegraphics{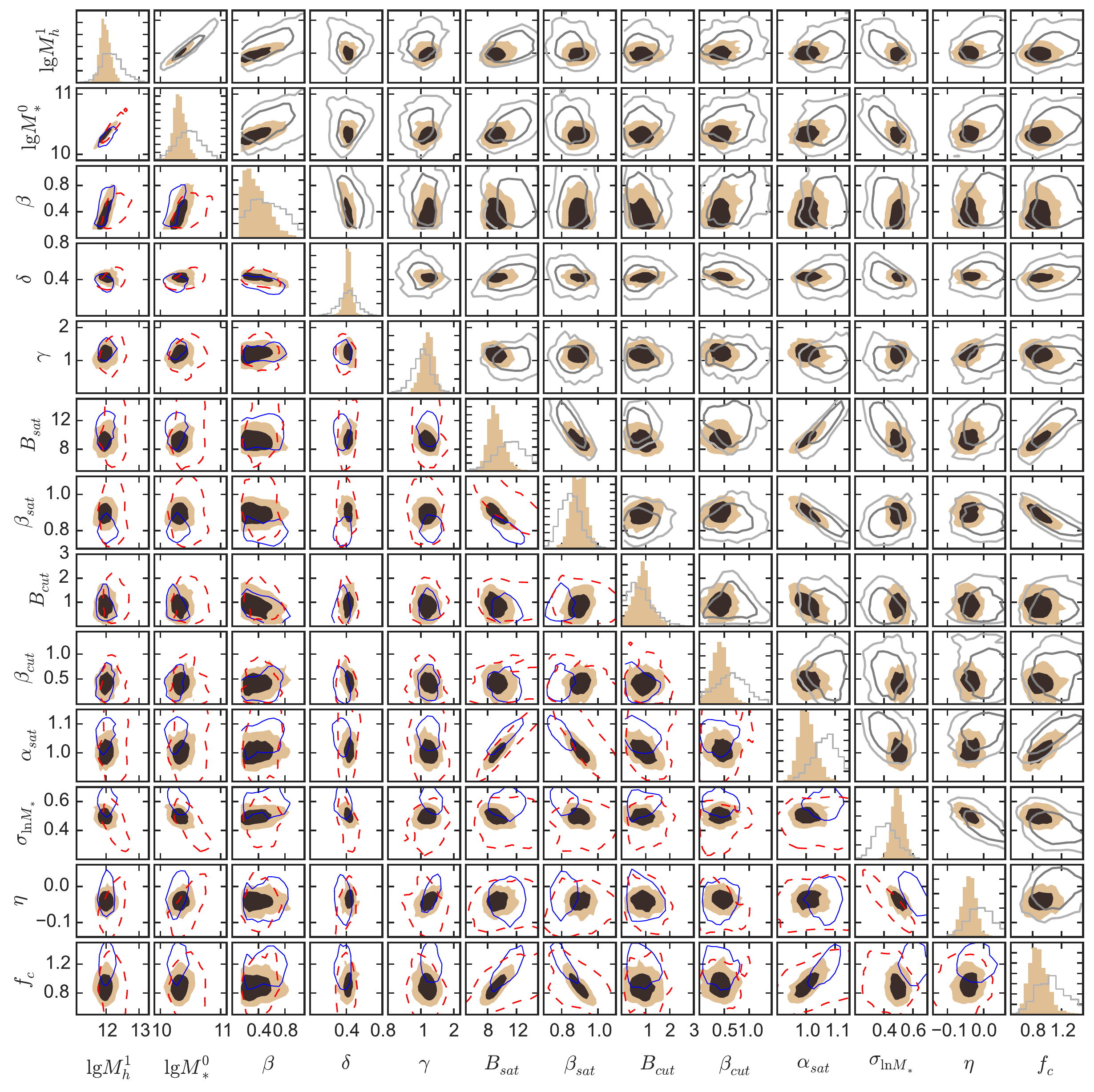}} \caption{Confidence
	regions from our {\ihod} analysis of the galaxy clustering and the g-g lensing data in the 2D planes
	that comprised of all the pair sets of model parameters. Histograms in the diagonal panels show 1D
    posterior distributions of individual parameters.  For comparison, constraints from the {\chod} analysis
    are only shown in the panels of the upper triangle.
    Contour levels run through confidence limits of $95\%$ (light brown/gray) and $68\%$ (dark brown/gray)
    inwards.  The tighter {\ihod} constraint compared to {\chod} is due to the combination of improved overall
    statistics from larger samples and additional information from the low mass galaxies.  In the panels of
    the lower triangle, blue solid and red dashed contours indicate
    the $95\%$ confidence regions of the two separate {\ihod} constraints using the clustering and the
    lensing data, respectively.
} \label{fig:glory}
\end{figure*}

\begin{figure*}
    \centering\resizebox{0.95\figtextwidth}{!}{\includegraphics{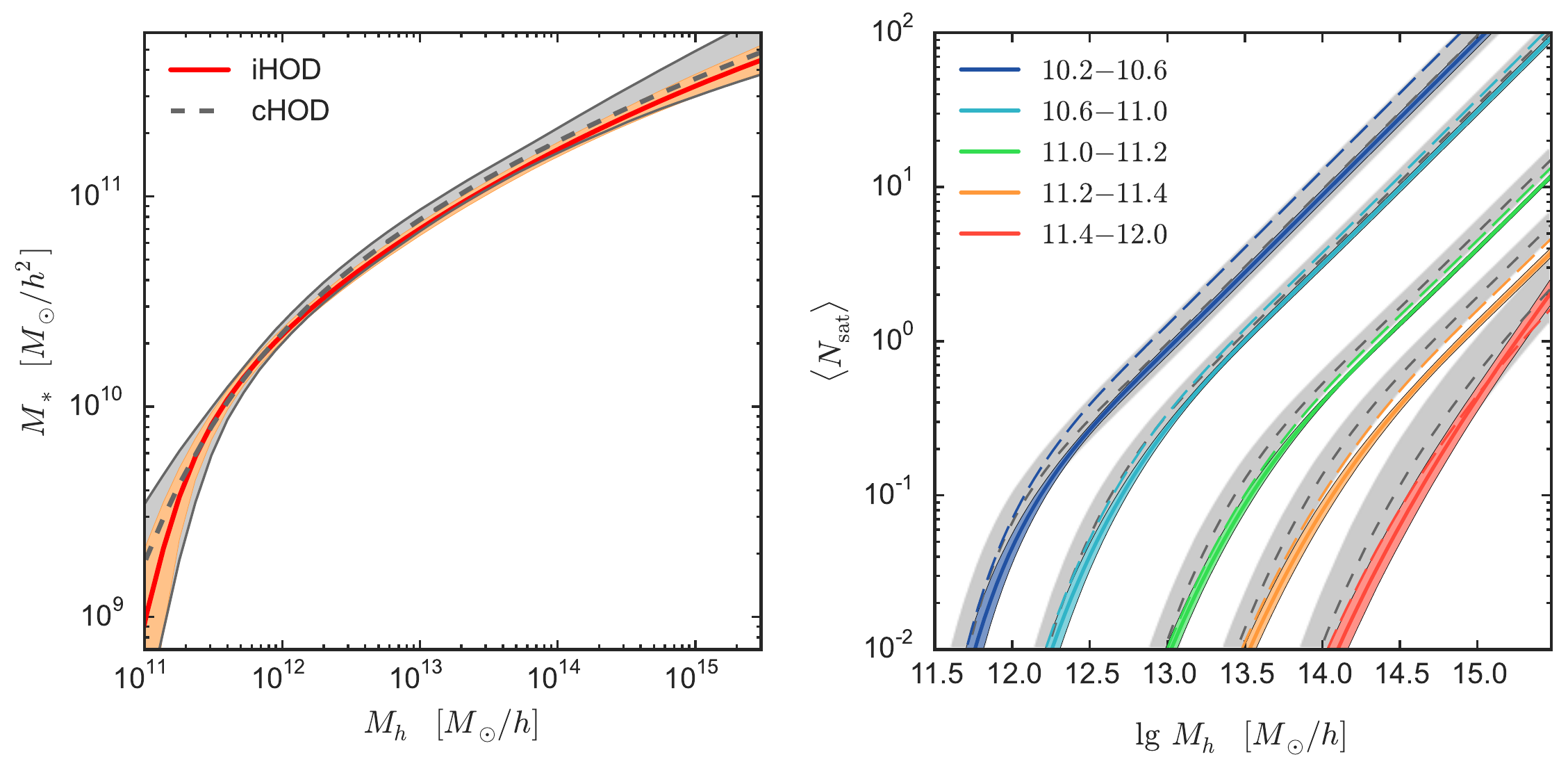}} \caption{
    Constraints on the SHMR for the central galaxies~(left) and the satellite HODs for four stellar mass
    bins~(right), as a more visually appealing and scientifically informative presentation of the confidence regions shown in
    Figure~\ref{fig:glory}. In the left panel, the red solid and the gray dashed lines show the
    inferred SHMR
    from the {\ihod} and {\chod} analysis, respectively, and the shaded regions represent the $68\%$ confidence
    range at fixed halo mass.
    In the right panel, solid coloured and dashed gray lines show the {\it expected}
    HODs for the {\it observed} galaxies from the best-fit {\ihod} and {\chod} model, respectively, with each
    shaded band indicating the $68\%$ confidence range at fixed halo mass; long--dashed coloured lines show the {\it
    parent} HODs predicted from the best-fit {\ihod} model. The difference between the solid and long--dashed
    coloured lines implies possible incompleteness in the stellar mass samples.
}\label{fig:cvi}
\end{figure*}

Armed with the capability of predicting the galaxy clustering and g-g lensing signals for any given {\chod} or
{\ihod} sample, we can infer the posterior probability distributions of the model parameters from the
measurements within a Bayesian framework, assuming a Gaussian likelihood model and a set of
uninformative priors on those parameters.

The observable vector in our likelihood model has two components:
\begin{itemize}
\item[1.] $w_p^i(r_p^j)$: $w_p$ profile of stellar mass sample $i$
    measured at projected radius $r_p^j$~(from $r_p^{\mathrm{fb}}$ to $20.0\hmpc$), for $i \in
    \{1\cdots8\}$~({\ihod}) or $\in \{1\cdots6\}$~({\chod}), and $j \in \{1\cdots n\}$, where $n$ is the
    number of data points used for each stellar mass samples.  Ranked by ascending order in the average stellar
    mass, the samples have $n \in \{17,16,15,14,14,13,13,10\}$~({\ihod}) or  $n \in
    \{16,15,14,14,13,10\}$~({\chod}), due to the different fibre-collision induced cutoffs on small scales.
    There are $112$~($82$) data points in total for the {\ihod}~({\chod}) analysis.
\item[2.] $\ds^i(r_p^k)$: $\ds$ profile of stellar mass sample $i$
    measured at radius $r_p^k$, ranging from $r_p^{\mathrm{cut}}$ to $20.0\hmpc$, where $r_p^{\mathrm{cut}}$
    is the small-scale lensing cutoff caused by the systematic uncertainties in estimating the ``boost
    factor'' for bright samples~(see Appendix for more details). We use the same set of stellar mass samples
    as clustering in each analysis.  The minimum fitting scale is $25\hkpc$ for the samples below
    $\lg\ms{=}10.6$, but increases as the sample gets brighter due to large $r_p^{\mathrm{cut}}$, reaching
    $0.12\hmpc$ for the highest stellar mass bin. Therefore, $k \in \{1\cdots n^\prime\}$, where
    $n^\prime \in \{18,18,18,18,17,17,16,15\}$~({\ihod}) and  $\in \{18,18,17,17,16,15\}$~({\chod}).
    There are $137$~($101$) $\ds$ data points in total for the {\ihod}~({\chod}) analysis.
\end{itemize}
We model the combinatorial vector $\mathbf{x}$ of the $w_p$ and the $\ds$ components as a multivariate
Gaussian~($N{=}256$ variables in total for {\ihod} and $N{=}190$ for {\chod}), which is fully specified by its
mean vector~($\bar{\mathbf{x}}$) and covariance matrix~($C$). The Gaussian likelihood is thus
\begin{equation}
    \mathcal{L}(\mathbf{x} | \boldsymbol{\theta}) =
    |C|^{-1/2}\exp\left(-\frac{(\mathbf{x}-\bar{\mathbf{x}})^TC^{-1}(\mathbf{x}-\bar{\mathbf{x}})}{2}\right),
\label{eqn:gauloglike}
\end{equation}
where
\begin{multline}
    \boldsymbol{\theta} \equiv \{\lg\mh^1,\lg\ms^0,\beta,\delta,\gamma,B_{\mathrm{sat}},\\
    \beta_{\mathrm{sat}},B_{\mathrm{cut}}, \beta_{\mathrm{cut}}, \alpha_{\mathrm{sat}}, \sig, \eta, f_c\}.
\end{multline}
We adopt flat priors on the model parameters, with a uniform distribution over a broad interval that covers
the entire possible range of each parameter~(see the 3rd column of Table~\ref{tab:pripos}). The final
covariance matrix $C$ is assembled by aligning the error matrices of $w_p$ and $\ds$ measured for individual
samples along the diagonal blocks of the full $N \times N$ matrix. We ignore the weak covariance between $w_p$
and $\ds$ (with the covariance being weak due to the fact that $\ds$ is dominated by shape noise), and between
any two measurements of the same type but for different stellar mass samples.

In the parameter inference stage, the posterior distribution is derived using the Markov Chain Monte
Carlo~(MCMC) algorithm \texttt{emcee}~\citep{foreman-mackey2013}, where an affine-invariant ensemble sampler
is utilised to fully explore the 13-D parameter space. For each MCMC chain of the {\ihod} and {\chod}
inferences, we perform $90,000$ iterations, $30,000$ of which belong to the burn-in period for adaptively
tuning the steps. To eliminate the tiny amount of residual correlation between adjacent iterations, we further
thin the chain by a factor of $10$ to obtain our final results.

\subsection{Posterior Probability Distributions}
\label{subsec:post}

\begin{table*}
\centering \caption{Description, prior specifications, and posterior constraints of the parameters in the
model. All the priors are uniform distributions running across the entire range of possible values for the
parameters, and the uncertainties are the $68\%$ confidence regions derived from the 1D posterior probability
distributions.}
\begin{tabular}{ccccc}
\hline
\hline
Parameter & Description & Uniform Prior Range & {\ihod} & {\chod}\\
\hline
 $\lg\mh^1$ & Characteristic halo mass of the SHMR & [9.5, 14.0]                   &$ 12.10_{-0.14}^{+0.17} $  & $ 12.32_{-0.29}^{+0.29} $   \\
 $\lg\ms^0$ & Characteristic stellar mass of the SHMR & [9.0, 13.0]                &$ 10.31_{-0.09}^{+0.10} $  & $ 10.47_{-0.21}^{+0.18} $   \\
 $\beta$    & Low-mass slope of the SHMR & [0.0, 2.0]                              &$  0.33_{-0.15}^{+0.21} $  & $  0.54_{-0.26}^{+0.29} $   \\
 $\delta$   & Controls high-mass slope of the SHMR & [0.0, 1.5]                    &$  0.42_{-0.04}^{+0.03} $  & $  0.42_{-0.09}^{+0.08} $   \\
 $\gamma$   & Controls intermediate-mass behaviour of the SHMR & [-0.1, 4.9]        &$  1.21_{-0.20}^{+0.18} $  & $  1.05_{-0.26}^{+0.24} $   \\
 $B_{\mathrm{sat}}$ & Normalises the scaling of $M_{\mathrm{sat}}$ & [0.01, 25.0]  &$  8.98_{-0.87}^{+1.18} $  & $ 11.22_{-1.99}^{+2.61} $   \\
 $\beta_{\mathrm{sat}}$ & Slope of the scaling of $M_{\mathrm{sat}}$ & [0.1, 1.8]  &$  0.90_{-0.05}^{+0.04} $  & $  0.85_{-0.05}^{+0.06} $   \\
 $B_{\mathrm{cut}}$ & Normalises the scaling of $M_{\mathrm{cut}}$ & [0.0, 6.0]    &$  0.86_{-0.37}^{+0.32} $  & $  0.73_{-0.44}^{+0.58} $   \\
 $\beta_{\mathrm{cut}}$ &Slope of the scaling of $M_{\mathrm{cut}}$ & [-0.05, 1.50]&$  0.41_{-0.15}^{+0.16} $  & $  0.63_{-0.33}^{+0.31} $   \\
 $\alpha_{\mathrm{sat}}$& Power-law slope of the satellite HOD & [0.5, 1.5]        &$  1.00_{-0.02}^{+0.03} $  & $  1.07_{-0.05}^{+0.05} $   \\
 $\sig$                 & Low-mass scatter in the SHMR & [0.01, 3.0]               &$  0.50_{-0.03}^{+0.04} $  & $  0.42_{-0.08}^{+0.07} $   \\
 $\eta$                 & Slope of the scaling of high-mass scatter & [-0.4, 0.4]  &$ -0.04_{-0.02}^{+0.02} $  & $ -0.01_{-0.03}^{+0.04} $   \\
 $f_c$    & Concentration ratio between satellites and dark matter & [0.1, 3.0]    &$  0.86_{-0.11}^{+0.14} $  & $  1.06_{-0.20}^{+0.32} $   \\
\hline
\end{tabular}
\label{tab:pripos}
\end{table*}
Figure~\ref{fig:glory} presents a summary of the inferences for both the {\ihod}~(brown filled) and
{\chod}~(gray open) analyses, showing
the 1D posterior distribution for each of the 13 model parameters~(diagonal panels), and the $95\%$ and $68\%$
confidence regions for all the parameter pairs~(off--diagonal panels). In the panels of the lower triangle, we
highlight the results from our fiducial model, i.e., the {\ihod} model, employing the clustering and g-g
lensing from the entire galaxy population above the mixture limit and self--consistently accounting for the
survey incompleteness in stellar mass. In those panels we also show the $95\%$ confidence regions from the
{\ihod} constraints that employ the clustering and the lensing signals separately. The constraint from using
the clustering alone~(blue solid contours) is significantly tighter than that from the g-g lensing~(red dashed
contours), due to the much higher S/N in the $w_p$ measurements. However, the g-g lensing is absolutely
essential in the joint {\ihod} analysis, helping break the degeneracy between the scatter parameters~($\sig$
and $\eta$) and the slope of the SHMR on the high mass end~($\delta$).  In each panel of the upper triangle,
we compare the constraints from the fiducial model~(filled contours) to that of the traditional approach,
i.e., the {\chod} model, which is limited to $54\%$ of the {\ihod} galaxies and assumes the samples to be
complete~(open contours).  While the two constraints are largely consistent with each other, the {\ihod}
analysis is able to obtain a much tighter constraint because of the larger span in stellar mass range and the
higher number of galaxies~(hence the higher $S/N$ of measurements illustrated in Figure~\ref{fig:compsn}).  In
particular, the {\ihod} analysis substantially improves the constraints on the pivot of the SHMR $(\lg\mh^1,
\lg\ms^0)$ and its high--mass end slope $\delta$. The five parameters that describe the satellite HOD benefit
the most from the inclusion of low stellar mass samples, with much stronger constraint in {\ihod} than in
{\chod}.  The $68\%$ confidence regions of the 1D posterior constraints are listed in Table~\ref{tab:pripos}.

The {\ihod} model favours a scatter of $\sig=0.50_{-0.03}^{+0.04}$ about the mean SHMR with a slightly negative
slope of $\eta=-0.04\pm{0.02}$, while the {\chod} model infers a constant scatter of
$\sig=0.42_{-0.08}^{+0.07}$.  However, the two constraints converge to a similar scatter on the high mass end,
where the {\ihod} model shrinks its mass--dependent scatter to meet the lower constant value preferred by the
{\chod} model. While the {\chod} does not constrain the concentration ratio $f_c$ well, the {\ihod} strongly
favours the scenario where the satellite distributions are $\sim 15\%$ less concentrated than the dark matter
within the same halos~($f_c=0.86_{-0.11}^{+0.14}$). This galaxy under-concentration agrees with observational
findings in \citet{wang2014} and \citet{budzynski2012}~(but see~\citealt{watson2012} for the degeneracy
between $f_c$ and the inner slope of the cluster density profiles).

The improvement on the overall constraints from {\ihod} to {\chod} is better illustrated in
Figure~\ref{fig:cvi}, where we translate the uncertainties of the individual model parameters to that of the
mean SHMR and the satellite HOD separately.  In the left panel, the red solid and gray dashed curves are the
mean SHMR inferred from the {\ihod} and {\chod} analyses, respectively.  The width of each shaded band, i.e.,
the $68\%$ uncertainty in the mean log-stellar mass at fixed halo mass, directly reflects the joint posterior
probability distribution of the five parameters~($\lg\mh^1, \lg\ms^0, \beta, \delta, \gamma$) that determine
the mean SHMR.  The two constraints are consistent with each other, especially on the high mass end. This is
very reassuring --- as we mentioned in Section~\ref{subsec:cvsi} the two models are identical to each other
for the high stellar samples, because the completeness approaches unity and is independent of halo mass.  As
expected from Figure~\ref{fig:glory}, the uncertainties in the mean SHMR are greatly reduced across all halo
masses in the {\ihod} constraint, with the most significant reduction happening above
$\mh{=}10^{13}\hmsol$~(by factor of two in log--stellar mass). Furthermore, both of the mean SHMRs are best
constrained at the intermediate mass range around the pivotal point $(\lg\mh^1, \lg\ms^0)$, due to the highest
model sensitivity and the highest $S/N$ at this stellar mass.

Similarly, in the right panel of Figure~\ref{fig:cvi} we show the constraints on the satellite occupation
number predicted for five of the stellar mass samples that were used in the {\chod} analysis. The gray dashed
lines show the median satellite occupation numbers as functions of halo mass inferred from the {\chod}
analysis, with their 1-$\sigma$ uncertainties shown as the corresponding gray bands. For comparison, we have
predicted two types of satellite HODs for these five {\chod} samples from the best-fit {\ihod} model, one is
for the {\it parent} satellite galaxies~(thin dashed lines), and the other for the {\it observed}
galaxies~(thick solid lines). To avoid clutter, we only show the uncertainty bands associated with the
inferred {\it observed} satellite HODs, all of which are considerably narrower than the gray bands, especially
for the two highest stellar mass samples. This improvement is very encouraging. As we will show later in
Section~\ref{subsec:csmf}, the occupation statistics of the satellite galaxies, long being regarded as a mere
nuance in the HOD modelling of galaxy clustering and HOD-based cosmological constraints~\citep{yoo2012}, is
constrained well enough in the {\ihod} analysis to provide important insights into the physical formation and
evolution of the satellite galaxies.

It is important to point out that, when calculating the HODs for the {\it observed} galaxies in
Figure~\ref{fig:cvi}, we have made use of the actual number of observed galaxies in each sample in order to
normalise the amplitude of $\avg{N_{\mathrm{sat}}(\mh)}$ using Equation~\eqref{eqn:nmhihod}, while the {\it
parent} HODs are predicted directly from the best--fit model parameters using Equation~\eqref{eqn:nmhchod}. As
mentioned in Section~\ref{subsec:pre}, the {\ihod} analysis constrains the host halo distribution of the
average galaxy in each sample by matching to the observed clustering and g-g lensing signals, and is thus
entirely agnostic of the observed amplitude of the stellar mass functions. Therefore, it is highly nontrivial
to discover in the right panel of Figure~\ref{fig:cvi} that, all the predicted {\it parent} HODs are lying
right above their {\it observed} counterparts, consistent with our expectation that the galaxy samples above
the mixture limit are subject to some mild level of incompleteness. We will inspect closely this consistency
later in Section~\ref{subsec:smf}.

Figure~\ref{fig:datavspred} compares the clustering and g-g lensing signals measured from data to that
predicted from the set of best-fit parameters derived from the median values of their corresponding  1D
posterior probability distributions. We refer to this set of ``median'' parameters as the ``best--fit''
parameters for simplicity, despite that it does not produce the maximum posterior probability or likelihood
value. The lowest stellar mass sample is subject to severe cosmic variance due to small volume, resulting in a
relatively poor $w_p$ fit on large scales. The model also under-predict the small scale clustering signal for
the $\lg\ms=[9.4, 9.8]$ sample, which is largely overrun by the CfA2 great wall~\citep{geller1989}.  Overall,
the model provides an excellent fit to the data over almost four decades in stellar mass, across distance
scales from the galactic disks to tens of Mega-parsecs.

\subsection{Systematic Uncertainties in the {\ihod} Model}
\label{subsec:syserr}

Observationally, the systematic errors in our fiducial {\ihod} analysis come from the uncertainties in the
stellar mass estimates and in the measurements of the projected correlation functions and the g-g lensing
signals. As mentioned in Section~\ref{subsec:mstar}, the estimation of $\ms$ is subject to various photometric
and model uncertainties that mostly affect the overall normalization of the derived stellar masses, with
little perturbation in the ranking order of individual galaxies within the catalogue. Therefore, the inferred
mapping between the stellar mass and the dark matter halos from our analysis can be straightforwardly
re-calibrated for other stellar mass estimators.

The uncertainties in the $w_p$ and the $\ds$ measurements are estimated internally from the data via jackknife
resampling. Compared to external estimates derived from multiple independent catalogues, the jackknife
estimate somewhat overestimates the correlation function errors on scales below
$r_p{\sim}2\hmpc$~\citep{norberg2009}, while on large scales it likely underestimates the errors because it
does not include the cosmic variance above the scales of individual subsamples.  Therefore, the jackknife
errors for the two lowest stellar mass/redshift samples are underestimated on large scales, because of the
small physical size of the jackknife patches. The error budget of these two stellar mass samples, however, is
still dominated by the uncertainties on small scales, so the fiducial constraint is insensitive to the
underestimation in the error covariance on large scales, despite the inadequate fit to the $w_p$ measurements
of the two lowest stellar samples on relevant scales.

The systematic errors in the $\ds$ measurements have additional contributions from the calibration biases
mainly related to the shear estimation and the photo-$z$ errors, each at a few per cent level~(see
Appendix~\ref{app:smallscale} for the scale-dependent lensing systematics). \citet{mandelbaum2013} addressed
these systematics by conducting a suite of ratio tests~\citep{mandelbaum2005}, i.e. comparisons of the signal
computed using the same lens samples, but with different sub-samples of the source catalogue. After applying
well-understood corrections, the systematic errors are sub-dominant compared to the statistical errors and
compared to the systematic uncertainties due to modelling assumptions, which we will discuss next.

The theoretical systematic errors in our fiducial {\ihod} analysis have four main sources: 1) simplified model
assumptions in the {\ihod} formalism, 2) uncertainties in the theoretical prediction of halo statistics like
the halo mass function and the halo bias function, 3) our ansatz of the weak $\mh$-dependence of the
detection rate at fixed $\ms$, and 4) uncertainties in the combined treatment of the mis-centering and
subhalo lensing.  We have discussed 3) in detail in Section~\ref{subsec:cvsi} and will comment more on its
potential impact in Section~\ref{subsec:smf}, and the impact from 4) has been discussed and addressed in
Section~\ref{subsec:subhalo}. Therefore, here we focus on the first two issues and discuss their potential
impacts on our model constraints in turn.

As a variant of the HOD formalism, the {\ihod} model inherits the systematic uncertainties associated with the
generic HOD models, which assume that the average galaxy content of halos depends solely on the halo mass,
predicted by the basic excursion set theory of structure formation~\citep{press1974, bardeen1986,
  bond1991}. However, since the halo assembly histories in cosmological simulations are affected by the large-scale
environment~\citep{wang2007, dalal2008}, some halo properties, including age, concentration, spin, richness,
and most importantly clustering, exhibit systematic differences between low and high density environments at
fixed halo mass~\citep{sheth2004, gao2005, wechsler2006, harker2006, zhu2006, hahn2007, jing2007, li2008,
    faltenbacher2010, croft2012}.  This effect, broadly termed ``halo assembly bias''~\citep{gao2007}, has
likely left its imprint on galaxy formation histories as well, resulting in a ``galaxy assembly bias''.
However, hydrodynamic simulations and SAMs predict only a small impact~(${<}10\%$) of the halo assembly
bias on galaxy clustering statistics~\citep{yoo2006, croton2007, zu2008}, while observationally a smoking-gun
detection of the galaxy assembly bias remains elusive~\citep{berlind2006, blanton2007, wang2013, hearin2014}.
Additionally, the halo assembly bias becomes only significant for halos below the characteristic non--linear
mass scale, where our error budget is dominated by statistical errors and cosmic variance. However, it is
worth noting that halo assembly bias may have a much greater impact on the HOD modelling of colour--selected
galaxy samples, and could dominate the error budget on all mass scales for both colours~\citep{zentner2014}.

The cause for the theoretical uncertainties in predictions of the halo mass and bias functions is three--fold:
i) the Universe may have a non-$\lcdm$ cosmology \citep[for halo statistics beyond $\lcdm$,
see][]{bhattacharya2011, cui2012b, ichiki2012, murray2013, zhang2013, loverdel2014} or a different set of
cosmological parameters \citep[especially $\sigma_8$; see][for details on the tension between different
probes]{planck2015} than the particular $\lcdm$ cosmology we adopt; ii) the predictions should be inaccurate
due to the unaccounted-for baryonic effects~\citep[at $5$ to $20$ per cent level depending on the mass scale
and the feedback models, see][]{cui2012, cusworth2014, velliscig2014}; and iii) the predictions may be poorly
calibrated on the high mass end where the halos are rare \citep[at the $5$ per cent level for
$\mh{=}10^{15}\hmsol$ at $z{=}0$, see][]{crocce2010, watson2013, bocquet2015}. Among the three types of
errors, the calibration errors affect few galaxies as the probability of any observed galaxies sitting in
a halo with mass above $10^{15}\hmsol$ is extremely low, and are thus negligible in our analysis. Errors due to wrong
cosmology, ignored baryonic effects, or the combination of both, change the halo mass and bias functions
together in a coherent manner that can be roughly mimicked by changing $\Omega_m$ and/or $\sigma_8$ within the $\lcdm$
model. In particular, changing $\Omega_m$ at fixed $\sigma_8$ shifts the predictions along the halo mass axis almost
uniformly~(albeit with a minor change in the functional shapes due to the change in the power spectrum), while
changing $\sigma_8$ at fixed $\Omega_m$ modifies the amplitude of the predictions more progressively at higher halo
mass~\citep[see figure 7 in ][for a pedagogical illustration]{zu2014}. We explore the impact of i) and ii) by
perturbing $\Omega_m$ and $\sigma_8$ using both theoretical arguments and mock tests below.

In our analysis on large scales, we are effectively using the halo bias vs.\ halo abundance relation
compressed from the halo mass and bias functions, i.e., the bias of $\mh$--thresholded halo samples
$b(\bar{n})$ as a function of the co-moving number density of that sample, $\bar{n}({>}\mh)$. Due to the
exponential decline of the abundance of massive halos, the theoretical uncertainties in this
$b(\bar{n})$--$\bar{n}({>}\mh)$ relation are only important at the high mass end, where the galaxy sample is
primarily composed of massive BCGs with small satellite fraction.  In the context of the peak
background--split theory of biased clustering~\citep{kaiser1984, cole1989, mo1996, sheth1999}, for very rare,
very highly biased peaks, $b(\bar{n})\,{\propto}\,\sigma_8^{-1}$, yielding
$w_p(\bar{n})\,{\propto}\,b^2(\bar{n})\xi_{mm}\,{\propto}\,b^2(\bar{n})\sigma_8^2\,{\simeq}\,\mathrm{constant}$,
and $\ds\,{\propto}\,\Omega_m b(\bar{n})\xi_{mm}\,{\propto}\,\Omega_m\sigma_8$, respectively. These two approximations hold
even in the presence of baryons. Therefore, the clustering is barely affected by the changes in $\Omega_m$ or
$\sigma_8$, while the lensing is linearly responsive to the change in the product of $\Omega_m$ and $\sigma_8$. Since the
galaxy clustering measurements have a better overall $S/N$ than the g-g lensing measurements, a $10\%$ change
in $\Omega_m\sigma_8$ should produce a shift in the SHMR smaller than $0.06$ dex in $\mh$~(if all of our constraints
came from large scales). To confirm this, we re-run the analysis using two different combinations of $(\Omega_m,
\sigma_8){=}(0.26, 0.80)$ and $(0.28, 0.80)$. For the first case where we change $\sigma_8$ alone, the shift in the
derived SHMR is indiscernible, whereas for the latter case where we increase the product of $\Omega_m$ and $\sigma_8$ by
$12\%$, the derived SHMR shifts to the lower halo mass by ${\sim}0.08$ dex, corresponding to an increase in
the stellar mass at $\mh{=}10^{13}\hmsol$ by only ${\sim}0.03$ dex, with little change in the parameters that
control the satellite HOD.

The remaining sources of theoretical systematic errors, including the uncertainties in the mean halo
mass--concentration relation $c_{\mathrm{dm}}(\mh)$~(we also ignore the dispersion in the halo concentration
at fixed $\mh$), calibration in the scale--dependence of halo bias $\zeta(r)$, and the deviation of the
cross-correlation coefficient $r_{\mathrm{cc}}(r)$ from unity, are expected to be much smaller than the four
main errors discussed above. Summing all the systematic errors in quadrature, we would find the total
systematic error comparable to or even dominate over the statistical uncertainties in the measurements.
However, as pointed out in \citet{coupon2015}, each of these systematic errors affects the inference
independently with different stellar mass and scale dependencies and we fit all of them jointly, the combined
impact of these systematic errors should be sub-dominant compared to the statistical errors.

\section{Stellar Mass Content within Halos}
\label{sec:smc}

In this section, we explore the implications of our inferred HOD parameters for the connection
between stellar and halo mass.

\subsection{Stellar Mass Functions}
\label{subsec:smf}

\begin{figure*}
    \centering\resizebox{0.95\figtextwidth}{!}{\includegraphics{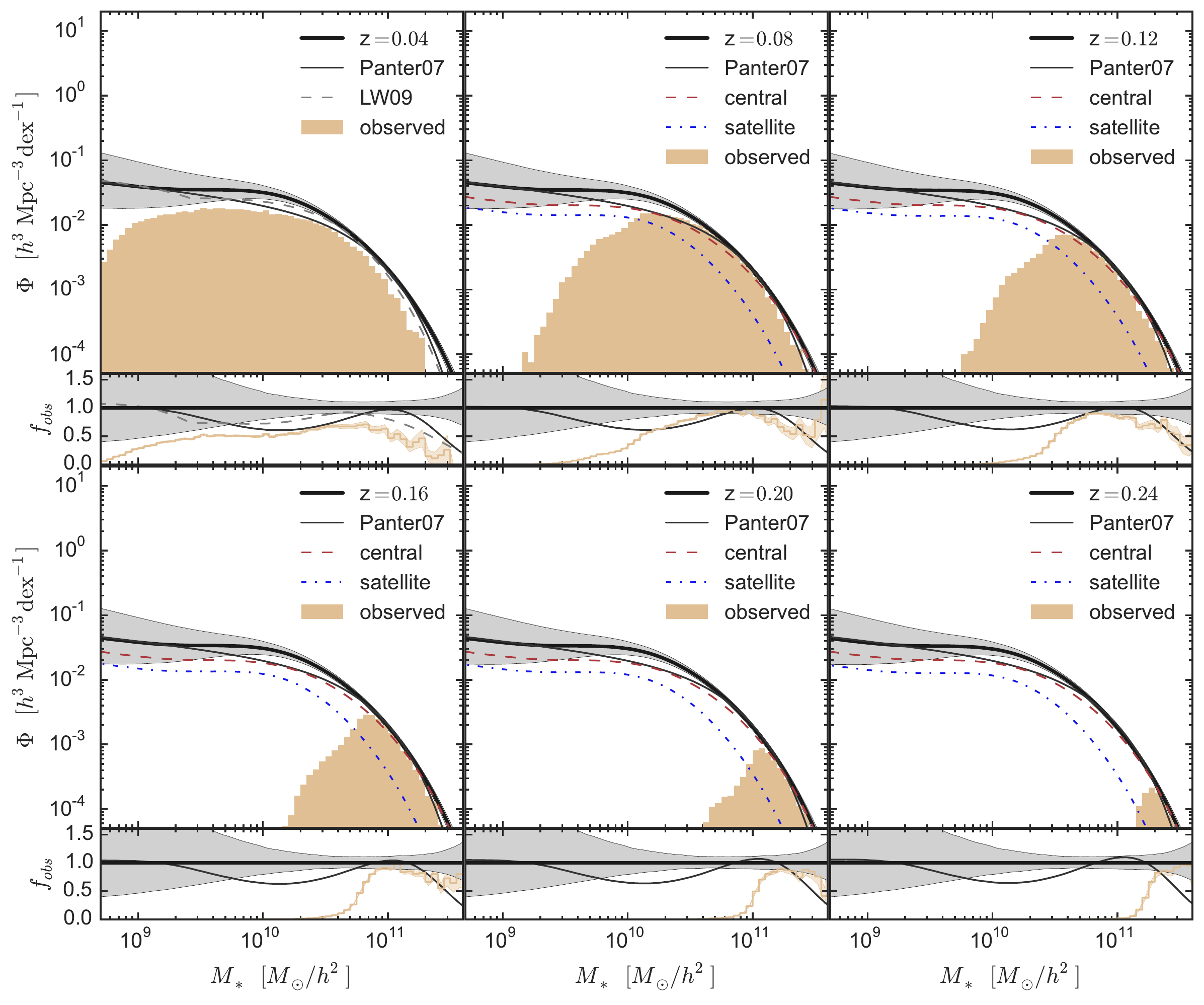}} \caption{
        Comparison between the observed~(brown histograms) and the {\it parent}~(thick solid curves with
        $68\%$ confidence range shown in gray bands) galaxy stellar mass functions predicted from the best-fit
        {\ihod} model at six different redshifts. In each upper sub-panel, the thin solid curve is the
        estimated SMF of SDSS DR3 galaxies from~\citet{panter2007}.  For the $z{=}0.04$ panel, we show another
        SDSS SMF estimated by~\citet{li2009} from the $z{<}0.05$ galaxies as the thin dashed curve~(after
        correction for the offset between the two stellar mass estimates); for the
        rest of the panels the predict parent SMF is decomposed into the contributions from the central and
        satellite galaxies, indicated by the red dashed and blue dot dashed lines, respectively.  The lower
        sub-panels show the ratios of the \citet{panter2007} (thin black) and the observed galaxy number
        density~(brown) to the predicted SMF from our model. The redshift evolution of the predicted SMF in
        the model is coming solely from the cosmic evolution of the halo mass function.}\label{fig:smf}
\end{figure*}

One of the most important consistency-checks for our fiducial {\ihod} model is to predict the {\it parent}
SMFs at each redshift, and compare to the {\it observed} versions as well as the intrinsic SMFs estimated
empirically, e.g., from the $V_{\mathrm{max}}$ method. As mentioned in Section~\ref{subsec:pre}, this check
echoes the philosophy of the abundance matching technique, but in practice is carried out in reverse. In
particular, the model uses the observed 2-point correlation function to infer the mapping from the stellar
mass content, including the central and satellite galaxies, to the dark matter halos, and the mapping in turn
gives an exact prediction~(via Equation~\ref{eqn:smf}) for the parent SMF at each redshift --- a
``correlation'' matching as opposed to ``abundance'' matching.

Figure~\ref{fig:smf} presents the results of this consistency check on SMFs at six different redshifts.  For
each redshift, the upper sub-panel compares the observed~(brown histogram), $V_{\mathrm{max}}$--estimated~(thin
black line), and our inferred parent~(thick black line with 1-$\sigma$ uncertainty band) SMFs.  The parent
SMF, predicted from the best-fit {\ihod} model at that redshift, is also decomposed into the central~(red
dashed) and satellite~(blue dot-dashed) contributions, except for the $z=0.04$ panel where we instead show
another measured SMF from~\citet{li2009}~(shifted by $0.1$ dex to remove the average bias between the two
stellar mass estimators; see figure A1 in their paper). The central galaxies dominate the satellite galaxies
in numbers across all stellar masses. The lower sub-panel shows the ratios of the observed~(brown line with
Poisson errors) and estimated SMFs~(thin black line) over the parent one, i.e., the observed fraction
$f_{\mathrm{obs}}$. The redshift evolution of the predicted parent SMFs is solely from the cosmic evolution of
the halo mass function, and the \citet{panter2007} SMF is the same in all panels. Although the stellar mass
estimates from~\citet{panter2007} were independently derived from the SDSS DR3 spectroscopy, the SMF~(also
used for the abundance matching in M10) provides a good fit to the observed SMF using the MPA/JHU stellar
masses, indicating little systematic offset between the two estimators.  Overall, at $\ms{>}5\times
10^{10}\hhmsol$ where the completeness of SDSS is expected to be high, the predicted parent SMFs show
remarkable consistency with the observed SMFs in both the shape {\it and} the amplitude at all redshifts. As
mentioned in Section~\ref{subsec:pre}, the {\ihod} analysis is completely agnostic to the total number density
of the galaxy catalogue~(i.e. the normalisation of the observed SMFs). Therefore, the agreement seen in
Figure~\ref{fig:smf} demonstrates that, in order to match the clustering and lensing signal measured in SDSS,
the placement of galaxies within the dark matter halos is uniquely determined so that the expected abundance
of the galaxies at each stellar mass, translated from the halo mass function predicted by the $\lcdm$
Universe, automatically agrees with the observed galaxy SMFs in SDSS.

For the two lowest redshifts, the observed SMFs at the high mass end are known to be notably incomplete due to
two photometric confusions: 1) to avoid saturation and excessive cross-talk in the spectrographs, some objects
were rejected because they either have saturated centres~(mostly bright active nuclei) or are blended with a
saturated star; and 2) the image deblending software sometimes over-deblended bright galaxies with large
angular extent~\citep{strauss2002}. A somewhat related issue is the over-subtraction of the sky background
mentioned in Section~\ref{subsec:mstar}, which results in suppressed flux estimates in bright galaxies.
Therefore, limited to the $z{<}0.05$ galaxies, the SMF measured by Li \& White has a substantially lower
amplitude at $\ms{>}3\times10^{10}\hhmsol$ than the Panter et al's SMF, the observed SMFs beyond $z{>}0.08$,
and the parent SMF predicted by our best-fit model.

We next consider the $f_{\mathrm{obs}}$ inferred from the ratio of the $V_{\mathrm{max}}$--estimated SMF to
the predicted parent SMF in the lower sub-panels of Fig.~\ref{fig:smf}. Starting with the low mass end,
$f_{\mathrm{obs}}$ is always above $60\%$, and reaches $100\%$ at above $6\times 10^{10}\hhmsol$ until
$2\times 10^{11}\hhmsol$, beyond which point the $V_{\mathrm{max}}$ method is expected to underestimate the
galaxy abundance due to the aforementioned photometric issues. Meanwhile at the high mass end, the {\ihod}
model predicts a much better match to the observed SMFs than the Panter et al's estimate does, reaching a
constant $f_{\mathrm{obs}}$ of $\sim 90\%$, especially at the two highest redshifts where the photometric
confusions that plague the lower redshifts are absent and the sample of very massive galaxies should be highly
complete.

At the mass scale around $10^{10}\hhmsol$, the $f_{\mathrm{obs}}$ inferred from the
$V_{\mathrm{max}}$--estimated SMF exhibits a dip at $60\%$--$70\%$ level in the lower sub-panels of
Figure~\ref{fig:smf}.  There are two main possible sources that could contribute to this dip, one theoretical
and one observational. On the one hand, the dip could be a manifestation of the systematic uncertainty in the
theoretical model. As seen from Figure~\ref{fig:mz}a, this mass range marks the transition of galaxies from a
predominantly blue, star--forming population below to a red, quiescent population above on the $\ms$--$z$
diagram. Given the strong correlation between galaxy colour and $\ms/L$, we expect that the ansatz we
have made, i.e., assuming a weak dependence of $f_{\mathrm{obs}}(\ms|z)$ on the host halo mass, to be the most
fragile at this transitional mass scale. For example, if the $\ms/L$ at fixed stellar mass is skewed higher in
more massive halos~\citep{taylor2015}, the observed $10^{10}\hhmsol$ galaxies would preferentially reside in
host halos less massive than does a typical $10^{10}\hhmsol$ galaxy in the Universe, and thus exhibit weaker
clustering and g-g lensing signals than their stellar mass suggests. As a result of this masquerade effect,
the {\ihod} model would place them inside smaller halos, resulting the over-prediction of their abundances. It
is worth noting that this theoretical systematic uncertainty is neither absent from the traditional HOD
modelling, nor would it be eliminated by adding the SMF data as an extra constraint.  On the other hand, the
photometric issues that affect the targeting of bright galaxies would also impact this intermediate mass
range, rendering the star-forming galaxy population underrepresented in the MGS at the low redshift. The
selection effect is easily discernible in Figure~\ref{fig:mz}a: at $\ms{\sim}10^{10}\hhmsol$, the galaxies at
redshifts lower than $0.04$ have a redder average $g{-}r$ colour than those at higher redshifts. Since the
majority of the detected galaxies with $\ms{\sim}10^{10}\hhmsol$ comes from $z{<}0.08$, they carry heavy
weight~(large values of $1/V_{\mathrm{max}}$) in the Panter et al's and Li \& White's SMFs. Therefore, the
under-representation of these galaxies at low redshift translates to potentially significant underestimate of
their abundances. In reality it is more likely that both the theoretical and observational sources contribute
to the dip at some level, and the true parent SMF should lie somewhere between the Panter et al's and our
predicted curves.

\subsection{Centrals: Stellar Mass to Halo Mass Relation}
\label{subsec:shmr}

\begin{figure}
    \centering\resizebox{0.48\figtextwidth}{!}{\includegraphics{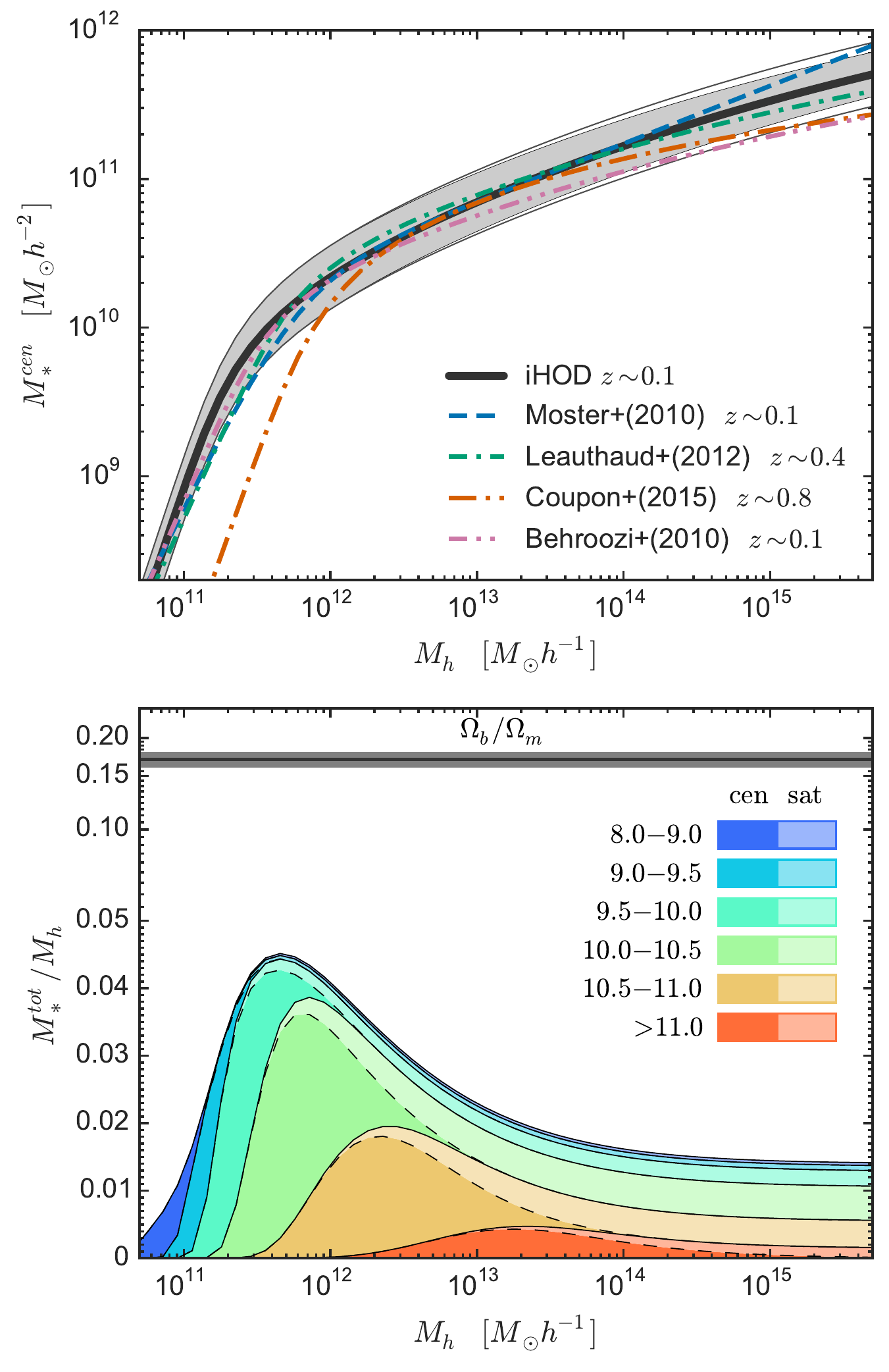}} \caption{ {\it Top panel:}
    Comparison between the SHMRs inferred from abundance--matching based methods and our best--fit {\ihod}
    model. The gray band shows the logarithmic scatter predicted by the best--fit model, while the two
    enveloping thin curves represent the range of constant scatter with $\eta{=}0$. Among the four external
    constraints used for comparison, the Moster et al. (2010) and the Behroozi et al. (2010) results are based
    on abundance--matching, while the other two are derived from fitting the clustering, lensing, and SMF
    jointly. {\it Bottom panel:} The total stellar mass to halo mass ratio as a function of halo mass.
    Coloured layers show the contribution from different stellar mass bins while separate portions of the same
    coloured layer indicate the relative contribution from centrals and satellites within each stellar mass
    bin. The horizontal band represents the cosmic baryon to dark matter ratio in the Universe. Note that the
    y-axis is logarithmic above $0.05$.} \label{fig:shmr}
\end{figure}

The most crucial piece for mapping the stellar content back to dark matter halos is  the SHMR of the central
galaxies. The top panel of Figure~\ref{fig:shmr} compares the SHMR and its logarithmic scatter derived from
our fiducial {\ihod} constraints~(black solid line with gray band) to that inferred from several methods,
including abundance matching. To highlight the shrinkage in scatter at high halo mass predicted by our
fiducial constraint, we also show two curves that delineate the boundaries of the constant scatter band with
$\eta=0$.  To avoid clutter, here we only show the scatter for the SHMR from our constraint.

Let us first compare to the constraint from~\citet{moster2010}~(red dashed), which employed the same cosmology
as our analysis and matches the halo abundance to the SDSS SMF measured by~\citet{panter2007}, albeit with a
different functional form for the SHMR and a fixed scatter. The two mean SHMRs show remarkable agreement over
$\lg\mh\sim[12$ -- $14]$, or $\ms\sim 2\times10^{10}$--$2\times10^{11}\hhmsol$, where the data are the most
constraining and the completeness is the highest. Above this mass range, the Moster et al's relation predicts
higher stellar mass for central galaxies than our mean SHMR for a given halo mass, i.e., a steeper SHMR.
Considering the lower amplitude of the $V_{\mathrm{max}}$--estimated SMF employed by the \citet{moster2010}
analysis for abundance matching, naively one might expect that the observed massive galaxies have to be pushed
to reside in progressively rarer and more massive halos, making their SHMR shallower instead. However, due to
the exponential decline of the halo mass function in this regime, the SHMR is much more sensitive to the
change in scatter than the small change in the detection rate. This discrepancy on the high mass end is mainly
caused by the smaller scatter assumed in the~\citet{moster2010} analysis~($\sig$$=0.345$, or $0.15$ in dex)
compared to that inferred by our model~($\sig{\sim}0.40$ at $\mh{=}10^{15}\hmsol$). Their smaller scatter
allows much fewer low--mass halos to be considered as hosts for the massive galaxies, thus requiring a steeper
slope in order to match the galaxy abundance.

The \citet{behroozi2010} analysis derives the constraints on the SHMR by abundance matching to the SDSS SMF
measured by~\citet{li2009} at $z<0.05$.  They carefully accounted for the effect of scatter in the SHMR by
varying both the scatters in the true stellar mass at fixed halo mass~(intrinsic scatter) and in the observed
stellar estimate at fixed true stellar mass~(measurement scatter). For the intrinsic scatter, a log-normal
prior centred at $0.16$ dex with a width of $0.04$ dex is placed~(i.e., $0.37\pm0.09$ in natural log), and
for the measurement scatter a fixed value of $0.07$ dex is applied for the SDSS data.  After abundance
matching, the posterior constraint on the intrinsic scatter is $0.15_{-0.02}^{+0.04}$ dex, consistent with
their input prior.  However, as clearly shown by the top left panel of Figure~\ref{fig:smf}, the
\citet{li2009} SMF is underestimated at the high mass end, thus shifting the mean SHMR horizontally toward
higher halo mass.  \citet{behroozi2010} also adopts a slightly higher $\sigma_8$ than M10 and our study, which
predicts more massive halos and further pushes the slope of the mean SHMR shallower than our constraint.

Unlike abundance matching, our fiducial constraint is able to break the degeneracy between the scatter and the
slope of the mean SHMR on the high mass end without assuming any external priors. The leverage comes from the
unique constraining power of combining clustering and lensing.  On large scales, the galaxy clustering signal
is proportional to $b_g^2 \ximm$, while the g-g lensing scales with $\Omega_m b_g \ximm$. On small scales, $w_p$ is
less a probe of the SHMR than $\ds$, which directly measures the average halo mass of individual galaxy
samples. For a fixed cosmology, simultaneous fit to $w_p$ and $\ds$ effectively constrains the clustering bias
as a function of halo mass, for which $\lcdm$ predicts a relatively steep slope on the high mass end~(see the
seventh and eighth columns of Table~\ref{tab:smbins}).  However, as shown by the open contours in the
$\delta$--$\sig$ panel in Figure~\ref{fig:glory}, the {\chod} constraint stills shows substantial degeneracy
between the slope parameter $\delta$ and the scatter, due to the inadequate $S/N$ in the measurements of $w_p$
and $\ds$. Thanks to the $84\%$ more galaxies used in the {\ihod} analysis, our fiducial analysis
derives much tighter constraint with minimum residual degeneracy~(filled contours).

The L12 result shown in Figure~\ref{fig:shmr} was derived from a slightly higher redshift
range~($0.22{<}z{<}0.48$) than the MGS used by us. Since we have adopted a similar model framework and
parameterization proposed in L12, the analysis of~\citet{leauthaud2012-a} is very similar to the {\chod}
analysis in this paper.  However, their analysis adds the measured SMF as part of the constraining data set
beside angular clustering and g-g lensing. In this regard, although allowing the scatter to vary freely, the
constraint in~\citet{leauthaud2012-a} is somewhat more related to the AM-based methods, because the smaller
uncertainties in SMF easily dominate the likelihood of the fit.  A straightforward comparison with the mean
SHMR of~\citet{leauthaud2012-a} is complicated by the differences in the redshift range of the galaxy samples
and the stellar mass estimators~(although the COSMOS SMF below $z{=}0.48$ agrees very well with that from
\citealt{panter2007}; see figure 5 of L12), but the two constraints appear to be qualitatively consistent with
each other above the characteristic halo mass. The \citet{coupon2015} constraint is very similar to L12,
derived from a joint lensing, clustering, and abundance analysis, but for galaxies at a much higher
redshift~($z{\sim}0.8$) and more massive than $10^{10}\hhmsol$.  Thus, the deviation between the
\citet{coupon2015} SHMR and ours at the low mass end can be attributed to the evolution of galaxy population
from $z{=}0.8$ to $z{=}0.1$. At the high end however, we do not see the steepening of the \citet{coupon2015}
SHMR at $z{\sim}0.8$, as hinted by the L12 constraints from $z{\sim}0.88$~(see figure 11 in L12).  The L12 and
the \citet{coupon2015} analyses also constrained the scatter in the SHMR, yielding $\sig{\simeq}0.474$~(i.e.,
$0.206$ dex) and $\sig{\simeq}0.506$~(i.e., $0.22$ dex; at $\mh{\simeq}10^{12}\hmsol$), respectively, in
excellent agreement with our constraint of $0.50_{-0.03}^{+0.04}$.  This level of agreement is not necessarily
expected, given that the scatter includes both intrinsic scatter and measurement uncertainty, the latter of
which could differ for the three different samples studied in these papers.

Below the characteristic halo mass, our mean SHMR has a slightly higher amplitude than all the other three
SHMRs, implying higher stellar-to-halo mass ratios for the low stellar mass galaxies. This difference echoes
the discrepancy seen in the comparison between the parent and measured SMFs at the intermediate and low
stellar mass scales. However, all those curves are still consistent with one another within the statistical
uncertainties in the inferred mean SHMRs below $\mh$$\sim$$10^{12}\hmsol$~(compare the difference to, e.g.,
the uncertainty band in the left panel of Figure~\ref{fig:cvi}).

\begin{figure*}
    \centering\resizebox{0.95\figtextwidth}{!}{\includegraphics{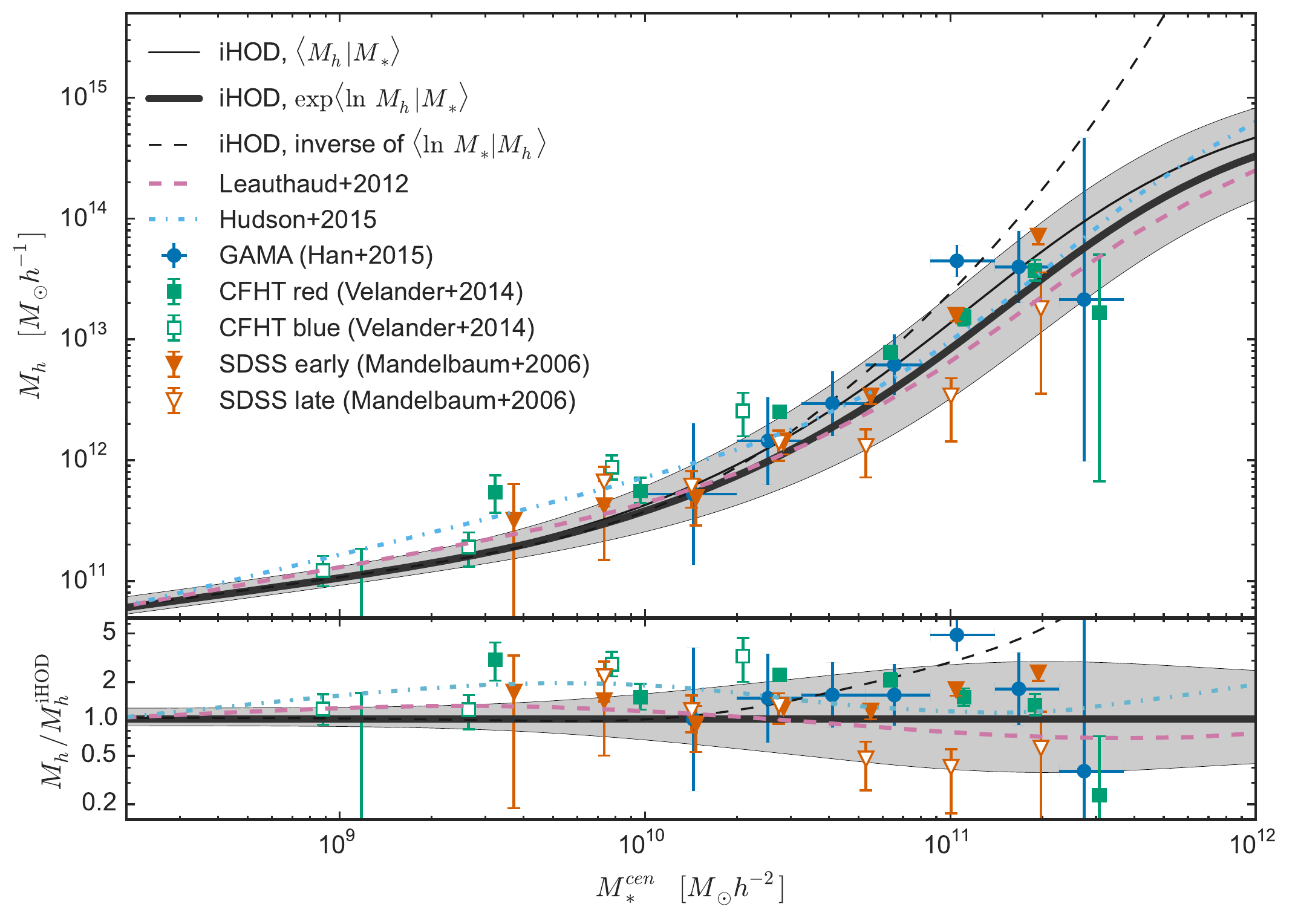}} \caption{Comparison
    between
	the halo-to-stellar mass relations inferred from the best-fit {\ihod} model and that from various
	other weak lensing measurements. {\it Top panel:} the dash black curve is just the inverse function of
	the best-fit SHMR, i.e., the same black curve in the top panel of Figure~\ref{fig:shmr} but with x-
	and y-axis swapped.  The thick black curve with a gray band indicate the mean and scatter~(not
	statistical uncertainty on the mean) of the distribution of logarithmic halo mass at fixed stellar
	mass, while the thin black curve shows the mean halo mass at fixed stellar mass. The magenta dashed
	and cyan dot-dashed curves are the results from
    L12 using COSMOS and \citet{hudson2015} using CFHTLenS, while the blue circles are from the
    maximum--likelihood weak lensing analysis of GAMA galaxies. The green and the red squares are the average
    weak lensing masses measured for CFHTLenS and SDSS galaxies, respectively, and within each survey the
    relations are constrained for red/early-type and blue/late-type galaxies separately using nearly identical
    HOD prescriptions. All the uncertainties shown for the weak lensing mass measurements are 1-$\sigma$.
    {\it Bottom panel:} Similar to the top panel, but showing the ratio of each quantity over the prediction
    by the best-fit {\ihod} model.
    }
    \label{fig:hsmr}
\end{figure*}

The efficiency of stellar mass assembly at the centre of halos, characterised by the central stellar-to-halo
mass ratio $\ms^{\mathrm{cen}}/\mh$, rises sharply as a function of halo mass at the low mass scale, probably
due to the increasing difficulty for the stellar feedback to drive cold gas out of the steepening
gravitational potential wells of halos~\citep{hopkins2012}.  The growth of the central galaxies shifts down
into a much lower gear beyond the characteristic mass ${\sim}10^{12}\hmsol$, decreasing
$\ms^{\mathrm{cen}}/\mh$ from its peak ${\sim}0.04$ at $\mh{\sim}10^{12}\hmsol$ to below $0.001$ at
$\mh{\sim}10^{15}\hmsol$. AGN feedback is believed to be one of the major culprits that quench the star
formation in those central galaxies within high mass halos. However, as shown by Figure~\ref{fig:h2p}, the
satellite ``cloud'' also begins to emerge within halos above $10^{12}\hmsol$, and a good indicator for the
efficiency of stellar mass assembly has to account for the stellar mass of satellites. To this end, the bottom
panel of Figure~\ref{fig:shmr} presents a more complete picture by showing the total stellar mass assembly
efficiency, $\ms^{\mathrm{tot}}/\mh$~(assuming $h{=}0.72$), integrated over all the predicted parent galaxies
above $10^{8}\hhmsol$. We also divide the total efficiency into contributions from six different stellar mass
bins marked in the legend. Within the layer of each stellar mass bin, we further split the contribution into
two components, the central~(shaded colour) and the satellite~(tinted colour), separated by a dashed line. For
comparison, the horizontal line on top indicates the universal baryon-to-matter mass ratio.  As expected, the
total efficiency is dominated by the central galaxies until its peak at around $\mh{\sim}10^{12}\hmsol$, then
it levels off and asymptote to a constant value~${\sim}0.015$ for galaxy groups and clusters, where the
satellite contribution dominates. This asymptotic value of the total stellar mass fraction is consistent with
observational studies by stacking galaxy catalogues or images inside samples of optical and/or X-ray
clusters~\citep{budzynski2014, bahcall2014}. Galaxies with $\ms$ around $10^{10}\hhmsol$ dominate the total
stellar mass content within halos of mass above $10^{12}\hmsol$.

The near--constant total stellar-to-halo mass ratio in group and cluster-size halos is very intriguing.  By
comparing the integrated stellar mass to the stacked weak lensing signal at different radii within the MaxBCG
clusters, Bahcall \& Kulier (2014) found that the total mass of clusters can be largely accounted for by the
total dark matter mass associated with all the subhalos that host the satellite galaxies.  Therefore, when the
fractional contribution from the central galaxy is negligible,
\begin{equation}
    \frac{\sum\ms}{\mh}\Big|_{\mh} \simeq \left[ \int \left(\frac{dN_{\mathrm{sat}}}{d\lg\ms}\Big|_{\mh}\right) \frac{\ms}{\mh^{\mathrm{sub}}} d\lg\ms
    \right] \Big/ N_{\mathrm{sat}}(\mh),
\end{equation}
where $(dN_{\mathrm{sat}}/d\lg\ms)|_{\mh}$ is the satellite stellar mass function conditioned at halo mass
$\mh$, and $\mh^{\mathrm{sub}}$ is the subhalo mass. If the satellite stellar-to-subhalo mass relation
$\ms/\mh^{\mathrm{sub}}$ were to roughly follow the SHMR of the centrals, as assumed by our subhalo lensing
model and the SHAM methods, the aggregate of the satellite galaxy population inside each halo is likely to be
a strong function of the halo mass when $dN_{\mathrm{sat}}/d\lg\ms$ differs from halo to halo.  In order for
any two halos with different masses to share the same total stellar-to-halo mass ratio, the stellar mass
functions within the two halos have to be somewhat self--similar --- all $dN_{\mathrm{sat}}/d\lg\ms$ have the
same shape near the knee of the SMF where most of the stellar mass is stored, with their amplitudes
$N_{\mathrm{sat}}$ linearly proportional to the halo mass. To test this hypothesis, we explore the conditional
SMFs of satellites below in Section~\ref{subsec:csmf}.

\begin{figure*}
    \centering\resizebox{0.95\figtextwidth}{!}{\includegraphics{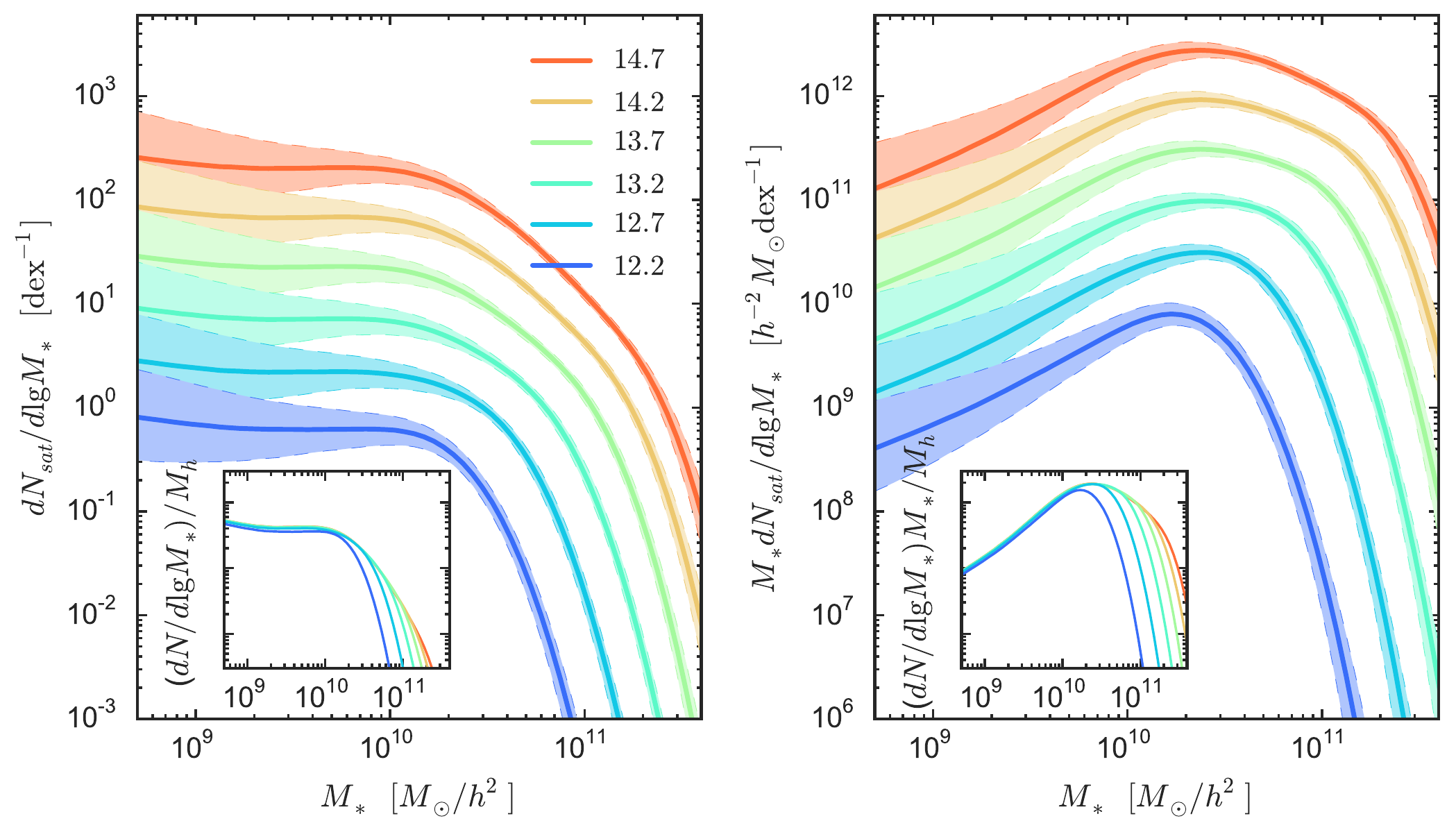}} \caption{ {\it Left panel}:
    Satellite stellar mass function $dN_{\mathrm{sat}}/d\ms$ conditioned at six different halo masses marked
    on the top right. The shaded regions indicate the $68\%$ uncertainties derived from the {\ihod}
    constraint. The inset panel shows the same set of curves, with each amplitude re-scaled by the
    corresponding halo mass, highlighting the departure of the conditional SMF from the Schechter functional
    form in high mass halos.  {\it Right panel}: Similar to the left panel, but for $\ms
    dN_{\mathrm{sat}}/d\ms$, the fractional contribution to the total stellar-to-halo mass ratio per dex in
    $\ms$. The satellite galaxies with stellar mass around a few times $10^{10}\hhmsol$ contribute the most to
    the total stellar mass content in each halo.
}\label{fig:csmf}
\end{figure*}

The SHMR, specifying the mean logarithmic stellar mass of central galaxies at fixed halo mass, is a convenient
choice from a theoretical modelling point of view, but observationally what is often measured is the opposite,
i.e., the mean log-halo mass or halo mass at given stellar mass. For instance, as the predecessor of the study
in this paper, \citet{mandelbaum2006} measured the average log-halo masses for a set of early and late-type
galaxy stellar mass samples in the SDSS, via the traditional HOD modelling of their g-g lensing signals. To
compare to the result from \citet{mandelbaum2006} and other similar studies in the literature, we compute the
distribution of the halo mass at fixed stellar mass as
\begin{equation}
    p(\mh|\ms^{\mathrm{cen}}) = \frac{p(\ms^{\mathrm{cen}}|\mh) p(\mh)}{p(\ms^{\mathrm{cen}})},
    \label{eqn:hsmr}
\end{equation}
where $p(\ms^{\mathrm{cen}}|\mh)$ is the SHMR of central galaxies~(similar to Equation~\ref{eqn:ncen}) and
$p(\mh)$ is the halo mass function normalised by the total number density of halos, while
$p(\ms^{\mathrm{cen}})$ is proportional to the SMF of central galaxies, calculated from
\begin{equation}
    p(\ms^{\mathrm{cen}}) = \int_0^{+\infty} p(\ms^{\mathrm{cen}}|\mh) p(\mh)\,\dd\mh.
    \label{eqn:pcms}
\end{equation}
The halo-to-stellar mass relation, $p(\mh|\ms^{\mathrm{cen}})$, computed from the best-fit SHMR and its
scatter shown in the top panel of Figure~\ref{fig:shmr} via Equations~\eqref{eqn:hsmr} and~\eqref{eqn:pcms},
is represented in Figure~\ref{fig:hsmr} in two forms, the mean log-halo mass at fixed stellar
mass~(thick black solid curve),
\begin{equation}
    \avg{\ln\mh|\ms} = \int p(\mh|\ms^{\mathrm{cen}})\ln\mh\,\dd\mh,
    \label{eqn:lnmhatms}
\end{equation}
and the mean halo mass at fixed stellar mass~(thin black solid curve),
\begin{equation}
    \avg{\mh|\ms} = \int p(\mh|\ms^{\mathrm{cen}})\mh\,\dd\mh,
\end{equation}
respectively. The difference between the two means are small at the low stellar mass end but grows to $0.1$
dex at high masses, and the gray band indicates the 1-$\sigma$ logarithmic scatter about the mean relation.
As seen in Figure~\ref{fig:shmr}, the original SHMR flattens on the high mass end, so the log-scatter in halo
mass at fixed stellar mass is much larger than on the low stellar mass end.  The 1-$\sigma$ error on the mean
relation~(not shown here, but see Figure~\ref{fig:cvi}) is comparable to the scatter at the lower stellar mass
end, but much smaller than the scatter above $\ms{\sim}10^{10}\hhmsol$.  Also shown on Figure~\ref{fig:hsmr}
is the black dashed curve from Equation~\eqref{eqn:shmr} using the best-fit parameters, i.e., the same thick
black curve shown in Figure~\ref{fig:shmr} except that the x- and y-axises are swapped.  The difference
between the black dashed and the two black solid curves can easily exceed $0.5$ dex for high stellar mass
samples, and reaches almost $1$ dex for galaxies with $\ms{>}5\times10^{11}\hhmsol$. Therefore, caution should
be exercised when trying to compare the theoretical SHMR with the direct observational estimates of halo
masses for any galaxy samples selected above $\ms{=}10^{10}\hhmsol$.  The magenta dashed and the cyan
dot-dashed curves in Figure~\ref{fig:hsmr} are the halo-to-stellar mass relations inferred from the
L12~(COSMOS) and the \citet{hudson2015}~(CFHTLenS) analyses, respectively, showing great consistency between
independent constraints from different surveys.  The circles and the squares indicate the average halo masses
inferred from individual galaxy stellar mass samples using the weak lensing measurements in
GAMA~\citep{han2015} and CFHTLenS~\citep{velander2014}, respectively.  The lenses used in the \citet{han2015}
analysis were central galaxies of groups and clusters that are almost volume-limited, while for the latter
two, the \citet{velander2014} constraint was separately derived for colour-segregated samples that are
flux-limited. Despite the differences in the methods and data sets, our relation is largely consistent with
both sets of measurements.  The triangles are the measurements from the earlier SDSS DR4
galaxies~\citep{mandelbaum2006}, also split into sub-samples of early- and late-type galaxies.  Our analysis
is a direct update from the~\citet{mandelbaum2006} study, using SDSS DR7 (vs.\ DR4) galaxies, updated MPA/JHU
stellar mass catalogue, improved photometric redshifts for the source sample, additional measurements (galaxy
clustering), and a more sophisticated HOD model, and the two constraints are fully consistent with each other.

Finally, to facilitate the comparison between our constraint and observational studies that measure the mean
(log)halo mass of individual galaxy samples, we provide an analytic fitting formula to the best-fit
halo-to-stellar mass relation shown in Figure~\ref{fig:hsmr}~(computed from the fiducial {\ihod} MCMC chain
via Equation~\ref{eqn:lnmhatms}),
\begin{multline}
    \avg{\lg\mh|\ms^{\mathrm{cen}}} = 4.41\,\left[1 + \exp\left(-1.82
            \left(\lg\ms^{\mathrm{cen}}-11.18\right)\right)\right]^{-1} \\ +
    11.12\,\sin \left(-0.12\,\left(\lg\ms^{\mathrm{cen}} - 23.37\right)\right),
    \label{eqn:hsmrfit}
\end{multline}
where $\mh$ and $\ms^{\mathrm{cen}}$ are in units of $\hmsol$ and $\hhmsol$, respectively. The above fitting
formula is accurate to within $0.15\%$ across the entire stellar mass range above $3\times10^{8}\hhmsol$.

\subsection{Satellites: Conditional Stellar Mass Function}
\label{subsec:csmf}

Instead of imposing some fixed functional form~(e.g., the Schechter function) for the conditional stellar mass
function~(CSMF), we compute the satellite CSMFs as the derivatives of the satellite HODs at fixed halo mass.
This model flexibility allows for a more thorough exploration of the shapes of the satellite CSMFs, especially
any potential mass--dependent deviations from the Schechter function at both the low and high stellar mass
ends. The left panel of Figure~\ref{fig:csmf} shows the satellite CSMFs inferred for halos of six different
masses marked on the legend, with each shaded band indicating the $68\%$ uncertainty from the {\ihod}
constraint.  For low mass halos, the form of the CSMFs resemble the Schechter function with a flattening low
$\ms$ portion and a sharp exponential cutoff at high $\ms$, but it starts to exhibit an extra power-law
portion at intermediate $\ms$ for halos more massive than $10^{13}\hmsol$. The inset panel shows the six CSMFs
scaled by their corresponding halo mass. All the scaled CSMFs would land on top of one another if the
satellite population inside halos follows an exact homology sequence. Since $\alpha_{\mathrm{sat}}$ is
strongly constrained to be around unity~($\alpha_{\mathrm{sat}}=1.000_{-0.02}^{+0.03}$), the number of
satellite galaxies above any stellar mass threshold scales linearly with halo mass above
$M_{\mathrm{cut}}$~(see Equation~\eqref{eqn:nsat}), leading to a self--similar behaviour across the low
$\ms$ range~(${<}10^{10}\hhmsol$). At high stellar mass range, however, the homology is broken and an excess
population of galaxies of successively higher stellar mass begin to emerge as satellites --- they used to be
the central galaxies of small halos that were later accreted by their current-day, more massive host halos ---
a physical picture from the hierarchical structure formation paradigm of the $\lcdm$ Universe, recovered by
the prediction of our fiducial model, a purely statistical framework that describes the clustering and lensing
of galaxies.

To understand the asymptotic behaviour of the total stellar-to-halo mass ratio mentioned in
Section~\ref{subsec:shmr}, we show the stellar mass--weighted CSMFs, $\ms dN_{\mathrm{sat}}/d\lg\ms$, in the
right panel of Figure~\ref{fig:csmf}. The peaks in those curves confirm the finding from the bottom panel of
Figure~\ref{fig:shmr} that galaxies of $\ms{\sim}2\times10^{10}\hhmsol$ contribute the most amount of stellar
mass per dex in $\ms$ to each halo. Again, in the inset panel we scale each quantity by the corresponding halo
mass, showing the fractional contribution to the total stellar-to-halo mass ratio per dex in $\ms$ for each
halo mass. The asymptotic behaviour of the total stellar mass fraction can be mainly attributed to two factors:
1) the self-similarity of satellites galaxies below $10^{11}\hhmsol$ ensures that the total ratio integrated
up to $\ms{\sim}10^{11}\hhmsol$ is the same for all halos above $5\times10^{13}\hmsol$; and 2) the satellite
galaxies above $10^{11}\hhmsol$ contribute negligibly to the total ratio, so any increase in halo mass brings
little change to the total stellar-to-halo mass ratio.

\section{Conclusions}
\label{sec:end}

We have developed a novel extension to the HOD framework --- namely the {\ihod} model --- to solve for the
mapping between the observed stellar mass distribution and the underlying dark matter halos, via modelling the
projected galaxy auto-correlation function $w_p$ and the g-g lensing $\ds$ signals. In particular, the {\ihod}
model has two main features:
\begin{itemize}
    \item It is able to predict $w_p$ and $\ds$ for all the galaxies above the mixture limit~(i.e., the
	minimum stellar mass for the quiescent galaxies to be detected above the flux limit). This flexibility
	allows us to include ${\simeq}84\%$ more galaxies than the traditional HOD method, which substantially improves
    the $S/N$ of the $w_p$ and $\ds$ measurements~(by $10\%$ to almost a factor of two depending on stellar
    mass).
    \item It takes into account the volume incompleteness of galaxy stellar mass samples in a statistically
	consistent way, eliminating the need to assume completeness or parameterize the redshift-dependent
	selection functions.
\end{itemize}
The crucial input to the {\ihod} model is the {\it shape} of the observed SMF at each redshift, which is
employed by the {\ihod} model to construct an HOD for the galaxies at that redshift slice using $p(\ms,\mh)$,
the 2D joint probability density distribution of a galaxy with stellar mass $\ms$ sitting in a halo of mass
$\mh$.  The clustering and lensing signals of any galaxy sample can then be calculated by combining the
signals predicted from the HODs of the individual redshift slices.

The 2D distribution $p(\ms,\mh)$ has two components, the mean and scatter of the SHMR for the central galaxies
and the global HOD for the satellite galaxies. We adopt a similar parameterization for the two components as
L11, but allow the power-law slope of the satellite HOD~($\alpha$) and the concentration ratio between the
satellite and dark matter profiles~($f_c$) to vary freely during the constraint. Furthermore, we also
re-calibrate the subhalo lensing model in our analysis against the state-of-the-art hydrodynamic simulation
MassiveBlack-II. Thanks to the greatly improved $S/N$ in the $w_p$ and $\ds$ measurements, our fiducial
{\ihod} analysis not only breaks the degeneracy between the slope and the scatter of the SHMR, therefore
placing stringent constraint on the link between the dark halos and their central galaxies, but also derives
the conditional SMF of satellite galaxies as a function of halo mass. The inferred SHMR is in good agreement
with constraints from the abundance matching methods and the joint clustering, lensing, and galaxy abundance
analyses using the L11 framework. For the SMF of satellite galaxies at fixed halo mass, the best-fit model
predicts a departure from the Schechter function in massive groups and clusters, probably a result of mergers
and accretion that convert central galaxies of small halos into satellites of more massive systems. It will be
interesting to compare the conditional SMFs of satellites statistically derived from the {\ihod} model, which
allows the satellites to have a different SHMR than the central galaxies, to that inferred from abundance
matching, which directly assigns satellite stellar masses to subhalos in the simulations assuming the same
SHMR as the central galaxies.

In principle, the {\ihod} model uses only the shape of the observed galaxy SMF as input and is agnostic of the
SMF amplitude --- using half of the galaxies randomly drawn from the original catalogue would yield the same
best-fit model parameters, albeit with larger uncertainties. However, the parent SMF predicted from the
best-fit {\ihod} model agrees remarkably well with the observed SMF in both the shape {\it and} the amplitude.
Since the amplitude of the predicted SMF is translated from the normalisation of the halo mass function
predicted by the $\lcdm$ cosmology, this agreement is highly non--trivial, demonstrating the efficacy of the
{\ihod} model --- in order to match the clustering and lensing signal measured in SDSS, the mapping between
the galaxy stellar content and the dark matter halos is uniquely determined so that the expected galaxy
abundance at each $\ms$, calculated by summing galaxies at that $\ms$ over all the dark matter halos,
automatically gives the observed galaxy SMF in SDSS.

In future work, we will extend the {\ihod} model to provide constraints on the quenching mechanisms that
transform galaxies from star-forming to quiescent. The most straightforward approach is to apply the same
analysis to just the quenched population~(i.e., the galaxies with $g{-}r{>}{0.8}$ above the mixture limit),
and compare the inferred $p_{\mathrm{red}}(\ms, \mh)$ to the best-fit $p(\ms, \mh)$ in this paper to predict
the quenched fraction as a function of $\ms$ and $\mh$~\cite[see e.g.,][for HOD studies of galaxies segregated
by colour]{zehavi2005, zehavi2011}. Alternatively, one can do a joint fit for the red and blue galaxies
simultaneously, assuming an ad hoc functional form for the quenched fraction~\citep[see,
e.g.,][]{tinker2013,rodriguezpuebla2015}.  However, as mentioned in Section~\ref{subsec:syserr}, the galaxy
assembly bias needs to be carefully treated or marginalised over to obtain a robust constraint when modelling
colour-selected samples.

Going beyond the goal of understanding galaxy formation, HOD modelling of galaxy populations is likely to have
a role in cosmological analyses of future large imaging surveys that seek to constrain dark energy using weak
lensing, such as the Large Synoptic Survey Telescope (LSST; \citealt{2009arXiv0912.0201L}),
Euclid~\citep{laureijs2011}, and WFIRST~\citep{spergel2015}.  While cosmological weak lensing was originally
identified as a very clean probe of dark energy due to its sensitivity to matter fluctuations, more recent
work has identified two major theoretical uncertainties: the effect of baryons on the matter power spectrum,
and the intrinsic alignments of galaxy shapes that violate the assumption that any coherent galaxy alignments
are due to weak lensing.  Leading proposals for the mitigation of these theoretical uncertainties include halo
modelling of the galaxy-shear cross-correlation (galaxy-galaxy lensing) in combination with galaxy clustering
and cosmic shear, instead of using cosmic shear alone
\citep[e.g.,][]{2010MNRAS.402.2127S,2011MNRAS.417.2020S,2013MNRAS.434..148S,2013PhRvD..87d3509Z}.  Thus, we
anticipate that the {\ihod} model will be an important contribution to such efforts, given that it will enable
the use of a larger galaxy sample in understanding these important contaminants to cosmic shear surveys, and
thus enable the use of tighter priors on the nuisance parameters for baryonic effects and/or intrinsic
alignments when extracting cosmological information from the cosmic shear signal.

As a general yet powerful statistical formalism, the {\ihod} model can be easily applied to galaxies at higher
redshifts, such as the existing datasets from the COSMOS survey~\citep{scoville2007} used by L12,
CFHTLenS~\citep{heymans2012}, the recently finished Baryon Oscillation Spectroscopic
Survey~\citep[BOSS;][]{eisenstein2011, dawson2013} and its near-term higher redshift successor eBOSS, and the
deeper surveys planned for future facilities such as the Dark Energy Spectroscopic
Instrument~\citep[DESI;][]{levi2013}, the Subaru Prime Focus Spectrograph~\citep{takada2012}, Euclid, and
WFIRST. Galaxies at higher redshifts, such as luminous red galaxies~\citep[LRGs;][]{eisenstein2001,
anderson2012} and emission line galaxies~\citep{comparat2015}, are usually targeted with some complicated
colour and flux cuts, so the galaxy samples are always volume-incomplete. For instance, \citet{hoshino2015}
found that the average number of LRG--type central galaxies in the massive halos does not asymptotically reach
unity, contradictory to the assumption in traditional HOD models~\citep{parejko2013}. Therefore, the {\ihod}
model would be an especially valuable tool in constraining the link between galaxies and dark matter halos at
high redshifts.

\section*{Acknowledgements}

We thank Neta Bahcall, David Weinberg, and Zheng Zheng for helpful discussions, and Jiaxin Han for providing
his weak lensing mass measurements. We thank Alexie Leauthaud for carefully reading an earlier version of the
manuscript and for giving detailed comments and suggestions that have greatly improved the manuscript.  We
also thank the MassiveBlack-II team for providing the simulated galaxy catalogue and the dark matter particle
data. Y.Z. acknowledges the hospitality of the CCAPP at the Ohio State University where he enjoyed a fruitful
discussion with participants of the ``Assembly Bias and the Galaxy-Halo Relation'' Workshop. R.M. and Y.Z. are
supported by the Department of Energy Early Career Program, and the Sloan Fellowship program.


{\footnotesize
\bibliographystyle{mn2e}

\begin{thebibliography}{215}
\expandafter\ifx\csname natexlab\endcsname\relax\def\natexlab#1{#1}\fi

\bibitem[{Abazajian {et~al}\mbox{.}(2009)Abazajian, {Adelman-McCarthy},
  Ag\"{u}eros, Allam, Allende~Prieto, An, Anderson, Anderson, Annis, Bahcall,
  {Bailer-Jones}, Barentine, Bassett, Becker, Beers, Bell, Belokurov, Berlind,
  Berman, Bernardi, Bickerton, Bizyaev, Blakeslee, Blanton, Bochanski, Boroski,
  Brewington, Brinchmann, Brinkmann, Brunner, Budav\'{a}ri, Carey, Carliles,
  Carr, Castander, Cinabro, Connolly, Csabai, Cunha, Czarapata, Davenport,
  de~Haas, Dilday, Doi, Eisenstein, Evans, Evans, Fan, Friedman, Frieman,
  Fukugita, G\"{a}nsicke, Gates, Gillespie, Gilmore, Gonzalez, Gonzalez,
  Grebel, Gunn, Gy\"{o}ry, Hall, Harding, Harris, Harvanek, Hawley, Hayes,
  Heckman, Hendry, Hennessy, Hindsley, Hoblitt, Hogan, Hogg, Holtzman, Hyde,
  Ichikawa, Ichikawa, Im, Ivezi\'{c}, Jester, Jiang, Johnson, Jorgensen,
  Juri\'{c}, Kent, Kessler, Kleinman, Knapp, Konishi, Kron, Krzesinski,
  Kuropatkin, Lampeitl, Lebedeva, Lee, Lee, French~Leger, L\'{e}pine, Li, Lima,
  Lin, Long, Loomis, Loveday, Lupton, Magnier, Malanushenko, Malanushenko,
  Mandelbaum, Margon, Marriner, {Mart\'{i}nez-Delgado}, Matsubara, {McGehee},
  {McKay}, Meiksin, Morrison, Mullally, Munn, Murphy, Nash, Nebot, Neilsen,
  Newberg, Newman, Nichol, Nicinski, {Nieto-Santisteban}, Nitta, Okamura,
  Oravetz, Ostriker, Owen, Padmanabhan, Pan, Park, Pauls, Peoples, Percival,
  Pier, Pope, Pourbaix, Price, Purger, Quinn, Raddick, Re~Fiorentin, Richards,
  Richmond, Riess, Rix, Rockosi, Sako, Schlegel, Schneider, Scholz, Schreiber,
  Schwope, Seljak, Sesar, Sheldon, Shimasaku, Sibley, Simmons, Sivarani,
  Allyn~Smith, Smith, Smol\v{c}i\'{c}, Snedden, Stebbins, Steinmetz, Stoughton,
  Strauss, {SubbaRao}, Suto, Szalay, Szapudi, Szkody, Tanaka, Tegmark, Teodoro,
  Thakar, Tremonti, Tucker, Uomoto, Vanden~Berk, Vandenberg, Vidrih, Vogeley,
  Voges, Vogt, Wadadekar, Watters, Weinberg, West, White, Wilhite, Wonders,
  Yanny, Yocum, York, Zehavi, Zibetti, \& Zucker}]{abazajian2009}
Abazajian K.~N. {et~al.}, 2009, \apjs, 182, 543

\bibitem[{{Anderson} {et~al}\mbox{.}(2012){Anderson}, {Aubourg}, {Bailey},
  {Bizyaev}, {Blanton}, {Bolton}, {Brinkmann}, {Brownstein}, {Burden},
  {Cuesta}, {da Costa}, {Dawson}, {de Putter}, {Eisenstein}, {Gunn}, {Guo},
  {Hamilton}, {Harding}, {Ho}, {Honscheid}, {Kazin}, {Kirkby}, {Kneib},
  {Labatie}, {Loomis}, {Lupton}, {Malanushenko}, {Malanushenko}, {Mandelbaum},
  {Manera}, {Maraston}, {McBride}, {Mehta}, {Mena}, {Montesano}, {Muna},
  {Nichol}, {Nuza}, {Olmstead}, {Oravetz}, {Padmanabhan},
  {Palanque-Delabrouille}, {Pan}, {Parejko}, {P{\^a}ris}, {Percival},
  {Petitjean}, {Prada}, {Reid}, {Roe}, {Ross}, {Ross}, {Samushia},
  {S{\'a}nchez}, {Schlegel}, {Schneider}, {Sc{\'o}ccola}, {Seo}, {Sheldon},
  {Simmons}, {Skibba}, {Strauss}, {Swanson}, {Thomas}, {Tinker}, {Tojeiro},
  {Maga{\~n}a}, {Verde}, {Wagner}, {Wake}, {Weaver}, {Weinberg}, {White}, {Xu},
  {Y{\`e}che}, {Zehavi}, \& {Zhao}}]{anderson2012}
{Anderson} L. {et~al.}, 2012, \mnras, 427, 3435

\bibitem[{Bahcall \& Kulier(2014)}]{bahcall2014}
Bahcall N.~A., Kulier A., 2014, \mnras, 439, 2505

\bibitem[{{Baldauf} {et~al}\mbox{.}(2010){Baldauf}, {Smith}, {Seljak}, \&
  {Mandelbaum}}]{baldauf2010}
{Baldauf} T., {Smith} R.~E., {Seljak} U., {Mandelbaum} R., 2010, \prd, 81,
  063531

\bibitem[{{Baldry} {et~al}\mbox{.}(2006){Baldry}, {Balogh}, {Bower},
  {Glazebrook}, {Nichol}, {Bamford}, \& {Budavari}}]{baldry2006}
{Baldry} I.~K., {Balogh} M.~L., {Bower} R.~G., {Glazebrook} K., {Nichol} R.~C.,
  {Bamford} S.~P., {Budavari} T., 2006, \mnras, 373, 469

\bibitem[{{Bardeen} {et~al}\mbox{.}(1986){Bardeen}, {Bond}, {Kaiser}, \&
  {Szalay}}]{bardeen1986}
{Bardeen} J.~M., {Bond} J.~R., {Kaiser} N., {Szalay} A.~S., 1986, \apj, 304, 15

\bibitem[{{Baugh}(2006)}]{baugh2006}
{Baugh} C.~M., 2006, Reports on Progress in Physics, 69, 3101

\bibitem[{Behroozi {et~al}\mbox{.}(2010)Behroozi, Conroy, \&
  Wechsler}]{behroozi2010}
Behroozi P.~S., Conroy C., Wechsler R.~H., 2010, \apj, 717, 379

\bibitem[{Bell \& de~Jong(2001)}]{bell2001}
Bell E.~F., de~Jong R.~S., 2001, \apj, 550, 212

\bibitem[{Benson(2010)}]{benson2010}
Benson A.~J., 2010, Physics Reports, 495, 33

\bibitem[{{Berlind} {et~al}\mbox{.}(2006){Berlind}, {Kazin}, {Blanton},
  {Pueblas}, {Scoccimarro}, \& {Hogg}}]{berlind2006}
{Berlind} A.~A., {Kazin} E., {Blanton} M.~R., {Pueblas} S., {Scoccimarro} R.,
  {Hogg} D.~W., 2006, astro-ph/0610524

\bibitem[{Berlind \& Weinberg(2002)}]{berlind2002}
Berlind A.~A., Weinberg D.~H., 2002, \apj, 575, 587

\bibitem[{Bernardi {et~al}\mbox{.}(2013)Bernardi, Meert, Sheth, Vikram,
  {Huertas-Company}, Mei, \& Shankar}]{bernardi2013}
Bernardi M., Meert A., Sheth R.~K., Vikram V., {Huertas-Company} M., Mei S.,
  Shankar F., 2013, \mnras, 436, 697

\bibitem[{{Bhattacharya} {et~al}\mbox{.}(2011){Bhattacharya}, {Heitmann},
  {White}, {Luki{\'c}}, {Wagner}, \& {Habib}}]{bhattacharya2011}
{Bhattacharya} S., {Heitmann} K., {White} M., {Luki{\'c}} Z., {Wagner} C.,
  {Habib} S., 2011, \apj, 732, 122

\bibitem[{{Blanton} \& {Berlind}(2007)}]{blanton2007}
{Blanton} M.~R., {Berlind} A.~A., 2007, \apj, 664, 791

\bibitem[{Blanton {et~al}\mbox{.}(2005)Blanton, Schlegel, Strauss, Brinkmann,
  Finkbeiner, Fukugita, Gunn, Hogg, Ivezi\'{c}, Knapp, Lupton, Munn, Schneider,
  Tegmark, \& Zehavi}]{blanton2005}
Blanton M.~R. {et~al.}, 2005, \aj, 129, 2562

\bibitem[{{Blazek} {et~al}\mbox{.}(2012){Blazek}, {Mandelbaum}, {Seljak}, \&
  {Nakajima}}]{BMS+12}
{Blazek} J., {Mandelbaum} R., {Seljak} U., {Nakajima} R., 2012, \jcap, 5, 41

\bibitem[{{Bocquet} {et~al}\mbox{.}(2015){Bocquet}, {Saro}, {Dolag}, \&
  {Mohr}}]{bocquet2015}
{Bocquet} S., {Saro} A., {Dolag} K., {Mohr} J.~J., 2015, arXiv:1502.07357

\bibitem[{{Bond} {et~al}\mbox{.}(1991){Bond}, {Cole}, {Efstathiou}, \&
  {Kaiser}}]{bond1991}
{Bond} J.~R., {Cole} S., {Efstathiou} G., {Kaiser} N., 1991, \apj, 379, 440

\bibitem[{{Booth} \& {Schaye}(2009)}]{booth2009}
{Booth} C.~M., {Schaye} J., 2009, \mnras, 398, 53

\bibitem[{{Bower} {et~al}\mbox{.}(2006){Bower}, {Benson}, {Malbon}, {Helly},
  {Frenk}, {Baugh}, {Cole}, \& {Lacey}}]{bower2006}
{Bower} R.~G., {Benson} A.~J., {Malbon} R., {Helly} J.~C., {Frenk} C.~S.,
  {Baugh} C.~M., {Cole} S., {Lacey} C.~G., 2006, \mnras, 370, 645

\bibitem[{Bruzual \& Charlot(2003)}]{bruzual2003}
Bruzual G., Charlot S., 2003, \mnras, 344, 1000

\bibitem[{Budzynski {et~al}\mbox{.}(2014)Budzynski, Koposov, {McCarthy}, \&
  Belokurov}]{budzynski2014}
Budzynski J.~M., Koposov S.~E., {McCarthy} I.~G., Belokurov V., 2014, \mnras,
  437, 1362

\bibitem[{Budzynski {et~al}\mbox{.}(2012)Budzynski, Koposov, {McCarthy},
  {McGee}, \& Belokurov}]{budzynski2012}
Budzynski J.~M., Koposov S.~E., {McCarthy} I.~G., {McGee} S.~L., Belokurov V.,
  2012, \mnras, 423, 104

\bibitem[{{Cacciato} {et~al}\mbox{.}(2009){Cacciato}, {van den Bosch}, {More},
  {Li}, {Mo}, \& {Yang}}]{cacciato2009}
{Cacciato} M., {van den Bosch} F.~C., {More} S., {Li} R., {Mo} H.~J., {Yang}
  X., 2009, \mnras, 394, 929

\bibitem[{{Cacciato} {et~al}\mbox{.}(2013){Cacciato}, {van den Bosch}, {More},
  {Mo}, \& {Yang}}]{cacciato2013}
{Cacciato} M., {van den Bosch} F.~C., {More} S., {Mo} H., {Yang} X., 2013,
  \mnras, 430, 767

\bibitem[{Cappellari {et~al}\mbox{.}(2012)Cappellari, {McDermid}, Alatalo,
  Blitz, Bois, Bournaud, Bureau, Crocker, Davies, Davis, de~Zeeuw, Duc,
  Emsellem, Khochfar, Krajnovi\'{c}, Kuntschner, Lablanche, Morganti, Naab,
  Oosterloo, Sarzi, Scott, Serra, Weijmans, \& Young}]{cappellari2012}
Cappellari M. {et~al.}, 2012, Nature, 484, 485

\bibitem[{{Cen} \& {Ostriker}(1999)}]{cen1999}
{Cen} R., {Ostriker} J.~P., 1999, \apj, 514, 1

\bibitem[{{Chabrier}(2003)}]{chabrier2003}
{Chabrier} G., 2003, \pasp, 115, 763

\bibitem[{{Cole} \& {Kaiser}(1989)}]{cole1989}
{Cole} S., {Kaiser} N., 1989, \mnras, 237, 1127

\bibitem[{{Comparat} {et~al}\mbox{.}(2015){Comparat}, {Richard}, {Kneib},
  {Ilbert}, {Gonzalez-Perez}, {Tresse}, {Zoubian}, {Arnouts}, {Brownstein},
  {Baugh}, {Delubac}, {Ealet}, {Escoffier}, {Ge}, {Jullo}, {Lacey}, {Ross},
  {Schlegel}, {Schneider}, {Steele}, {Tasca}, {Yeche}, {Lesser}, {Jiang},
  {Jing}, {Fan}, {Fan}, {Ma}, {Nie}, {Wang}, {Wu}, {Zhang}, {Zhou}, {Zhou}, \&
  {Zou}}]{comparat2015}
{Comparat} J. {et~al.}, 2015, \aap, 575, A40

\bibitem[{Conroy(2013)}]{conroy2013}
Conroy C., 2013, \araa, 51, 393

\bibitem[{Conroy \& van Dokkum(2012)}]{conroy2012}
Conroy C., van Dokkum P.~G., 2012, \apj, 760, 71

\bibitem[{Conroy {et~al}\mbox{.}(2006)Conroy, Wechsler, \&
  Kravtsov}]{conroy2006}
Conroy C., Wechsler R.~H., Kravtsov A.~V., 2006, \apj, 647, 201

\bibitem[{{Cooray} \& {Sheth}(2002)}]{cooray2002}
{Cooray} A., {Sheth} R., 2002, \physrep, 372, 1

\bibitem[{{Coupon} {et~al}\mbox{.}(2015){Coupon}, {Arnouts}, {van Waerbeke},
  {Moutard}, {Ilbert}, {van Uitert}, {Erben}, {Garilli}, {Guzzo}, {Heymans},
  {Hildebrandt}, {Hoekstra}, {Kilbinger}, {Kitching}, {Mellier}, {Miller},
  {Scodeggio}, {Bonnett}, {Branchini}, {Davidzon}, {De Lucia}, {Fritz}, {Fu},
  {Hudelot}, {Hudson}, {Kuijken}, {Leauthaud}, {Le F{\`e}vre}, {McCracken},
  {Moscardini}, {Rowe}, {Schrabback}, {Semboloni}, \& {Velander}}]{coupon2015}
{Coupon} J. {et~al.}, 2015, \mnras, 449, 1352

\bibitem[{Cowie {et~al}\mbox{.}(1996)Cowie, Songaila, Hu, \& Cohen}]{cowie1996}
Cowie L.~L., Songaila A., Hu E.~M., Cohen J.~G., 1996, \aj, 112, 839

\bibitem[{{Crocce} {et~al}\mbox{.}(2010){Crocce}, {Fosalba}, {Castander}, \&
  {Gazta{\~n}aga}}]{crocce2010}
{Crocce} M., {Fosalba} P., {Castander} F.~J., {Gazta{\~n}aga} E., 2010, \mnras,
  403, 1353

\bibitem[{{Croft} {et~al}\mbox{.}(2012){Croft}, {Matteo}, {Khandai},
  {Springel}, {Jana}, \& {Gardner}}]{croft2012}
{Croft} R.~A.~C., {Matteo} T.~D., {Khandai} N., {Springel} V., {Jana} A.,
  {Gardner} J.~P., 2012, \mnras, 425, 2766

\bibitem[{{Croton} {et~al}\mbox{.}(2007){Croton}, {Gao}, \&
  {White}}]{croton2007}
{Croton} D.~J., {Gao} L., {White} S.~D.~M., 2007, \mnras, 374, 1303

\bibitem[{{Cui} {et~al}\mbox{.}(2012{\natexlab{a}}){Cui}, {Baldi}, \&
  {Borgani}}]{cui2012b}
{Cui} W., {Baldi} M., {Borgani} S., 2012{\natexlab{a}}, \mnras, 424, 993

\bibitem[{{Cui} {et~al}\mbox{.}(2012{\natexlab{b}}){Cui}, {Borgani}, {Dolag},
  {Murante}, \& {Tornatore}}]{cui2012}
{Cui} W., {Borgani} S., {Dolag} K., {Murante} G., {Tornatore} L.,
  2012{\natexlab{b}}, \mnras, 423, 2279

\bibitem[{{Cusworth} {et~al}\mbox{.}(2014){Cusworth}, {Kay}, {Battye}, \&
  {Thomas}}]{cusworth2014}
{Cusworth} S.~J., {Kay} S.~T., {Battye} R.~A., {Thomas} P.~A., 2014, \mnras,
  439, 2485

\bibitem[{{Dalal} {et~al}\mbox{.}(2008){Dalal}, {White}, {Bond}, \&
  {Shirokov}}]{dalal2008}
{Dalal} N., {White} M., {Bond} J.~R., {Shirokov} A., 2008, \apj, 687, 12

\bibitem[{Dav\'{e} {et~al}\mbox{.}(2012)Dav\'{e}, Finlator, \&
  Oppenheimer}]{dave2012}
Dav\'{e} R., Finlator K., Oppenheimer B.~D., 2012, \mnras, 421, 98

\bibitem[{{Dawson} {et~al}\mbox{.}(2013){Dawson}, {Schlegel}, {Ahn},
  {Anderson}, {Aubourg}, {Bailey}, {Barkhouser}, {Bautista}, {Beifiori},
  {Berlind}, {Bhardwaj}, {Bizyaev}, {Blake}, {Blanton}, {Blomqvist}, {Bolton},
  {Borde}, {Bovy}, {Brandt}, {Brewington}, {Brinkmann}, {Brown}, {Brownstein},
  {Bundy}, {Busca}, {Carithers}, {Carnero}, {Carr}, {Chen}, {Comparat},
  {Connolly}, {Cope}, {Croft}, {Cuesta}, {da Costa}, {Davenport}, {Delubac},
  {de Putter}, {Dhital}, {Ealet}, {Ebelke}, {Eisenstein}, {Escoffier}, {Fan},
  {Filiz Ak}, {Finley}, {Font-Ribera}, {G{\'e}nova-Santos}, {Gunn}, {Guo},
  {Haggard}, {Hall}, {Hamilton}, {Harris}, {Harris}, {Ho}, {Hogg}, {Holder},
  {Honscheid}, {Huehnerhoff}, {Jordan}, {Jordan}, {Kauffmann}, {Kazin},
  {Kirkby}, {Klaene}, {Kneib}, {Le Goff}, {Lee}, {Long}, {Loomis}, {Lundgren},
  {Lupton}, {Maia}, {Makler}, {Malanushenko}, {Malanushenko}, {Mandelbaum},
  {Manera}, {Maraston}, {Margala}, {Masters}, {McBride}, {McDonald}, {McGreer},
  {McMahon}, {Mena}, {Miralda-Escud{\'e}}, {Montero-Dorta}, {Montesano},
  {Muna}, {Myers}, {Naugle}, {Nichol}, {Noterdaeme}, {Nuza}, {Olmstead},
  {Oravetz}, {Oravetz}, {Owen}, {Padmanabhan}, {Palanque-Delabrouille}, {Pan},
  {Parejko}, {P{\^a}ris}, {Percival}, {P{\'e}rez-Fournon},
  {P{\'e}rez-R{\`a}fols}, {Petitjean}, {Pfaffenberger}, {Pforr}, {Pieri},
  {Prada}, {Price-Whelan}, {Raddick}, {Rebolo}, {Rich}, {Richards}, {Rockosi},
  {Roe}, {Ross}, {Ross}, {Rossi}, {Rubi{\~n}o-Martin}, {Samushia},
  {S{\'a}nchez}, {Sayres}, {Schmidt}, {Schneider}, {Sc{\'o}ccola}, {Seo},
  {Shelden}, {Sheldon}, {Shen}, {Shu}, {Slosar}, {Smee}, {Snedden}, {Stauffer},
  {Steele}, {Strauss}, {Streblyanska}, {Suzuki}, {Swanson}, {Tal}, {Tanaka},
  {Thomas}, {Tinker}, {Tojeiro}, {Tremonti}, {Vargas Maga{\~n}a}, {Verde},
  {Viel}, {Wake}, {Watson}, {Weaver}, {Weinberg}, {Weiner}, {West}, {White},
  {Wood-Vasey}, {Yeche}, {Zehavi}, {Zhao}, \& {Zheng}}]{dawson2013}
{Dawson} K.~S. {et~al.}, 2013, \aj, 145, 10

\bibitem[{De~Lucia \& Blaizot(2007)}]{delucia2007}
De~Lucia G., Blaizot J., 2007, \mnras, 375, 2

\bibitem[{De~Lucia {et~al}\mbox{.}(2006)De~Lucia, Springel, White, Croton, \&
  Kauffmann}]{delucia2006}
De~Lucia G., Springel V., White S. D.~M., Croton D., Kauffmann G., 2006,
  \mnras, 366, 499

\bibitem[{{Di Matteo} {et~al}\mbox{.}(2005){Di Matteo}, {Springel}, \&
  {Hernquist}}]{dimatteo2005}
{Di Matteo} T., {Springel} V., {Hernquist} L., 2005, \nat, 433, 604

\bibitem[{{Dutton} \& {Macci{\`o}}(2014)}]{dutton2014}
{Dutton} A.~A., {Macci{\`o}} A.~V., 2014, \mnras, 441, 3359

\bibitem[{{Einasto}(1965)}]{einasto1965}
{Einasto} J., 1965, Trudy Astrofizicheskogo Instituta Alma-Ata, 5, 87

\bibitem[{{Eisenstein} {et~al}\mbox{.}(2001){Eisenstein}, {Annis}, {Gunn},
  {Szalay}, {Connolly}, {Nichol}, {Bahcall}, {Bernardi}, {Burles}, {Castander},
  {Fukugita}, {Hogg}, {Ivezi{\'c}}, {Knapp}, {Lupton}, {Narayanan}, {Postman},
  {Reichart}, {Richmond}, {Schneider}, {Schlegel}, {Strauss}, {SubbaRao},
  {Tucker}, {Vanden Berk}, {Vogeley}, {Weinberg}, \& {Yanny}}]{eisenstein2001}
{Eisenstein} D.~J. {et~al.}, 2001, \aj, 122, 2267

\bibitem[{{Eisenstein} {et~al}\mbox{.}(2011){Eisenstein}, {Weinberg}, {Agol},
  {Aihara}, {Allende Prieto}, {Anderson}, {Arns}, {Aubourg}, {Bailey},
  {Balbinot}, \& et~al.}]{eisenstein2011}
{Eisenstein} D.~J. {et~al.}, 2011, \aj, 142, 72

\bibitem[{{Faltenbacher} \& {White}(2010)}]{faltenbacher2010}
{Faltenbacher} A., {White} S.~D.~M., 2010, \apj, 708, 469

\bibitem[{{Faucher-Gigu{\`e}re} {et~al}\mbox{.}(2011){Faucher-Gigu{\`e}re},
  {Kere{\v s}}, \& {Ma}}]{faucher2011}
{Faucher-Gigu{\`e}re} C.-A., {Kere{\v s}} D., {Ma} C.-P., 2011, \mnras, 417,
  2982

\bibitem[{{Feldmann} {et~al}\mbox{.}(2006){Feldmann}, {Carollo}, {Porciani},
  {Lilly}, {Capak}, {Taniguchi}, {Le F{\`e}vre}, {Renzini}, {Scoville},
  {Ajiki}, {Aussel}, {Contini}, {McCracken}, {Mobasher}, {Murayama}, {Sanders},
  {Sasaki}, {Scarlata}, {Scodeggio}, {Shioya}, {Silverman}, {Takahashi},
  {Thompson}, \& {Zamorani}}]{2006MNRAS.372..565F}
{Feldmann} R. {et~al.}, 2006, \mnras, 372, 565

\bibitem[{Ferreras {et~al}\mbox{.}(2013)Ferreras, La~Barbera, de~la Rosa,
  Vazdekis, de~Carvalho, {Falc\'{o}n-Barroso}, \& Ricciardelli}]{ferreras2013}
Ferreras I., La~Barbera F., de~la Rosa I.~G., Vazdekis A., de~Carvalho R.~R.,
  {Falc\'{o}n-Barroso} J., Ricciardelli E., 2013, \mnras, 429, L15

\bibitem[{{Fontanot} {et~al}\mbox{.}(2009){Fontanot}, {De Lucia}, {Monaco},
  {Somerville}, \& {Santini}}]{fontanot2009}
{Fontanot} F., {De Lucia} G., {Monaco} P., {Somerville} R.~S., {Santini} P.,
  2009, \mnras, 397, 1776

\bibitem[{{Foreman-Mackey} {et~al}\mbox{.}(2013){Foreman-Mackey}, Hogg, Lang,
  \& Goodman}]{foreman-mackey2013}
{Foreman-Mackey} D., Hogg D.~W., Lang D., Goodman J., 2013, \pasp, 125, 306

\bibitem[{{Fukugita} {et~al}\mbox{.}(1998){Fukugita}, {Hogan}, \&
  {Peebles}}]{fukugita1998}
{Fukugita} M., {Hogan} C.~J., {Peebles} P.~J.~E., 1998, \apj, 503, 518

\bibitem[{Fukugita {et~al}\mbox{.}(1996)Fukugita, Ichikawa, Gunn, Doi,
  Shimasaku, \& Schneider}]{fukugita1996}
Fukugita M., Ichikawa T., Gunn J.~E., Doi M., Shimasaku K., Schneider D.~P.,
  1996, \aj, 111, 1748

\bibitem[{Gallazzi \& Bell(2009)}]{gallazzi2009}
Gallazzi A., Bell E.~F., 2009, \apjs, 185, 253

\bibitem[Gao et al.(2008)]{gao2008} Gao, L., Navarro, J.~F.,
  Cole, S., et al.\ 2008, \mnras, 387, 536

\bibitem[{{Gao} {et~al}\mbox{.}(2005){Gao}, {Springel}, \& {White}}]{gao2005}
{Gao} L., {Springel} V., {White} S.~D.~M., 2005, \mnras, 363, L66

\bibitem[{{Gao} \& {White}(2007)}]{gao2007}
{Gao} L., {White} S.~D.~M., 2007, \mnras, 377, L5

\bibitem[{Geller \& Huchra(1989)}]{geller1989}
Geller M.~J., Huchra J.~P., 1989, Science, 246, 897

\bibitem[{George {et~al}\mbox{.}(2012)George, Leauthaud, Bundy, Finoguenov, Ma,
  Rykoff, Tinker, Wechsler, Massey, \& Mei}]{george2012}
George M.~R. {et~al.}, 2012, \apj, 757, 2

\bibitem[{{George} {et~al}\mbox{.}(2011){George}, {Leauthaud}, {Bundy},
  {Finoguenov}, {Tinker}, {Lin}, {Mei}, {Kneib}, {Aussel}, {Behroozi}, {Busha},
  {Capak}, {Coccato}, {Covone}, {Faure}, {Fiorenza}, {Ilbert}, {Le Floc'h},
  {Koekemoer}, {Tanaka}, {Wechsler}, \& {Wolk}}]{george2011}
{George} M.~R. {et~al.}, 2011, \apj, 742, 125

\bibitem[{Gunn {et~al}\mbox{.}(1998)Gunn, Carr, Rockosi, Sekiguchi, Berry,
  Elms, de~Haas, Ivezi\'{c}, Knapp, Lupton, Pauls, Simcoe, Hirsch, Sanford,
  Wang, York, Harris, Annis, Bartozek, Boroski, Bakken, Haldeman, Kent, Holm,
  Holmgren, Petravick, Prosapio, Rechenmacher, Doi, Fukugita, Shimasaku, Okada,
  Hull, Siegmund, Mannery, Blouke, Heidtman, Schneider, Lucinio, \&
  Brinkman}]{gunn1998}
Gunn J.~E. {et~al.}, 1998, \aj, 116, 3040

\bibitem[{Gunn {et~al}\mbox{.}(2006)Gunn, Siegmund, Mannery, Owen, Hull, Leger,
  Carey, Knapp, York, Boroski, Kent, Lupton, Rockosi, Evans, Waddell, Anderson,
  Annis, Barentine, Bartoszek, Bastian, Bracker, Brewington, Briegel,
  Brinkmann, Brown, Carr, Czarapata, Drennan, Dombeck, Federwitz, Gillespie,
  Gonzales, Hansen, Harvanek, Hayes, Jordan, Kinney, Klaene, Kleinman, Kron,
  Kresinski, Lee, Limmongkol, Lindenmeyer, Long, Loomis, {McGehee}, Mantsch,
  Neilsen, Neswold, Newman, Nitta, Peoples, Pier, Prieto, Prosapio, Rivetta,
  Schneider, Snedden, \& Wang}]{gunn2006}
Gunn J.~E. {et~al.}, 2006, \aj, 131, 2332

\bibitem[{Guo {et~al}\mbox{.}(2012)Guo, Zehavi, \& Zheng}]{guo2012}
Guo H., Zehavi I., Zheng Z., 2012, \apj, 756, 127

\bibitem[{{Guo} {et~al}\mbox{.}(2014){Guo}, {Zheng}, {Zehavi}, {Xu},
  {Eisenstein}, {Weinberg}, {Bahcall}, {Berlind}, {Comparat}, {McBride},
  {Ross}, {Schneider}, {Skibba}, {Swanson}, {Tinker}, {Tojeiro}, \&
  {Wake}}]{guo2014}
{Guo} H. {et~al.}, 2014, \mnras, 441, 2398

\bibitem[{Guo {et~al}\mbox{.}(2011)Guo, White, {Boylan-Kolchin}, De~Lucia,
  Kauffmann, Lemson, Li, Springel, \& Weinmann}]{guo2011}
Guo Q. {et~al.}, 2011, \mnras, 413, 101

\bibitem[{Guo {et~al}\mbox{.}(2010)Guo, White, Li, \&
  {Boylan-Kolchin}}]{guo2010}
Guo Q., White S., Li C., {Boylan-Kolchin} M., 2010, \mnras, 404, 1111

\bibitem[{{Guzik} \& {Seljak}(2001)}]{guzik2001}
{Guzik} J., {Seljak} U., 2001, \mnras, 321, 439

\bibitem[{{Guzik} \& {Seljak}(2002)}]{guzik2002}
{Guzik} J., {Seljak} U., 2002, \mnras, 335, 311

\bibitem[{{Hahn} {et~al}\mbox{.}(2007){Hahn}, {Porciani}, {Carollo}, \&
  {Dekel}}]{hahn2007}
{Hahn} O., {Porciani} C., {Carollo} C.~M., {Dekel} A., 2007, \mnras, 375, 489

\bibitem[{{Hamilton} \& {Tegmark}(2004)}]{hamilton2004}
{Hamilton} A.~J.~S., {Tegmark} M., 2004, \mnras, 349, 115

\bibitem[{{Han} {et~al}\mbox{.}(2015){Han}, {Eke}, {Frenk}, {Mandelbaum},
  {Norberg}, {Schneider}, {Peacock}, {Jing}, {Baldry}, {Bland-Hawthorn},
  {Brough}, {Brown}, {Liske}, {Loveday}, \& {Robotham}}]{han2015}
{Han} J. {et~al.}, 2015, \mnras, 446, 1356

\bibitem[{{Hansen} {et~al}\mbox{.}(2009){Hansen}, {Sheldon}, {Wechsler}, \&
  {Koester}}]{hansen2009}
{Hansen} S.~M., {Sheldon} E.~S., {Wechsler} R.~H., {Koester} B.~P., 2009, \apj,
  699, 1333

\bibitem[{{Harker} {et~al}\mbox{.}(2006){Harker}, {Cole}, {Helly}, {Frenk}, \&
  {Jenkins}}]{harker2006}
{Harker} G., {Cole} S., {Helly} J., {Frenk} C., {Jenkins} A., 2006, \mnras,
  367, 1039

\bibitem[{{Hearin} \& {Watson}(2013)}]{hearin2013}
{Hearin} A.~P., {Watson} D.~F., 2013, \mnras, 435, 1313

\bibitem[{{Hearin} {et~al}\mbox{.}(2014){Hearin}, {Watson}, \& {van den
  Bosch}}]{hearin2014}
{Hearin} A.~P., {Watson} D.~F., {van den Bosch} F.~C., 2014, arXiv:1404.6524

\bibitem[{{Hernquist} \& {Katz}(1989)}]{hernquist1989}
{Hernquist} L., {Katz} N., 1989, \apjs, 70, 419

\bibitem[{{Heymans} {et~al}\mbox{.}(2012){Heymans}, {Van Waerbeke}, {Miller},
  {Erben}, {Hildebrandt}, {Hoekstra}, {Kitching}, {Mellier}, {Simon},
  {Bonnett}, {Coupon}, {Fu}, {Harnois D{\'e}raps}, {Hudson}, {Kilbinger},
  {Kuijken}, {Rowe}, {Schrabback}, {Semboloni}, {van Uitert}, {Vafaei}, \&
  {Velander}}]{heymans2012}
{Heymans} C. {et~al.}, 2012, \mnras, 427, 146

\bibitem[{{Hirata} \& {Seljak}(2003)}]{2003MNRAS.343..459H}
{Hirata} C., {Seljak} U., 2003, \mnras, 343, 459

\bibitem[{Hopkins {et~al}\mbox{.}(2012)Hopkins, Quataert, \&
  Murray}]{hopkins2012}
Hopkins P.~F., Quataert E., Murray N., 2012, \mnras, 421, 3522

\bibitem[{{Hoshino} {et~al}\mbox{.}(2015){Hoshino}, {Leauthaud}, {Lackner},
  {Hikage}, {Rozo}, {Rykoff}, {Mandelbaum}, {More}, {More}, {Saito}, \&
  {Vulcani}}]{hoshino2015}
{Hoshino} H. {et~al.}, 2015, arXiv:1503.05200

\bibitem[{Hudson {et~al}\mbox{.}(2015)Hudson, Gillis, Coupon, Hildebrandt,
  Erben, Heymans, Hoekstra, Kitching, Mellier, Miller, Van~Waerbeke, Bonnett,
  Fu, Kuijken, Rowe, Schrabback, Semboloni, van Uitert, \&
  Velander}]{hudson2015}
Hudson M.~J. {et~al.}, 2015, \mnras, 447, 298

\bibitem[{{Ichiki} \& {Takada}(2012)}]{ichiki2012}
{Ichiki} K., {Takada} M., 2012, \prd, 85, 063521

\bibitem[{{Jing} {et~al}\mbox{.}(1998){Jing}, {Mo}, \& {B{\"o}rner}}]{jing1998}
{Jing} Y.~P., {Mo} H.~J., {B{\"o}rner} G., 1998, \apj, 494, 1

\bibitem[{{Jing} {et~al}\mbox{.}(2007){Jing}, {Suto}, \& {Mo}}]{jing2007}
{Jing} Y.~P., {Suto} Y., {Mo} H.~J., 2007, \apj, 657, 664

\bibitem[{{Kaiser}(1984)}]{kaiser1984}
{Kaiser} N., 1984, \apjl, 284, L9

\bibitem[{{Kang} {et~al}\mbox{.}(2005){Kang}, {Jing}, {Mo}, \&
  {B{\"o}rner}}]{kang2005}
{Kang} X., {Jing} Y.~P., {Mo} H.~J., {B{\"o}rner} G., 2005, \apj, 631, 21

\bibitem[{{Katz} {et~al}\mbox{.}(1996){Katz}, {Weinberg}, \&
  {Hernquist}}]{katz1996}
{Katz} N., {Weinberg} D.~H., {Hernquist} L., 1996, \apjs, 105, 19

\bibitem[{Kauffmann {et~al}\mbox{.}(2003)Kauffmann, Heckman, White, Charlot,
  Tremonti, Brinchmann, Bruzual, Peng, Seibert, Bernardi, Blanton, Brinkmann,
  Castander, Cs\'{a}bai, Fukugita, Ivezic, Munn, Nichol, Padmanabhan, Thakar,
  Weinberg, \& York}]{kauffmann2003}
Kauffmann G. {et~al.}, 2003, \mnras, 341, 33

\bibitem[{{Kere{\v s}} {et~al}\mbox{.}(2005){Kere{\v s}}, {Katz}, {Weinberg},
  \& {Dav{\'e}}}]{keres2005}
{Kere{\v s}} D., {Katz} N., {Weinberg} D.~H., {Dav{\'e}} R., 2005, \mnras, 363,
  2

\bibitem[{Kerscher {et~al}\mbox{.}(2000)Kerscher, Szapudi, \&
  Szalay}]{kerscher2000}
Kerscher M., Szapudi I., Szalay A.~S., 2000, The Astrophysical Journal Letters,
  535, L13

\bibitem[{Khandai {et~al}\mbox{.}(2014)Khandai, Di~Matteo, Croft, Wilkins,
  Feng, Tucker, {DeGraf}, \& Liu}]{khandai2014}
Khandai N., Di~Matteo T., Croft R., Wilkins S.~M., Feng Y., Tucker E., {DeGraf}
  C., Liu M., 2014, {ArXiv} e-prints, 1402, 888

\bibitem[{{Kravtsov} {et~al}\mbox{.}(2004){Kravtsov}, {Berlind}, {Wechsler},
  {Klypin}, {Gottl{\"o}ber}, {Allgood}, \& {Primack}}]{kravtsov2004}
{Kravtsov} A.~V., {Berlind} A.~A., {Wechsler} R.~H., {Klypin} A.~A.,
  {Gottl{\"o}ber} S., {Allgood} B., {Primack} J.~R., 2004, \apj, 609, 35

\bibitem[{{Kulier} \& {Ostriker}(2015)}]{kulier2015}
{Kulier} A., {Ostriker} J.~P., 2015, arXiv:1503.07533

\bibitem[{Landy \& Szalay(1993)}]{landy1993}
Landy S.~D., Szalay A.~S., 1993, \apj, 412, 64

\bibitem[{Lang {et~al}\mbox{.}(2010)Lang, Hogg, Mierle, Blanton, \&
  Roweis}]{lang2010}
Lang D., Hogg D.~W., Mierle K., Blanton M., Roweis S., 2010, \aj, 139, 1782

\bibitem[{{Laporte} \& {White}(2014)}]{laporte2014}
{Laporte} C.~F.~P., {White} S.~D.~M., 2014, arXiv:1409.1924

\bibitem[{{Laureijs} {et~al}\mbox{.}(2011){Laureijs}, {Amiaux}, {Arduini},
  {Augu{\`e}res}, {Brinchmann}, {Cole}, {Cropper}, {Dabin}, {Duvet}, {Ealet},
  \& et~al.}]{laureijs2011}
{Laureijs} R. {et~al.}, 2011, arXiv:1110.3193

\bibitem[{Leauthaud {et~al}\mbox{.}(2012{\natexlab{a}})Leauthaud, George,
  Behroozi, Bundy, Tinker, Wechsler, Conroy, Finoguenov, \&
  Tanaka}]{leauthaud2012}
Leauthaud A. {et~al.}, 2012{\natexlab{a}}, \apj, 746, 95

\bibitem[{Leauthaud {et~al}\mbox{.}(2011)Leauthaud, Tinker, Behroozi, Busha, \&
  Wechsler}]{leauthaud2011}
Leauthaud A., Tinker J., Behroozi P.~S., Busha M.~T., Wechsler R.~H., 2011,
  \apj, 738, 45

\bibitem[{Leauthaud {et~al}\mbox{.}(2012{\natexlab{b}})Leauthaud, Tinker,
  Bundy, Behroozi, Massey, Rhodes, George, Kneib, Benson, Wechsler, Busha,
  Capak, Cort\^{e}s, Ilbert, Koekemoer, Le~F\`{e}vre, Lilly, {McCracken},
  Salvato, Schrabback, Scoville, Smith, \& Taylor}]{leauthaud2012-a}
Leauthaud A. {et~al.}, 2012{\natexlab{b}}, \apj, 744, 159

\bibitem[{{Levi} {et~al}\mbox{.}(2013){Levi}, {Bebek}, {Beers}, {Blum}, {Cahn},
  {Eisenstein}, {Flaugher}, {Honscheid}, {Kron}, {Lahav}, {McDonald}, {Roe},
  {Schlegel}, \& {representing the DESI collaboration}}]{levi2013}
{Levi} M. {et~al.}, 2013, arXiv:1308.0847

\bibitem[{Li \& White(2009)}]{li2009}
Li C., White S. D.~M., 2009, \mnras, 398, 2177

\bibitem[{Li {et~al}\mbox{.}(2014)Li, Shan, Mo, Kneib, Yang, Luo, van~den
  Bosch, Erben, Moraes, \& Makler}]{li2014}
Li R. {et~al.}, 2014, \mnras, 438, 2864

\bibitem[{{Li} {et~al}\mbox{.}(2008){Li}, {Mo}, \& {Gao}}]{li2008}
{Li} Y., {Mo} H.~J., {Gao} L., 2008, \mnras, 389, 1419

\bibitem[{{LoVerde}(2014)}]{loverdel2014}
{LoVerde} M., 2014, \prd, 90, 083530

\bibitem[{{LSST Science Collaborations} \& {LSST
  Project}(2009)}]{2009arXiv0912.0201L}
{LSST Science Collaborations}, {LSST Project}, 2009, ArXiv e-prints
  (0912.0201), \url{http://www.lsst.org/lsst/scibook}

\bibitem[{{Ma} \& {Fry}(2000)}]{ma2000}
{Ma} C.-P., {Fry} J.~N., 2000, \apj, 543, 503

\bibitem[{{Madau} \& {Dickinson}(2014)}]{madau2014}
{Madau} P., {Dickinson} M., 2014, \araa, 52, 415

\bibitem[{{Mandelbaum} {et~al}\mbox{.}(2012){Mandelbaum}, {Hirata},
  {Leauthaud}, {Massey}, \& {Rhodes}}]{2012MNRAS.420.1518M}
{Mandelbaum} R., {Hirata} C.~M., {Leauthaud} A., {Massey} R.~J., {Rhodes} J.,
  2012, \mnras, 420, 1518

\bibitem[{{Mandelbaum} {et~al}\mbox{.}(2005){Mandelbaum}, {Hirata}, {Seljak},
  {Guzik}, {Padmanabhan}, {Blake}, {Blanton}, {Lupton}, \&
  {Brinkmann}}]{2005MNRAS.361.1287M}
{Mandelbaum} R. {et~al.}, 2005, \mnras, 361, 1287

\bibitem[{{Mandelbaum} {et~al}\mbox{.}(2006){Mandelbaum}, {Seljak}, {Cool},
  {Blanton}, {Hirata}, \& {Brinkmann}}]{mandelbaum2006b}
{Mandelbaum} R., {Seljak} U., {Cool} R.~J., {Blanton} M., {Hirata} C.~M.,
  {Brinkmann} J., 2006, \mnras, 372, 758

\bibitem[{Mandelbaum {et~al}\mbox{.}(2006)Mandelbaum, Seljak, Kauffmann,
  Hirata, \& Brinkmann}]{mandelbaum2006}
Mandelbaum R., Seljak U., Kauffmann G., Hirata C.~M., Brinkmann J., 2006,
  \mnras, 368, 715

\bibitem[{{Mandelbaum} {et~al}\mbox{.}(2013){Mandelbaum}, {Slosar}, {Baldauf},
  {Seljak}, {Hirata}, {Nakajima}, {Reyes}, \& {Smith}}]{mandelbaum2013}
{Mandelbaum} R., {Slosar} A., {Baldauf} T., {Seljak} U., {Hirata} C.~M.,
  {Nakajima} R., {Reyes} R., {Smith} R.~E., 2013, \mnras, 432, 1544

\bibitem[{Mandelbaum {et~al}\mbox{.}(2005)Mandelbaum, Tasitsiomi, Seljak,
  Kravtsov, \& Wechsler}]{mandelbaum2005}
Mandelbaum R., Tasitsiomi A., Seljak U., Kravtsov A.~V., Wechsler R.~H., 2005,
  \mnras, 362, 1451

\bibitem[{{McGaugh} {et~al}\mbox{.}(2010){McGaugh}, {Schombert}, {de Blok}, \&
  {Zagursky}}]{mcgaugh2010}
{McGaugh} S.~S., {Schombert} J.~M., {de Blok} W.~J.~G., {Zagursky} M.~J., 2010,
  \apjl, 708, L14

\bibitem[{Miyatake {et~al}\mbox{.}(2013)Miyatake, More, Mandelbaum, Takada,
  Spergel, Kneib, Schneider, Brinkmann, \& Brownstein}]{miyatake2013}
Miyatake H. {et~al.}, 2013, arXiv:1311.1480

\bibitem[{{Mo} {et~al}\mbox{.}(1996){Mo}, {Jing}, \& {White}}]{mo1996}
{Mo} H.~J., {Jing} Y.~P., {White} S.~D.~M., 1996, \mnras, 282, 1096

\bibitem[{{More} {et~al}\mbox{.}(2014){More}, {Miyatake}, {Mandelbaum},
  {Takada}, {Spergel}, {Brownstein}, \& {Schneider}}]{more2014}
{More} S., {Miyatake} H., {Mandelbaum} R., {Takada} M., {Spergel} D.,
  {Brownstein} J., {Schneider} D.~P., 2014, arXiv:1407.1856

\bibitem[{Moster {et~al}\mbox{.}(2010)Moster, Somerville, Maulbetsch, van~den
  Bosch, Macci\`{o}, Naab, \& Oser}]{moster2010}
Moster B.~P., Somerville R.~S., Maulbetsch C., van~den Bosch F.~C., Macci\`{o}
  A.~V., Naab T., Oser L., 2010, \apj, 710, 903

\bibitem[{{Murray} {et~al}\mbox{.}(2013){Murray}, {Power}, \&
  {Robotham}}]{murray2013}
{Murray} S.~G., {Power} C., {Robotham} A.~S.~G., 2013, \mnras, 434, L61

\bibitem[{{Nakajima} {et~al}\mbox{.}(2012){Nakajima}, {Mandelbaum}, {Seljak},
  {Cohn}, {Reyes}, \& {Cool}}]{2012MNRAS.420.3240N}
{Nakajima} R., {Mandelbaum} R., {Seljak} U., {Cohn} J.~D., {Reyes} R., {Cool}
  R., 2012, \mnras, 420, 3240

\bibitem[{Navarro {et~al}\mbox{.}(1997)Navarro, Frenk, \& White}]{navarro1997}
Navarro J.~F., Frenk C.~S., White S. D.~M., 1997, \apj, 490, 493

\bibitem[{{Newman} {et~al}\mbox{.}(2015){Newman}, {Ellis}, \&
  {Treu}}]{newman2015}
{Newman} A.~B., {Ellis} R.~S., {Treu} T., 2015, arXiv:1503.05282

\bibitem[{Newman {et~al}\mbox{.}(2013)Newman, Treu, Ellis, Sand, Nipoti,
  Richard, \& Jullo}]{newman2013}
Newman A.~B., Treu T., Ellis R.~S., Sand D.~J., Nipoti C., Richard J., Jullo
  E., 2013, \apj, 765, 24

\bibitem[{Norberg {et~al}\mbox{.}(2009)Norberg, Baugh, Gazta\~{n}aga, \&
  Croton}]{norberg2009}
Norberg P., Baugh C.~M., Gazta\~{n}aga E., Croton D.~J., 2009, \mnras, 396, 19

\bibitem[{{Norman} \& {Bryan}(1999)}]{norman1999}
{Norman} M.~L., {Bryan} G.~L., 1999, in Astrophysics and Space Science Library,
  Vol. 240, Numerical Astrophysics, {Miyama} S.~M., {Tomisaka} K., {Hanawa} T.,
  eds., p.~19

\bibitem[{{Oppenheimer} \& {Dav{\'e}}(2006)}]{oppenheimer2006}
{Oppenheimer} B.~D., {Dav{\'e}} R., 2006, \mnras, 373, 1265

\bibitem[{Oppenheimer \& Dav\'{e}(2008)}]{oppenheimer2008}
Oppenheimer B.~D., Dav\'{e} R., 2008, \mnras, 387, 577

\bibitem[{{O'Shea} {et~al}\mbox{.}(2004){O'Shea}, {Bryan}, {Bordner}, {Norman},
  {Abel}, {Harkness}, \& {Kritsuk}}]{oshea2004}
{O'Shea} B.~W., {Bryan} G., {Bordner} J., {Norman} M.~L., {Abel} T., {Harkness}
  R., {Kritsuk} A., 2004, astro-ph/0403044

\bibitem[{{Padmanabhan} {et~al}\mbox{.}(2008){Padmanabhan}, {Schlegel},
  {Finkbeiner}, {Barentine}, {Blanton}, {Brewington}, {Gunn}, {Harvanek},
  {Hogg}, {Ivezi{\'c}}, {Johnston}, {Kent}, {Kleinman}, {Knapp}, {Krzesinski},
  {Long}, {Neilsen}, {Nitta}, {Loomis}, {Lupton}, {Roweis}, {Snedden},
  {Strauss}, \& {Tucker}}]{padmanabhan2008}
{Padmanabhan} N. {et~al.}, 2008, \apj, 674, 1217

\bibitem[{Panter {et~al}\mbox{.}(2007)Panter, Jimenez, Heavens, \&
  Charlot}]{panter2007}
Panter B., Jimenez R., Heavens A.~F., Charlot S., 2007, \mnras, 378, 1550

\bibitem[{{Parejko} {et~al}\mbox{.}(2013){Parejko}, {Sunayama}, {Padmanabhan},
  {Wake}, {Berlind}, {Bizyaev}, {Blanton}, {Bolton}, {van den Bosch},
  {Brinkmann}, {Brownstein}, {da Costa}, {Eisenstein}, {Guo}, {Kazin}, {Maia},
  {Malanushenko}, {Maraston}, {McBride}, {Nichol}, {Oravetz}, {Pan},
  {Percival}, {Prada}, {Ross}, {Ross}, {Schlegel}, {Schneider}, {Simmons},
  {Skibba}, {Tinker}, {Tojeiro}, {Weaver}, {Wetzel}, {White}, {Weinberg},
  {Thomas}, {Zehavi}, \& {Zheng}}]{parejko2013}
{Parejko} J.~K. {et~al.}, 2013, \mnras, 429, 98

\bibitem[{{Peacock} \& {Smith}(2000)}]{peacock2000}
{Peacock} J.~A., {Smith} R.~E., 2000, \mnras, 318, 1144

\bibitem[{{Peebles} \& {Hauser}(1974)}]{peebles1974}
{Peebles} P.~J.~E., {Hauser} M.~G., 1974, \apjs, 28, 19

\bibitem[{Peng {et~al}\mbox{.}(2012)Peng, Lilly, Renzini, \&
  Carollo}]{peng2012}
Peng Y.-j., Lilly S.~J., Renzini A., Carollo M., 2012, \apj, 757, 4

\bibitem[{{Planck Collaboration} {et~al}\mbox{.}(2015){Planck Collaboration},
  {Ade}, {Aghanim}, {Arnaud}, {Ashdown}, {Aumont}, {Baccigalupi}, {Banday},
  {Barreiro}, {Bartlett}, \& et~al.}]{planck2015}
{Planck Collaboration} {et~al.}, 2015, arXiv:1502.01589

\bibitem[{Prada {et~al}\mbox{.}(2012)Prada, Klypin, Cuesta, {Betancort-Rijo},
  \& Primack}]{prada2012}
Prada F., Klypin A.~A., Cuesta A.~J., {Betancort-Rijo} J.~E., Primack J., 2012,
  \mnras, 3206

\bibitem[{{Press} \& {Schechter}(1974)}]{press1974}
{Press} W.~H., {Schechter} P., 1974, \apj, 187, 425

\bibitem[{{Putman} {et~al}\mbox{.}(2012){Putman}, {Peek}, \&
  {Joung}}]{putman2012}
{Putman} M.~E., {Peek} J.~E.~G., {Joung} M.~R., 2012, \araa, 50, 491

\bibitem[{Reddick {et~al}\mbox{.}(2013)Reddick, Wechsler, Tinker, \&
  Behroozi}]{reddick2013}
Reddick R.~M., Wechsler R.~H., Tinker J.~L., Behroozi P.~S., 2013, \apj, 771,
  30

\bibitem[{{Reyes} {et~al}\mbox{.}(2012){Reyes}, {Mandelbaum}, {Gunn},
  {Nakajima}, {Seljak}, \& {Hirata}}]{2012MNRAS.425.2610R}
{Reyes} R., {Mandelbaum} R., {Gunn} J.~E., {Nakajima} R., {Seljak} U., {Hirata}
  C.~M., 2012, \mnras, 425, 2610

\bibitem[{{Rodr{\'{\i}}guez-Puebla}
  {et~al}\mbox{.}(2015){Rodr{\'{\i}}guez-Puebla}, {Avila-Reese}, {Yang},
  {Foucaud}, {Drory}, \& {Jing}}]{rodriguezpuebla2015}
{Rodr{\'{\i}}guez-Puebla} A., {Avila-Reese} V., {Yang} X., {Foucaud} S.,
  {Drory} N., {Jing} Y.~P., 2015, \apj, 799, 130

\bibitem[{Salim {et~al}\mbox{.}(2007)Salim, Rich, Charlot, Brinchmann, Johnson,
  Schiminovich, Seibert, Mallery, Heckman, Forster, Friedman, Martin,
  Morrissey, Neff, Small, Wyder, Bianchi, Donas, Lee, Madore, Milliard, Szalay,
  Welsh, \& Yi}]{salim2007}
Salim S. {et~al.}, 2007, \apjs, 173, 267

\bibitem[{{Schmidt} {et~al}\mbox{.}(2009){Schmidt}, {Rozo}, {Dodelson}, {Hui},
  \& {Sheldon}}]{2009PhRvL.103e1301S}
{Schmidt} F., {Rozo} E., {Dodelson} S., {Hui} L., {Sheldon} E., 2009, \prl,
  103, 051301

\bibitem[{{Schneider} \& {Bridle}(2010)}]{2010MNRAS.402.2127S}
{Schneider} M.~D., {Bridle} S., 2010, \mnras, 402, 2127

\bibitem[{{Scoccimarro} {et~al}\mbox{.}(2001){Scoccimarro}, {Sheth}, {Hui}, \&
  {Jain}}]{scoccimarro2001}
{Scoccimarro} R., {Sheth} R.~K., {Hui} L., {Jain} B., 2001, \apj, 546, 20

\bibitem[{{Scoville} {et~al}\mbox{.}(2007){Scoville}, {Aussel}, {Brusa},
  {Capak}, {Carollo}, {Elvis}, {Giavalisco}, {Guzzo}, {Hasinger}, {Impey},
  {Kneib}, {LeFevre}, {Lilly}, {Mobasher}, {Renzini}, {Rich}, {Sanders},
  {Schinnerer}, {Schminovich}, {Shopbell}, {Taniguchi}, \&
  {Tyson}}]{scoville2007}
{Scoville} N. {et~al.}, 2007, \apjs, 172, 1

\bibitem[{{Seljak}(2000)}]{seljak2000}
{Seljak} U., 2000, \mnras, 318, 203

\bibitem[{{Semboloni} {et~al}\mbox{.}(2013){Semboloni}, {Hoekstra}, \&
  {Schaye}}]{2013MNRAS.434..148S}
{Semboloni} E., {Hoekstra} H., {Schaye} J., 2013, \mnras, 434, 148

\bibitem[{{Semboloni} {et~al}\mbox{.}(2011){Semboloni}, {Hoekstra}, {Schaye},
  {van Daalen}, \& {McCarthy}}]{2011MNRAS.417.2020S}
{Semboloni} E., {Hoekstra} H., {Schaye} J., {van Daalen} M.~P., {McCarthy}
  I.~G., 2011, \mnras, 417, 2020

\bibitem[{Shankar {et~al}\mbox{.}(2014)Shankar, Guo, Bouillot, Rettura, Meert,
  Buchan, Kravtsov, Bernardi, Sheth, Vikram, Marchesini, Behroozi, Zheng,
  Maraston, Ascaso, Lemaux, Capozzi, {Huertas-Company}, Gal, Lubin, Conselice,
  Carollo, \& Cattaneo}]{shankar2014}
Shankar F. {et~al.}, 2014, The Astrophysical Journal Letters, 797, L27

\bibitem[{Shankar {et~al}\mbox{.}(2006)Shankar, Lapi, Salucci, De~Zotti, \&
  Danese}]{shankar2006}
Shankar F., Lapi A., Salucci P., De~Zotti G., Danese L., 2006, \apj, 643, 14

\bibitem[{{Sheldon} {et~al}\mbox{.}(2004){Sheldon}, {Johnston}, {Frieman},
  {Scranton}, {McKay}, {Connolly}, {Budav{\' a}ri}, {Zehavi}, {Bahcall},
  {Brinkmann}, \& {Fukugita}}]{2004AJ....127.2544S}
{Sheldon} E.~S. {et~al.}, 2004, \aj, 127, 2544

\bibitem[{{Sheth} \& {Tormen}(1999)}]{sheth1999}
{Sheth} R.~K., {Tormen} G., 1999, \mnras, 308, 119

\bibitem[{{Sheth} \& {Tormen}(2004)}]{sheth2004}
{Sheth} R.~K., {Tormen} G., 2004, \mnras, 350, 1385

\bibitem[{{Shull} {et~al}\mbox{.}(2012){Shull}, {Smith}, \&
  {Danforth}}]{shull2012}
{Shull} J.~M., {Smith} B.~D., {Danforth} C.~W., 2012, \apj, 759, 23

\bibitem[{{Simet} \& {Mandelbaum}(2015)}]{2015MNRAS.449.1259S}
{Simet} M., {Mandelbaum} R., 2015, \mnras, 449, 1259

\bibitem[{{Singh} {et~al}\mbox{.}(2014){Singh}, {Mandelbaum}, \&
  {More}}]{2014arXiv1411.1755S}
{Singh} S., {Mandelbaum} R., {More} S., 2014, arXiv:1411.1755

\bibitem[{Smith {et~al}\mbox{.}(2003)Smith, Peacock, Jenkins, White, Frenk,
  Pearce, Thomas, Efstathiou, \& Couchman}]{smith2003}
Smith R.~E. {et~al.}, 2003, \mnras, 341, 1311{\textendash}1332

\bibitem[{{Somerville} {et~al}\mbox{.}(2008){Somerville}, {Hopkins}, {Cox},
  {Robertson}, \& {Hernquist}}]{somerville2008}
{Somerville} R.~S., {Hopkins} P.~F., {Cox} T.~J., {Robertson} B.~E.,
  {Hernquist} L., 2008, \mnras, 391, 481

\bibitem[{{Spergel} {et~al}\mbox{.}(2015){Spergel}, {Gehrels}, {Baltay},
  {Bennett}, {Breckinridge}, {Donahue}, {Dressler}, {Gaudi}, {Greene}, {Guyon},
  {Hirata}, {Kalirai}, {Kasdin}, {Macintosh}, {Moos}, {Perlmutter}, {Postman},
  {Rauscher}, {Rhodes}, {Wang}, {Weinberg}, {Benford}, {Hudson}, {Jeong},
  {Mellier}, {Traub}, {Yamada}, {Capak}, {Colbert}, {Masters}, {Penny},
  {Savransky}, {Stern}, {Zimmerman}, {Barry}, {Bartusek}, {Carpenter}, {Cheng},
  {Content}, {Dekens}, {Demers}, {Grady}, {Jackson}, {Kuan}, {Kruk}, {Melton},
  {Nemati}, {Parvin}, {Poberezhskiy}, {Peddie}, {Ruffa}, {Wallace}, {Whipple},
  {Wollack}, \& {Zhao}}]{spergel2015}
{Spergel} D. {et~al.}, 2015, arXiv:1503.03757

\bibitem[{{Springel} {et~al}\mbox{.}(2005){Springel}, {White}, {Jenkins},
  {Frenk}, {Yoshida}, {Gao}, {Navarro}, {Thacker}, {Croton}, {Helly},
  {Peacock}, {Cole}, {Thomas}, {Couchman}, {Evrard}, {Colberg}, \&
  {Pearce}}]{springel2005}
{Springel} V. {et~al.}, 2005, \nat, 435, 629

\bibitem[{Strauss {et~al}\mbox{.}(2002)Strauss, Weinberg, Lupton, Narayanan,
  Annis, Bernardi, Blanton, Burles, Connolly, Dalcanton, Doi, Eisenstein,
  Frieman, Fukugita, Gunn, Ivezi\'{c}, Kent, Kim, Knapp, Kron, Munn, Newberg,
  Nichol, Okamura, Quinn, Richmond, Schlegel, Shimasaku, {SubbaRao}, Szalay,
  Vanden~Berk, Vogeley, Yanny, Yasuda, York, \& Zehavi}]{strauss2002}
Strauss M.~A. {et~al.}, 2002, \aj, 124, 1810

\bibitem[{{Swanson} {et~al}\mbox{.}(2008){Swanson}, {Tegmark}, {Hamilton}, \&
  {Hill}}]{swanson2008}
{Swanson} M.~E.~C., {Tegmark} M., {Hamilton} A.~J.~S., {Hill} J.~C., 2008,
  \mnras, 387, 1391

\bibitem[{{Takada} {et~al}\mbox{.}(2014){Takada}, {Ellis}, {Chiba}, {Greene},
  {Aihara}, {Arimoto}, {Bundy}, {Cohen}, {Dor{\'e}}, {Graves}, {Gunn},
  {Heckman}, {Hirata}, {Ho}, {Kneib}, {F{\`e}vre}, {Lin}, {More}, {Murayama},
  {Nagao}, {Ouchi}, {Seiffert}, {Silverman}, {Sodr{\'e}}, {Spergel}, {Strauss},
  {Sugai}, {Suto}, {Takami}, \& {Wyse}}]{takada2012}
{Takada} M. {et~al.}, 2014, \pasj, 66, 1

\bibitem[{{Takahashi} {et~al}\mbox{.}(2012){Takahashi}, {Sato}, {Nishimichi},
  {Taruya}, \& {Oguri}}]{takahashi2012}
{Takahashi} R., {Sato} M., {Nishimichi} T., {Taruya} A., {Oguri} M., 2012,
  \apj, 761, 152

\bibitem[{{Taylor} {et~al}\mbox{.}(2015){Taylor}, {Hopkins}, {Baldry},
  {Bland-Hawthorn}, {Brown}, {Colless}, {Driver}, {Norberg}, {Robotham},
  {Alpaslan}, {Brough}, {Cluver}, {Gunawardhana}, {Kelvin}, {Liske},
  {Conselice}, {Croom}, {Foster}, {Jarrett}, {Lara-Lopez}, \&
  {Loveday}}]{taylor2015}
{Taylor} E.~N. {et~al.}, 2015, \mnras, 446, 2144

\bibitem[{{Teyssier}(2002)}]{teyssier2002}
{Teyssier} R., 2002, \aap, 385, 337

\bibitem[{Tinker {et~al}\mbox{.}(2008)Tinker, Kravtsov, Klypin, Abazajian,
  Warren, Yepes, Gottl\"{o}ber, \& Holz}]{tinker2008}
Tinker J., Kravtsov A.~V., Klypin A., Abazajian K., Warren M., Yepes G.,
  Gottl\"{o}ber S., Holz D.~E., 2008, \apj, 688, 709{\textendash}728

\bibitem[{{Tinker} {et~al}\mbox{.}(2013){Tinker}, {Leauthaud}, {Bundy},
  {George}, {Behroozi}, {Massey}, {Rhodes}, \& {Wechsler}}]{tinker2013}
{Tinker} J.~L., {Leauthaud} A., {Bundy} K., {George} M.~R., {Behroozi} P.,
  {Massey} R., {Rhodes} J., {Wechsler} R.~H., 2013, \apj, 778, 93

\bibitem[{Tinker {et~al}\mbox{.}(2010)Tinker, Robertson, Kravtsov, Klypin,
  Warren, Yepes, \& Gottl\"{o}ber}]{tinker2010}
Tinker J.~L., Robertson B.~E., Kravtsov A.~V., Klypin A., Warren M.~S., Yepes
  G., Gottl\"{o}ber S., 2010, \apj, 724, 878

\bibitem[{Tinker {et~al}\mbox{.}(2005)Tinker, Weinberg, Zheng, \&
  Zehavi}]{tinker2005}
Tinker J.~L., Weinberg D.~H., Zheng Z., Zehavi I., 2005, \apj, 631, 41

\bibitem[{Vale \& Ostriker(2004)}]{vale2004}
Vale A., Ostriker J.~P., 2004, \mnras, 353, 189

\bibitem[{van Daalen \& Schaye(2015)}]{vandaalen2015}
van Daalen M.~P., Schaye J., 2015, arXiv:1501.05950

\bibitem[{van~den Bosch {et~al}\mbox{.}(2013)van~den Bosch, More, Cacciato, Mo,
  \& Yang}]{vandenbosch2013}
van~den Bosch F.~C., More S., Cacciato M., Mo H., Yang X., 2013, \mnras, 430,
  725

\bibitem[{van~den Bosch {et~al}\mbox{.}(2005)van~den Bosch, Tormen, \&
  Giocoli}]{vandenbosch2005}
van~den Bosch F.~C., Tormen G., Giocoli C., 2005, \mnras, 359, 1029

\bibitem[{{van den Bosch} {et~al}\mbox{.}(2007){van den Bosch}, {Yang}, {Mo},
  {Weinmann}, {Macci{\`o}}, {More}, {Cacciato}, {Skibba}, \&
  {Kang}}]{vandenbosch2007}
{van den Bosch} F.~C. {et~al.}, 2007, \mnras, 376, 841

\bibitem[{van Dokkum \& Conroy(2010)}]{vandokkum2010}
van Dokkum P.~G., Conroy C., 2010, Nature, 468, 940

\bibitem[{Velander {et~al}\mbox{.}(2014)Velander, van Uitert, Hoekstra, Coupon,
  Erben, Heymans, Hildebrandt, Kitching, Mellier, Miller, Van~Waerbeke,
  Bonnett, Fu, Giodini, Hudson, Kuijken, Rowe, Schrabback, \&
  Semboloni}]{velander2014}
Velander M. {et~al.}, 2014, \mnras, 437, 2111

\bibitem[{{Velliscig} {et~al}\mbox{.}(2014){Velliscig}, {van Daalen}, {Schaye},
  {McCarthy}, {Cacciato}, {Le Brun}, \& {Dalla Vecchia}}]{velliscig2014}
{Velliscig} M., {van Daalen} M.~P., {Schaye} J., {McCarthy} I.~G., {Cacciato}
  M., {Le Brun} A.~M.~C., {Dalla Vecchia} C., 2014, \mnras, 442, 2641

\bibitem[{{Vogelsberger} {et~al}\mbox{.}(2013){Vogelsberger}, {Genel},
  {Sijacki}, {Torrey}, {Springel}, \& {Hernquist}}]{vogelsberger2013}
{Vogelsberger} M., {Genel} S., {Sijacki} D., {Torrey} P., {Springel} V.,
  {Hernquist} L., 2013, \mnras, 436, 3031

\bibitem[{{Vogelsberger} {et~al}\mbox{.}(2014){Vogelsberger}, {Genel},
  {Springel}, {Torrey}, {Sijacki}, {Xu}, {Snyder}, {Bird}, {Nelson}, \&
  {Hernquist}}]{vogelsberger2014}
{Vogelsberger} M. {et~al.}, 2014, \nat, 509, 177

\bibitem[{{von der Linden} {et~al}\mbox{.}(2007){von der Linden}, {Best},
  {Kauffmann}, \& {White}}]{vonderlinden2007}
{von der Linden} A., {Best} P.~N., {Kauffmann} G., {White} S.~D.~M., 2007,
  \mnras, 379, 867

\bibitem[{{Wang} {et~al}\mbox{.}(2007){Wang}, {Mo}, \& {Jing}}]{wang2007}
{Wang} H.~Y., {Mo} H.~J., {Jing} Y.~P., 2007, \mnras, 375, 633

\bibitem[{{Wang} {et~al}\mbox{.}(2013){Wang}, {Weinmann}, {De Lucia}, \&
  {Yang}}]{wang2013}
{Wang} L., {Weinmann} S.~M., {De Lucia} G., {Yang} X., 2013, \mnras, 433, 515

\bibitem[{Wang {et~al}\mbox{.}(2014)Wang, Sales, Henriques, \&
  White}]{wang2014}
Wang W., Sales L.~V., Henriques B. M.~B., White S. D.~M., 2014, \mnras, 442,
  1363

\bibitem[{Watson {et~al}\mbox{.}(2012)Watson, Berlind, {McBride}, Hogg, \&
  Jiang}]{watson2012}
Watson D.~F., Berlind A.~A., {McBride} C.~K., Hogg D.~W., Jiang T., 2012, \apj,
  749, 83

\bibitem[{{Watson} {et~al}\mbox{.}(2013){Watson}, {Iliev}, {D'Aloisio},
  {Knebe}, {Shapiro}, \& {Yepes}}]{watson2013}
{Watson} W.~A., {Iliev} I.~T., {D'Aloisio} A., {Knebe} A., {Shapiro} P.~R.,
  {Yepes} G., 2013, \mnras, 433, 1230

\bibitem[{{Wechsler} {et~al}\mbox{.}(2006){Wechsler}, {Zentner}, {Bullock},
  {Kravtsov}, \& {Allgood}}]{wechsler2006}
{Wechsler} R.~H., {Zentner} A.~R., {Bullock} J.~S., {Kravtsov} A.~V., {Allgood}
  B., 2006, \apj, 652, 71

\bibitem[{{Weinberg} {et~al}\mbox{.}(2004){Weinberg}, {Dav{\'e}}, {Katz}, \&
  {Hernquist}}]{weinberg2004}
{Weinberg} D.~H., {Dav{\'e}} R., {Katz} N., {Hernquist} L., 2004, \apj, 601, 1

\bibitem[{{Yang} {et~al}\mbox{.}(2003){Yang}, {Mo}, \& {van den
  Bosch}}]{yang2003}
{Yang} X., {Mo} H.~J., {van den Bosch} F.~C., 2003, \mnras, 339, 1057

\bibitem[{{Yang} {et~al}\mbox{.}(2009){Yang}, {Mo}, \& {van den
  Bosch}}]{yang2009}
{Yang} X., {Mo} H.~J., {van den Bosch} F.~C., 2009, \apj, 695, 900

\bibitem[{Yang {et~al}\mbox{.}(2007)Yang, Mo, van~den Bosch, Pasquali, Li, \&
  Barden}]{yang2007}
Yang X., Mo H.~J., van~den Bosch F.~C., Pasquali A., Li C., Barden M., 2007,
  \apj, 671, 153

\bibitem[{Yoo \& Seljak(2012)}]{yoo2012}
Yoo J., Seljak U., 2012, \prd, 86, 083504

\bibitem[{Yoo {et~al}\mbox{.}(2006)Yoo, Tinker, Weinberg, Zheng, Katz, \&
  Dav\'{e}}]{yoo2006}
Yoo J., Tinker J.~L., Weinberg D.~H., Zheng Z., Katz N., Dav\'{e} R., 2006,
  \apj, 652, 26

\bibitem[{{York} {et~al}\mbox{.}(2000){York}, {Adelman}, {Anderson},
  {Anderson}, {Annis}, {Bahcall}, {Bakken}, {Barkhouser}, {Bastian}, {Berman},
  {Boroski}, {Bracker}, {Briegel}, {Briggs}, {Brinkmann}, {Brunner}, {Burles},
  {Carey}, {Carr}, {Castander}, {Chen}, {Colestock}, {Connolly}, {Crocker},
  {Csabai}, {Czarapata}, {Davis}, {Doi}, {Dombeck}, {Eisenstein}, {Ellman},
  {Elms}, {Evans}, {Fan}, {Federwitz}, {Fiscelli}, {Friedman}, {Frieman},
  {Fukugita}, {Gillespie}, {Gunn}, {Gurbani}, {de Haas}, {Haldeman}, {Harris},
  {Hayes}, {Heckman}, {Hennessy}, {Hindsley}, {Holm}, {Holmgren}, {Huang},
  {Hull}, {Husby}, {Ichikawa}, {Ichikawa}, {Ivezi{\'c}}, {Kent}, {Kim},
  {Kinney}, {Klaene}, {Kleinman}, {Kleinman}, {Knapp}, {Korienek}, {Kron},
  {Kunszt}, {Lamb}, {Lee}, {Leger}, {Limmongkol}, {Lindenmeyer}, {Long},
  {Loomis}, {Loveday}, {Lucinio}, {Lupton}, {MacKinnon}, {Mannery}, {Mantsch},
  {Margon}, {McGehee}, {McKay}, {Meiksin}, {Merelli}, {Monet}, {Munn},
  {Narayanan}, {Nash}, {Neilsen}, {Neswold}, {Newberg}, {Nichol}, {Nicinski},
  {Nonino}, {Okada}, {Okamura}, {Ostriker}, {Owen}, {Pauls}, {Peoples},
  {Peterson}, {Petravick}, {Pier}, {Pope}, {Pordes}, {Prosapio},
  {Rechenmacher}, {Quinn}, {Richards}, {Richmond}, {Rivetta}, {Rockosi},
  {Ruthmansdorfer}, {Sandford}, {Schlegel}, {Schneider}, {Sekiguchi}, {Sergey},
  {Shimasaku}, {Siegmund}, {Smee}, {Smith}, {Snedden}, {Stone}, {Stoughton},
  {Strauss}, {Stubbs}, {SubbaRao}, {Szalay}, {Szapudi}, {Szokoly}, {Thakar},
  {Tremonti}, {Tucker}, {Uomoto}, {Vanden Berk}, {Vogeley}, {Waddell}, {Wang},
  {Watanabe}, {Weinberg}, {Yanny}, {Yasuda}, \& {SDSS
  Collaboration}}]{york2000}
{York} D.~G. {et~al.}, 2000, \aj, 120, 1579

\bibitem[{{Zehavi} {et~al}\mbox{.}(2011){Zehavi}, {Zheng}, {Weinberg},
  {Blanton}, {Bahcall}, {Berlind}, {Brinkmann}, {Frieman}, {Gunn}, {Lupton},
  {Nichol}, {Percival}, {Schneider}, {Skibba}, {Strauss}, {Tegmark}, \&
  {York}}]{zehavi2011}
{Zehavi} I. {et~al.}, 2011, \apj, 736, 59

\bibitem[{Zehavi {et~al}\mbox{.}(2005)Zehavi, Zheng, Weinberg, Frieman,
  Berlind, Blanton, Scoccimarro, Sheth, Strauss, Kayo, Suto, Fukugita,
  Nakamura, Bahcall, Brinkmann, Gunn, Hennessy, Ivezi\'{c}, Knapp, Loveday,
  Meiksin, Schlegel, Schneider, Szapudi, Tegmark, Vogeley, York, \&
  Collaboration}]{zehavi2005}
Zehavi I. {et~al.}, 2005, \apj, 630, 1

\bibitem[{{Zentner} {et~al}\mbox{.}(2014){Zentner}, {Hearin}, \& {van den
  Bosch}}]{zentner2014}
{Zentner} A.~R., {Hearin} A.~P., {van den Bosch} F.~C., 2014, \mnras, 443, 3044

\bibitem[{{Zentner} {et~al}\mbox{.}(2013){Zentner}, {Semboloni}, {Dodelson},
  {Eifler}, {Krause}, \& {Hearin}}]{2013PhRvD..87d3509Z}
{Zentner} A.~R., {Semboloni} E., {Dodelson} S., {Eifler} T., {Krause} E.,
  {Hearin} A.~P., 2013, \prd, 87, 043509

\bibitem[{{Zhang} {et~al}\mbox{.}(2013){Zhang}, {Zhang}, {Yang}, \&
  {Cui}}]{zhang2013}
{Zhang} Y., {Zhang} P., {Yang} X., {Cui} W., 2013, \prd, 87, 023521

\bibitem[{Zhao {et~al}\mbox{.}(2009)Zhao, Jing, Mo, \& B\"{o}rner}]{zhao2009}
Zhao D.~H., Jing Y.~P., Mo H.~J., B\"{o}rner G., 2009, \apj, 707,
  354{\textendash}369

\bibitem[{Zheng {et~al}\mbox{.}(2005)Zheng, Berlind, Weinberg, Benson, Baugh,
  Cole, Dav\'{e}, Frenk, Katz, \& Lacey}]{zheng2005}
Zheng Z. {et~al.}, 2005, \apj, 633, 791

\bibitem[{Zheng \& Weinberg(2007)}]{zheng2007}
Zheng Z., Weinberg D.~H., 2007, \apj, 659, 1

\bibitem[{{Zhu} {et~al}\mbox{.}(2006){Zhu}, {Zheng}, {Lin}, {Jing}, {Kang}, \&
  {Gao}}]{zhu2006}
{Zhu} G., {Zheng} Z., {Lin} W.~P., {Jing} Y.~P., {Kang} X., {Gao} L., 2006,
  \apjl, 639, L5

\bibitem[{{Zu} {et~al}\mbox{.}(2014){Zu}, {Weinberg}, {Rozo}, {Sheldon},
  {Tinker}, \& {Becker}}]{zu2014}
{Zu} Y., {Weinberg} D.~H., {Rozo} E., {Sheldon} E.~S., {Tinker} J.~L., {Becker}
  M.~R., 2014, \mnras, 439, 1628

\bibitem[{{Zu} {et~al}\mbox{.}(2008){Zu}, {Zheng}, {Zhu}, \& {Jing}}]{zu2008}
{Zu} Y., {Zheng} Z., {Zhu} G., {Jing} Y.~P., 2008, \apj, 686, 41

\end{thebibliography}

}

\appendix

\section{Small-scale lensing cutoffs}\label{app:smallscale}

In this appendix, we describe our choice of minimum scale for modelling of the lensing signals.
Our estimator  for $\Delta\Sigma$, Eq.~\eqref{eq:ds-estimator}, is designed such that our signal
includes a correction for the zero shear contributed by ``source'' galaxies that are at the lens
redshift.  This correction is often written separately as a ``boost factor''
\citep[e.g.,][]{2004AJ....127.2544S,2005MNRAS.361.1287M},
\begin{equation}
B(r_p) = \frac{\sum_{ls} w_{ls}(r_p)}{\sum_{rs} w_{rs}(r_p)},
\end{equation}
where the summed weights are over lens-source and random lens-source pairs.  $B(r_p)=1$ if there are
no galaxies associated with the lens that are included as sources and if lensing magnification is
negligible.  Generally, since both clustering and magnification are large at small $r_p$, $B(r_p)$
is a monotonically decreasing function.

As shown in \cite{2005MNRAS.361.1287M}, small-scale systematics due to deblending and sky
subtraction can appear as an {\em increase} in $B(r_p)$ on the smallest scales, followed by the
expected decrease on scales where those systematics no longer operate.  In the decreasing region, we
know that source detection is no longer 100\% effective, but what we do not know is the actual level
of inefficiency, since $B(r_p)$ gives a single constraint on two completely generate quantities
(amount of contamination by physically-associated galaxies, and inefficiency in the detection of
both real sources and physically-associated ones).  To properly correct for physically-associated
sources, we would need to know both of these quantities.  We would also have to know how the shear
estimates and photo-$z$ estimates of the sources in those regions may have been modified due to the
systematic that is causing difficulties with source detection.  Due to the lack of sufficient
information to model the signal on scales where $B(r_p)$ indicates small-scale systematics, we do
not plot or attempt to model $\Delta\Sigma$ there.  The quoted minimum scales used for the lensing
modelling comes from this consideration.

There are two additional systematics that could, in principle, operate on small scales above our
minimum cutoff.  The first systematic is the intrinsic alignment of galaxy shapes, pointing
coherently in the radial direction with respect to the lens.  While intrinsic alignments of bright
red galaxies are well-established in the literature \citep[e.g., most
recently,][]{2014arXiv1411.1755S}, attempts to measure any radial alignments of faint galaxies in
mixed blue and red source populations with respect to the positions of nearby bright galaxies have
thus far only resulted in null detections \citep{BMS+12} with relatively tight upper limits.  Hence
we do not consider this to be an important systematic for this work.

Lensing magnification has also been considered as a systematic since, if present, it means that our
normalization by $\sum_{rs} w_{rs}$ to correct for physically-associated sources is incorrect.  As
discussed in \citet{2005MNRAS.361.1287M} and \citet{2009PhRvL.103e1301S}, the amount of
magnification for a flux- and size-selected sample depends on both the slope of the number counts
near the flux limit and the slope of the apparent size distribution near its limit.  Moreover, it
depends on the slopes of these distributions weighted by whatever per-object weights are used for
the lensing analysis.  \citet{2015MNRAS.449.1259S} calculated the appropriate slopes for the source
catalogue used in this paper, and based on the numbers in their table 1, the ratio of observed
number counts with magnification to the un-magnified number counts is very near 1
($n_{\text{obs}}/n \approx 1-0.03\kappa$).  Hence, even though the best-fitting model for our
eight stellar mass samples have convergences that range from 0.0014 to 0.063 at $0.1h^{-1}$Mpc, the
maximum effect of magnification on the observed counts and therefore the size of the systematic
error in the boost factor and lensing calibration at that scale is $-0.2$ per cent.  We
therefore neglect this source of error.

\end{document}